\RequirePackage{fix-cm}
\documentclass[smallextended]{svjour3}       
\smartqed  
\usepackage{graphicx}
\usepackage{natbib}
\usepackage[title]{appendix}
\usepackage{amssymb}
\usepackage{amsmath}
\usepackage[colorlinks=true,citecolor=blue,urlcolor=blue,breaklinks]{hyperref}

\DeclareMathOperator*{\argmin}{arg\,min}
\newcommand{\aap}{{A\&A\ }}
\newcommand{\aaps}{{A\&AS\ }}
\newcommand{\apj}{{ApJ\ }}
\newcommand{\apjl}{{ApJL\ }}
\newcommand{\apjs}{{ApJS\ }}
\newcommand{\mnras}{{MNRAS\ }}

\newcommand{\solphys}{{Sol Phys\ }}
\newcommand{\ao}{{Appl Opt\ }}
\newcommand{\ssr}{{Space Sci Rev\ }}

\newcommand{\TP}{\textcolor{blue}{\rm{TP}}}
\newcommand{\TN}{\textcolor{red}{\rm{TN}}}
\newcommand{\FP}{\textcolor{red}{\rm{FP}}}
\newcommand{\FN}{\textcolor{blue}{\rm{FN}}}

\journalname{Living Reviews in Solar Physics}
\begin{document}

\title{Machine learning in solar physics}
\author{Andr\'es Asensio Ramos \and Mark C.~M.~Cheung \and Iulia Chifu \and Ricardo Gafeira}

\authorrunning{A. Asensio Ramos et al.} 

\institute{A. Asensio Ramos \at
        Instituto de Astrof\'{\i}sica de Canarias, 38205, 
        La Laguna, Tenerife, Spain \\        
        Departamento de Astrof\'{\i}sica, Universidad de La Laguna, 
        38205 La Laguna, Tenerife, Spain \\
        \email{aasensio@iac.es}
        \and
        M. C. M. Cheung \at
        CSIRO, Space \& Astronomy, PO Box 76, Epping, NSW 1710, Australia\\
        \email{mark.cheung@csiro.au}
        \and
        I. Chifu \at
        Institute for Astrophysics and Geophysics, University of G\"ottingen, Friedrich-Hund-Platz 1, 37077 G\"ottingen, Germany \\
        \email{iulia.chifu@uni-goettingen.de}
        \and
        R. Gafeira \at
        Instituto de Astrof\'{\i}sica e Ci\^encias do Espa\c{c}o, Departamento de F\'{\i}sica, Universidade de Coimbra, OGAUC, Rua do Observat\'orio s/n, 3040-004 Coimbra, Portugal \\
        \email{gafeira@uc.pt}
}

\date{Received: date / Accepted: date}

\maketitle

\begin{abstract}
The application of machine learning in solar physics has the potential to greatly 
enhance our understanding of the complex processes that take place in the atmosphere 
of the Sun. By using techniques such as deep learning, we are now in the position to analyze 
large amounts of data from solar observations and identify patterns and trends 
that may not have been apparent using traditional methods. This can help us
improve our understanding of explosive events like solar flares, which can have a strong effect on the
Earth environment. Predicting hazardous events on Earth becomes crucial for our technological society.
Machine learning can also improve our understanding of the inner workings of the sun itself
by allowing us to go deeper into the data and to propose more complex models
to explain them. Additionally, the use of machine learning can help to automate the analysis of 
solar data, reducing the need for manual labor and increasing the efficiency of 
research in this field.
\end{abstract}

\setcounter{tocdepth}{3}
\tableofcontents

\section{Introduction}
\label{intro}
Astrophysics, and solar physics in particular, is an observational science in which we cannot
change the experimental conditions, we simply observe. Therefore, the only way
of learning is by confronting observations with state-of-the-art theoretical modeling. The models are then tuned until the observations
are explained and conclusions are drawn from this comparison. As a consequence, our understanding 
of the universe is based on the availability of data. 

The amount of data available until the final decades of  
the 20th century was very reduced and could easily be stored in relatively standard
storage media, from notebooks, books or small computing centers. The scarcity of data forced researchers to use strongly informed generative models based
on our theoretical advances, with a heavy use of inductive biases\footnote{Set of explicit or implicit assumptions 
made by an algorithm to properly generalize what is learned from a finite set of observation into a 
general model.}. This is necessary to allow generalization
of the conclusions. From a probabilistic point of view, generative models are a way to describe
the joint probability $p(x,y)$, where $x$ are the observations and $y$ are the parameters
of the model. The ever-increasing quality of the observations allowed
researchers to propose more and more complex physical scenarios to be compared with observations. 

Solar physics is rapidly entering into the \emph{big data} era, an era dominated by the availability of
data, which cannot fit in current computers and have to be stored, in many cases, in
a distributed manner. The storage and access to this data is a technological challenge
and has not been completely solved in our field. For example, access to the curated
Solar Dynamics Observatory dataset of \cite{2019ApJS..242....7G} implies downloading 6.5 TB of data. 
Unless a dedicated connection is used, the transfer and local storage of all this 
data is hard. On the
other hand, the estimated amount of data in an excellent observing day for the multi-instrument
telescopes Daniel K. Inouye Solar Telescope (DKIST) and European Solar Telescope (EST)
easily reaches the PB regime.

\begin{figure}
    \centering
    \includegraphics[width=0.9\textwidth]{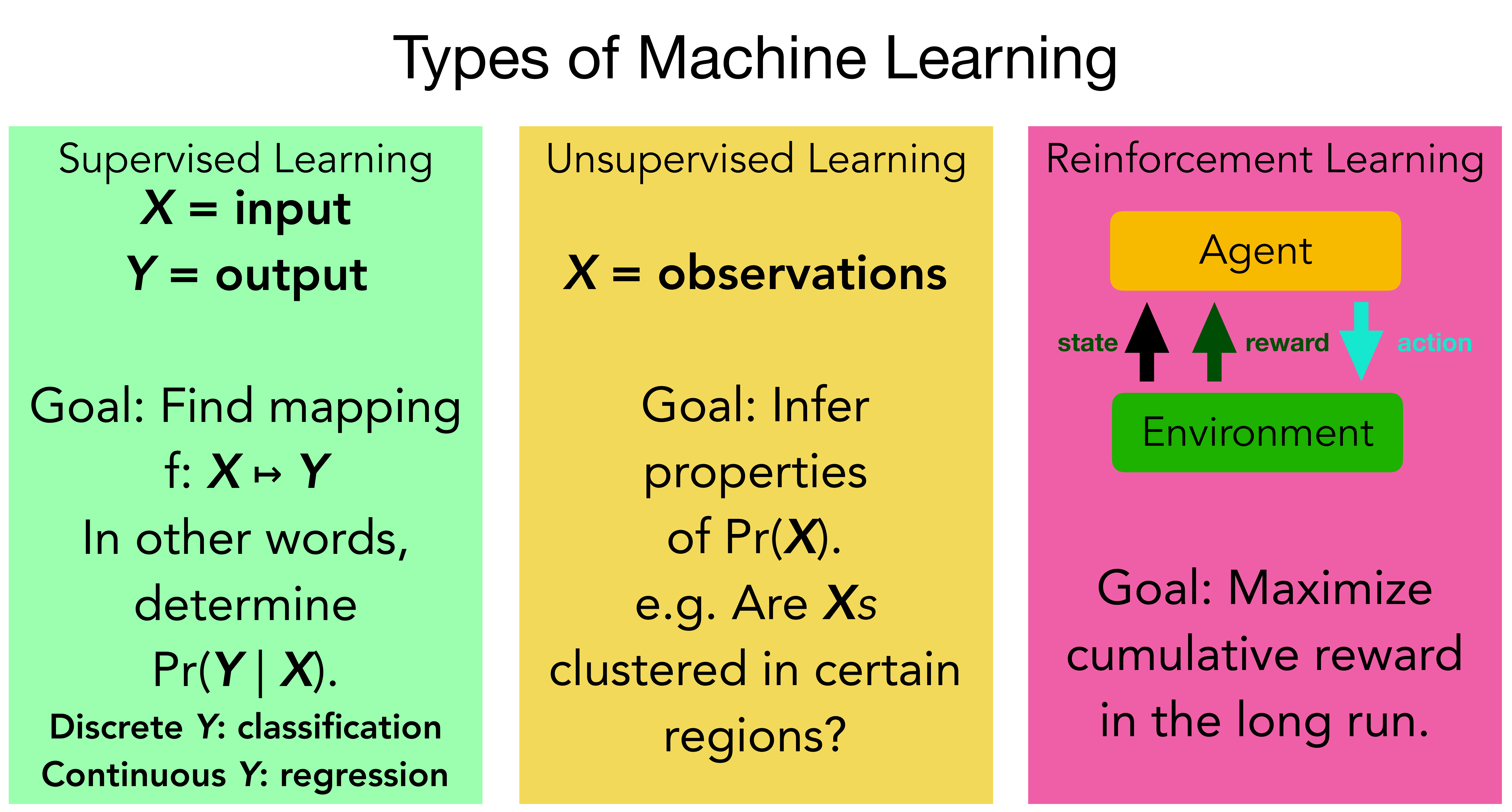}
    \caption{Categories of machine learning (ML): supervised, unsupervised and reinforcement learning. Supervised learning and unsupervised learning have deep roots in the field of statistical learning \citep[][]{Hastie:2009}, while
    reinforcement learning has strong connections with control theory.}
    \label{fig:ml_types}
\end{figure}

Having access to large datasets is not very useful unless one can extract 
relevant information out from them. Such large datasets have made it impossible to have people
looking at the data and search for interesting correlations. For this reason, the
field of machine learning (ML) has recently bloomed as a very attractive way of 
using our computing power to extract conclusions from data. The access to a large amount of data is opening 
up the possibility of using discriminative models to directly learn from the data. From
a probabilistic point of view, these models try to directly model the distribution $p(y|x)$. They
do not put emphasis on understanding the generation process of the data $x$, but 
on directly inferring properties from observations. The machine learning revolution that
we are witnessing in solar physics is fundamentally based on discriminative models. The
large databases that we have available are allowing us to directly learn from data, or 
use data for speeding up certain complex operations.


Machine learning methods are often divided into three main classes: supervised, unsupervised, and reinforcement 
learning (see Fig.~\ref{fig:ml_types}). Supervised and unsupervised learning have deep roots in the field of 
statistics known as statistical learning \citep*[see the textbook by][]{Hastie:2009}, which is concerned with 
model fitting, parameter estimation, and learning about the structure of data. Reinforcement learning 
is, however, strongly based on control theory \citep[e.g.,][]{10.5555/519085}. The term ML became popular in 
the era of Big Data. To scale statistical learning algorithms to effectively utilize large data sets, new 
algorithms and the appropriate software and hardware stack were developed in tandem. For instance, the 
development of general purpose graphical processing units (GPUs) accelerated the development of computer 
vision techniques. This in turn drove the development of GPU hardware with higher throughput, and the 
development of ML programming frameworks, such as Tensorflow \citep{tensorflow2015-whitepaper}, 
PyTorch \citep{pytorch19} or JAX \citep{jax2018github}. In turn, these developments 
facilitated the development of models with greater expressivity and applicability across scientific and 
engineering domains.

In the following sections, we briefly introduce the goals of supervised, unsupervised and reinforcement 
learning. Most applications of ML in solar physics pertains to the first two classes. 
The applications of reinforcement learning in solar physics have received little attention 
but it can bring substantial improvements in observational planning and other complex control tasks like adaptive optics. 
Functional optimization is used in all three classes of ML, so we will 
discuss optimization too.

\subsection{Supervised learning}
Supervised learning is the task of learning a mapping between inputs (the collection thereof is often 
denoted by $\mathbf{X}$) and outputs (denoted by $\mathbf{Y}$; also called targets) for which examples of input-output pairs are available. From a probabilistic perspective, the goal of
supervised learning is to model the conditional distribution $p(y|x)$.

\begin{figure}
    \centering
    \includegraphics[width=0.9\textwidth]{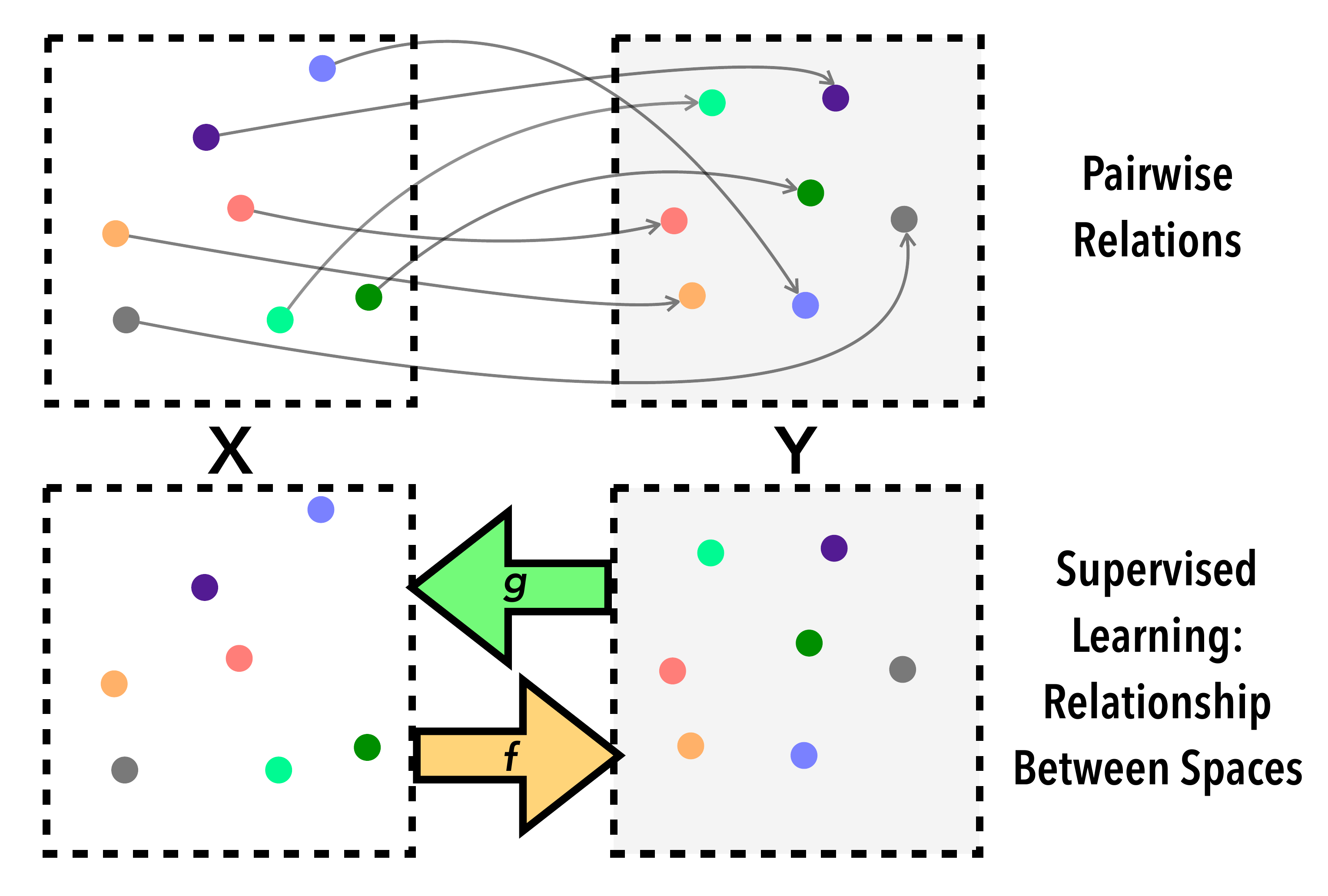}
    \caption{Schematic representation of supervised learning, with the 
    feature space $X$ and the target space $Y$. The aim is to define or learn mappings
    $f$ and $g$ between the two spaces by taking advantage of the information encoded in
    the pairwise relations were known in advance for a specific sample of the two spaces.}
    \label{fig:mapping}
\end{figure}

Supervised learning is especially suited for the physical sciences because it can be used to infer parameters $\mathbf{Y}$ from the inputs $\mathbf{X}$. To illustrate with a concrete example from solar physics, suppose we have measured the 
Stokes IQUV parameters of a magnetically sensitive line over a region of interest on the Sun. The goal here is to infer the physical properties of the plasma producing the radiation. This process is commonly known as \emph{inversion}.

For each spatial location on the Sun, the input data is 
$\mathbf{x} = (\mathrm{I},\mathrm{Q},\mathrm{U},\mathrm{V})$, where each Stokes parameter is a function of the wavelength (so $x$ is a tuple with size $N=4\times N_{\lambda}$, with $N_\lambda$ being
the number of measured wavelength points). It is convenient to collect the set of Stokes profiles measured at all $M$ locations of interest as a matrix $\mathbf{X} = (\mathbf{x}_0, \mathbf{x}_1, ..., \mathbf{x}_i, ..., \mathbf{x}_M)^T $, i.e. each row of $\mathbf{X}$ is a sample of input data. In Fig.~\ref{fig:mapping}, each sample (row) of input data $\mathbf{X}$ is schematically denoted as a point residing in a subspace. In ML parlance, this is called the input \emph{feature space}, denoted by $X$. 

In our example, the physical parameters of interest include the magnetic field strength $B$, the orientation of the magnetic field vector (inclination and azimuth, $\phi$, $\theta$ respectively), the ambient plasma temperature ($T$), etc. For each spatial location 
on the solar surface, we have $y=(B,\phi,\theta,T, ...)$. The traditional approach to inferring these physical 
parameters is to perform an iterative inversion with the help of a physics-based forward model. Except under special conditions (generally not valid on the Sun), the forward model is a nonlinear radiative transfer calculation. The forward model $g: Y \longrightarrow X$, often known in advance and based on physical
arguments, allows us to compute the predicted Stokes profiles $x\in X$ for any $y \in Y$. In Fig.~\ref{fig:mapping}, $g$ is denoted by the arrow going from subspaces $Y$ to $X$. 

An iterative inversion begins with an initial estimate of $y$. The physics model is used to compute $x_{\rm pred} = g(y)$, which is compared to the target (observed) $x$ with a chosen penalty function. A common function is the mean-squared error, which
is motivated by the assumption of Gaussian noise with diagonal covariance in the 
observations. In general, a suitable probabilistic approach can take into account other sources of
noise or regularization (see Sect.~\ref{sec:regularization}). The physical parameters are adjusted at each iteration to minimize the discrepancy between predicted and observed data. To guide the updates of $y$ in a way to reduce the penalty function, the curvature of the penalty function with respect to the parameters is often used \citep[see][]{2016LRSP...13....4D}. The forward model $g$ is used to generate pairs of $(y,x_{\rm pred})$ until a pair is found such that $x_{\rm pred} \approx x$ to some tolerance. This entire procedure is then repeated for each $x \in \mathbf{X}$. In this approach, the inversion provides a pairwise mapping between each sample pair. Although it works, it has two major drawbacks. First of all, it is
inefficient because the inference of a pair $x_i \longrightarrow y_i$ is performed completely independently of other pairs in the data. Secondly, the approach does not let us efficiently compute how a perturbation of the input impacts the output. 

Whereas the traditional iterative optimization procedure gives a mapping between \emph{individual pairs} in the input and output feature spaces, supervised learning aims to provide the mapping between the two spaces, i.e. $f: X\longrightarrow Y $ (see Fig.~\ref{fig:mapping}). Supervised learning does so by using the data $\mathbf{X}$ and $\mathbf{Y}$ globally (not individual rows) to fit a model approximating $f$. This is usually posed as an optimization problem of the following form:

\begin{equation}
    f^{\#} = {\rm argmin}~L ( f(\mathbf{X}), \mathbf{Y}).
\end{equation}
\noindent In other words, find an optimal function $f$ such that the loss function $L$ (which compares the predicted and observed values) is minimized\footnote{A simple widespread loss function is the mean squared error (MSE) between predicted and measured $y$, which assumes that the residual between the predictions and
the measurements follows a Gaussian distribution.}. Usually, the function $f$ is expressed in terms of parameters 
(e.g., weights and biases in a neural network, as explained in Sect.~\ref{sec:ann}), and fitting is performed to adjust the parameters of $f$ to minimize $L$. This step is called \emph{model training}.

Note that the forward model $g$ is not needed when doing model training. As long as pairwise data linking $X$ and $Y$ is available, 
the forward model is not a prerequisite for supervised learning. In fact, in many applications, 
both $f$ and $g$ are unknown prior to fitting. In cases where $g$ is known (e.g., our Stokes spectropolarimetry example) 
and is used to generate observables $\mathbf{X}$ from $\mathbf{Y}$, supervised learning amounts to learning the inverse mapping $f = g^{-1}$.

Having presented the goal of supervised learning, we introduce some necessary nomenclature to aid discussion throughout this review article. 

\subsubsection{Classification vs. regression}
In a supervised learning setting, the target variable $\mathbf{Y}$ may be a continuous variable or 
may be discrete (e.g., the set of non-negative integers $\mathcal{Z}^+$). These two types of 
problems are called regression and classification, respectively. Both regression and classification
can be tuned to deal with the same problem. For instance, let us consider the problem of
flare prediction. The following is a \emph{classification problem}: predict whether the Sun will 
produce a flare of class M or higher. The ground truth values 
are binary (Yes or No). Another way to pose a similar question involves a \emph{regression problem}:
predict the peak X-ray flux within the next 24 hours. This is a regression problem since the 
peak X-ray flux is a continuous variable. Classification problems are often simpler to solve
using machine learning.

\subsubsection{Data partitioning}
Data partitioning or data splitting is the act of splitting $\mathbf{X}$ and $\mathbf{Y}$ into training, testing 
and validation sets. Members of the training set $(\mathbf{X}_{\rm train}, \mathbf{Y}_{\rm train})$ are used at time of 
model fitting, and the loss function gives a scalar computed over this set.
After training, the loss function is evaluated over the test set $(\mathbf{X}_{\rm test}, \mathbf{Y}_{\rm test})$ and
is compared with the value for the training set. The case $L_{\rm test} > L_{\rm train}$ is a sign of possible overfitting,
so that the model is ``memorizing'' the training set and not generalizing correctly. 

The test set is reserved for the evaluation of the performance of the final trained model(s) on certain chosen metrics. 
If the model contains hyperparameters (for instance, the width of a densely connected layer), a \emph{validation set} 
is carved out from the training set for exactly this purpose \citep{RussellNorvig}. This further partitioning 
of the (non-test) data into a training and validation set allows one to evaluate the performance 
of the model during training time without building bias toward fitting the test set.

Data partitioning is one of the most important decisions in a machine learning project. Depending on the 
goal, the partitioning strategy will vary. For instance, should flares from the same AR be in both test 
and training sets? The appropriate partitioning strategy is highly dependent on the nature of the 
scientific/engineering objective as well as the nature of the underlying system under consideration. 
To successfully apply ML to solar physics problems, it is desirable not to apply it
blindly but take into account the existing knowledge.

\subsubsection{Encoders and decoders}
In ML parlance, the forward model $g$ is sometimes called the~\emph{encoder}, and the optimal function $f$ is
known as the \emph{decoder}. The two functions in 
series, i.e., $g(f(x)$ or $f(g(y))$, is called autoencoder. As will be discussed later, autoencoders are 
useful in a number of applications, including data denoising. They are useful, as well, when both $g$ and $f$ are 
not known a priori. However, when one of them is known (e.g., 
a physics model for $g$), autoencoders can be used to directly learn the other mapping from data. 
This way of combining machine learning and physical information turns out
to be extremely powerful.

\subsection{Unsupervised learning}
Unsupervised learning is the task of discovering patterns in the data $\mathbf{X}$. Unlike supervised 
learning, this task does not require matching target values $\mathbf{Y}$. In other words, unsupervised 
learning is about characterizing the structure of the probability density function $P(\mathbf{X})$. 
For instance, a common question addressed using unsupervised learning is in regard to clustering of 
data points. Unless the components of $\mathbf{x}$ are independent and identically distributed 
(i.i.d.), $P(\mathbf{X})$ will have local minima and maxima, with the latter indicating clustering 
of data in parts of the input space.


Unsupervised learning can be very useful for a global understanding of the observations. From a probabilistic perspective, the task is to model the prior distribution of observations $p(x)$.
As such, unsupervised models do not make use of any labeling, just purely observations. 
As an illustration, we return to the example application of spectropolarimetry. Due to the physics (e.g., Zeeman 
or Hanle effects), the values of the Stokes IQUV parameters emergent from the Sun's atmosphere are not independent 
across the spectral (i.e., wavelength) dimension. Furthermore, the four Stokes parameters are correlated 
with each other as a consequence of the laws of physics 
(e.g., $I_\lambda^2 \geq V_\lambda^2 + U_\lambda^2 + Q_\lambda^2$ where the subscript denotes 
the monochromatic intensity at wavelength $\lambda$). Even if a spectrogram has $N_\lambda$ wavelength positions, the number of 
degrees of freedom in a Stokes IQUV measurement is significantly less than $4N_\lambda$. If we knew the exact details 
of the underlying plasma (e.g., turbulence properties, whether there is subpixel structure), the number of degrees 
of freedom would be known a priori. In the absence of such insights, unsupervised learning can help with 
dimensionality reduction, by automatically finding the correlations and exploiting them.


\subsection{Reinforcement learning}
Although reinforcement learning (RL) often involves techniques used in supervised and unsupervised 
learning, it is considered a separate field of ML. The goal of RL is to explore how autonomous agents (e.g.,
a robot) interact with an environment (e.g., the world), and how to effectively train such agents to 
achieve desired objectives \citep{Sutton98}.
In an RL setting, an agent has an internal state. The agent is exposed to input (stimuli) from its 
environment (e.g., an image of the scene surrounding the agent). Based on policies available to the
agent, it carries out an action which can change the agent's state and its environment. The objective 
of the agent is to maximize its cumulative rewards, as determined by a suitable reward function.
RL has found extensive applications in robotics, the automotive industry, and gaming. As of writing, 
the authors are aware of a single application of RL in solar physics, that is 
discussed in Section \ref{sec:deep_flare}. 
However, we envisage RL will eventually be used for 
complex solar physics-related applications, such as observation planning or
better adaptive optics systems \citep{2022A&A...664A..71N}.

\begin{figure}
    \centering
    \includegraphics[width=0.75\textwidth]{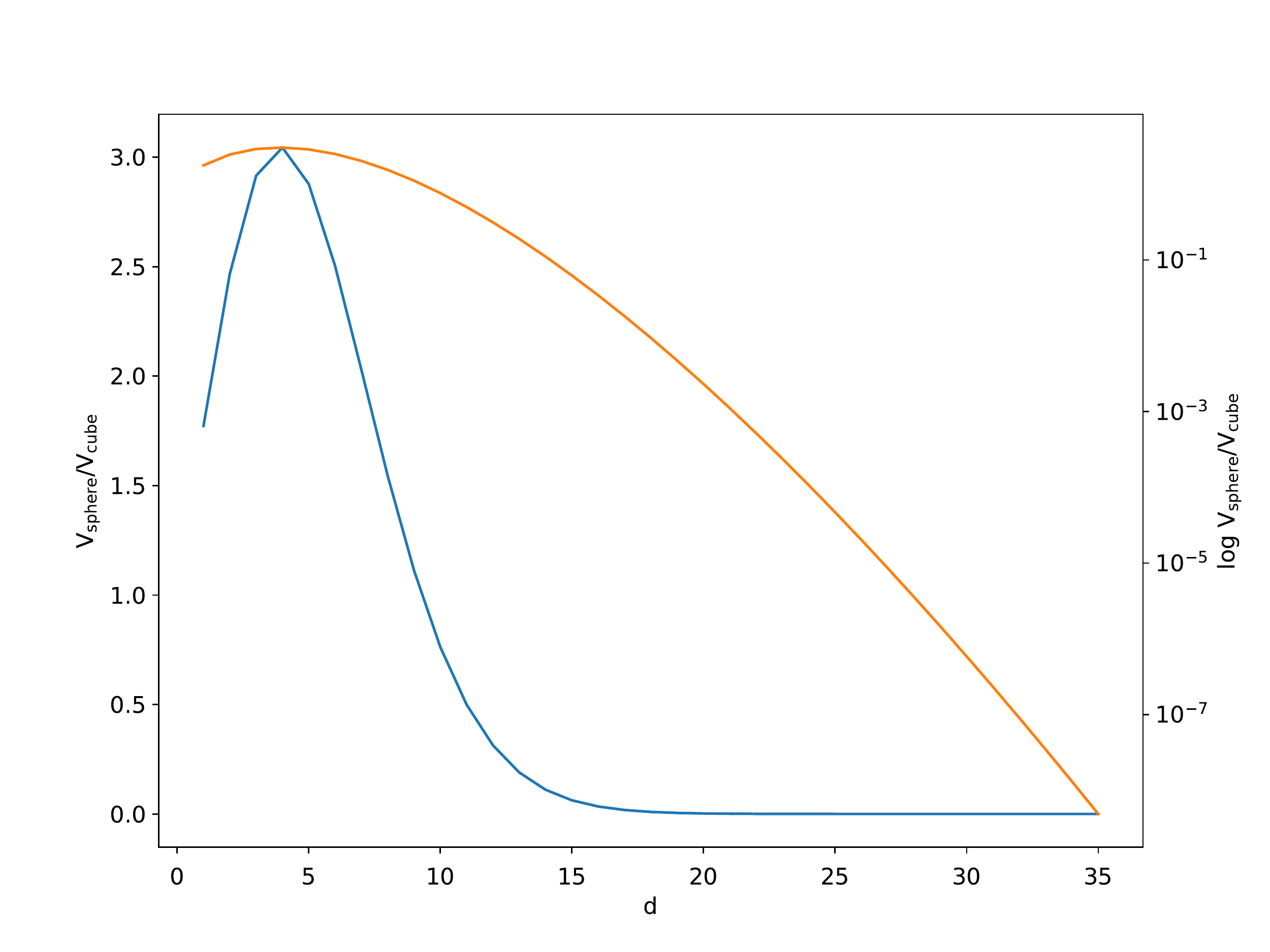}
    \caption{Curse of dimensionality for an Euclidean space of dimension $d$. This
    figure shows the volume ratio between an hypersphere of radius $r$ and
    the hypercube in which the hypersphere is circumscribed. It shows that
    the volume resides in the external parts of the space. Both linear and
    logarithmic scales are shown for clarity.}
    \label{fig:curse}
\end{figure}

\section{Some ideas about dimensionality}
Data living in very high dimensions present difficulties when analyzing and
understanding their statistical properties. The efficiency of typical
statistical and computational methods usually degrades very fast
when the dimensionality of the problem increases, thus making
the analysis of the observed data cumbersome or, sometimes,
unfeasible. This fact is often referred to as the \emph{curse of dimensionality}. 
The fundamental reason for that lies in the fact that, for high dimensional spaces,
almost all the volume of the space tends to accumulate in
the borders of the space. We can easily visualize this in
an Euclidean space of dimension $d$. The ratio between 
the volume of the hypersphere of radius $r$ centered at the origin and that of the 
hypercube of side length $2r$ centered at the origin in which the
hypersphere is inscribed is given by:
\begin{equation}
    \frac{V_\mathrm{hypersphere}}{V_\mathrm{hypercube}} = \frac{\pi^d}{d 2^{d-1} \Gamma(d/2)}.
\end{equation}
As shown in Fig.~\ref{fig:curse}, the ratio exponentially goes down to zero when $d$ increases, so that the volume very quickly accumulates on the borders of the
hypercube. As a consequence, any sampling of a high-dimensional
space rapidly becomes useless. The enormous success of ML in recent years, thanks
to the deep learning revolution, is rooted in the ability of deep learning
to overcome the curse of dimensionality.

The advent of computers has permitted us to face
the analysis of increasingly complex data. These data usually 
exhibit an intricate behavior, and in order to understand the 
underlying physics that produces such effects, we have been forced to
develop very complicated models. Ideally, these models have to
be based on physical grounds, but there seems to be no way of
knowing in advance how complicated this model has to be to
correctly reproduce the observed behavior.
Despite their inherent complexity, the analysis of large data
sets, such as those produced by modern instrumentation, indicates
that not all measured data points are equally relevant for the
understanding of the underlying phenomena (one of the simplest example is
the limited information carried out by spectral points in the
continuum versus spectral points sampling spectral lines). In other words, it is
clear that the reason why many simplified physical models are 
successful in reproducing a large amount of observations is because
the data itself is not truly high dimensional. Based on this premise,
it makes sense to develop and apply methods that are capable of
reducing the dimensionality of the observed data sets while still
preserving their fundamental properties. Mathematically, the idea is that while the original data may have a very large dimensionality, 
they are in fact confined to a small manifold of that
high-dimensional space. In this case, we can consider that the data
‘‘lives’’ in a subspace of low dimension (the so-called intrinsic
dimension) that is embedded in the high-dimensional space. This
lower dimension manifold is not simple to describe in general, simply
because it is highly nonlinear and unknown.

Instead of fully characterizing the manifold, the simpler task of estimating 
its intrinsic dimensionality is of interest
to understand the complexity of the models used to extract information. 
Additionally, it is a check that the machine learning
method used to analyze the data is able to overcome the curse
of dimensionality. One of the simplest examples is the one in which
the data consists of spectral, or more in general, Stokes profiles, that
encode the polarization state of light. A deep analysis of objects of much larger 
dimensionality, like images of the Sun, has never been carried out.
As shown later in this review, some of the recent deep learning methods 
have been successfully applied to solar images. This implicitly demonstrates that solar 
images also lie in a manifold of reduced dimensionality when compared with 
the potential dimensionality of all possible images.

The estimation of the dimensionality of the
manifold of Stokes profiles for photospheric lines was pursued by
\cite{asensio_dimension07}. They used an estimator of the dimensionality 
based on the maximum likelihood principle previously developed by
\cite{levina_bickel05}. When applied to Fe \textsc{i} lines in the
visible and the infrared, they reached the following conclusions. First, 
the dimensionality of the infrared lines is slightly larger than those in the visible, 
thus suggesting that there is more variability in the Stokes profiles in the infrared. This 
is probably a consequence of the fact that Doppler shifts and Zeeman splittings in the 
infrared are slightly larger than in the visible, producing more deformations in the line 
profile. Second, the dimensionality of circular polarization is larger than that 
of Stokes $I$, a consequence of the fact that Stokes $V$
is much more sensitive to variations in the magnetic field than Stokes $I$.
Finally, they quantitatively proved the idea that adding more spectral lines
increases the amount of information available \citep[see, e.g.,][]{semel81,socas_navarro_inversion04}.
Adding more spectral lines monotonically increases the dimensionality
of the manifold but clearly not in proportion to the number of added
spectral lines. There is a lot of redundant information already encoded in all spectral lines 
and only small details can be better seen in one spectral line or another.

Given that Stokes profiles sampled at $N_\lambda$ wavelength points 
are demonstrated to be lying in a manifold of dimension $d \ll N_\lambda$, it makes sense to exploit this property for different purposes. The two more obvious ones
are denoising and compression. Uncorrelated additive noise typically assumed to be present in 
spectropolarimetric observations, spans the full space of $N_\lambda$ dimensions.
It is advantageous to find a suitable representation in which the signal is separated
from the noise by exploiting the fact that the signal lies in a manifold
of reduced dimensionality. Similarly, representing the spectropolarimetric data
with a reduced set of numbers leads to an important compression factor, which
turns out to be important for data storage and transfer via telemetry.
The methods presented in the following sections have been successfully used in 
solar physics for these purposes.

\section{Linear models: unsupervised}
The availability of data to analyze, their large sizes, and the difficulties
in extracting physical information from the observations has led to 
the widespread application of unsupervised machine learning methods.
These methods allow us to extract relevant information from the
observations directly, typically focusing on the regularity that can
be explained in a posteriori. Clustering or classification is perhaps one
of the most obvious tasks of unsupervised machine learning methods. We start
first by describing linear models for unsupervised ML. These methods, despite their
limitations, have been extremely successful in science and, specifically, in 
solar physics.

\subsection{Principal component analysis}
Principal Components Analysis \citep[PCA;][]{loeve1955}, also known as 
the Kar\-hu\-nen--Lo\`eve transformation,
is perhaps one of the most used algorithms in multivariate statistics\footnote{PCA is available 
on the \texttt{scikit-learn} Python package.}.
Briefly, its main use is to obtain an orthogonal basis on which 
the data can be efficiently expressed. This basis has the property that the 
largest amount of variance is explained with 
the least number of basis vectors. It is useful to reduce the dimensionality of data sets 
that depend on a very large number of parameters and one of its most
straightforward applications is denoising.

PCA can be seen as the solution to a linear regression problem in which
both the weights and the basis functions are inferred from the data.
Since this is an ill-defined problem, PCA imposes the additional restriction of
orthogonality for the basis functions. Summarizing, PCA is a way to decompose
any observed signal as a weighted sum of empirical orthogonal basis functions. It has
been extensively used in the field of spectropolarimetry for denoising and, in general,
dimensionality reduction purposes.

Let us assume that the wavelength variation of the Stokes profiles of a 
particular spectral line is described by the quantity $S_{ij}$. The index $i$ represents the 
wavelength position while the index $j$ = \{I, Q, U, V\} labels the Stokes parameter. 
Each Stokes parameter is a vector of length $N_\lambda$, corresponding to the number of 
sampled wavelength points. Assume that the spectral line is observed in many locations in the field
of view, that we term $N_\mathrm{obs}$. By stacking all observations, one can build the observation
matrix $\mathbf{O}$, which is of size $N_\mathrm{obs} \times N_\lambda$. As well, 
all Stokes parameters can be stacked together and one ends up with a matrix of
size $N_\mathrm{obs} \times 4N_\lambda$.
The principal components can then be found by computing the eigenvectors of this matrix of observations. 
This means that the PCA procedure reduces to the diagonalization of the matrix $\mathbf{O}$. Since 
we often have that $N_\mathrm{obs} \gg N_\lambda$, this matrix is not square and
one needs to use the singular value decomposition \citep[SVD; see, e.g.,][]{numerical_recipes86} 
to diagonalize $\mathrm{O}$ and compute its singular vectors. The SVD decomposition
reads as follows:
\begin{equation}
    \mathbf{O} = \mathbf{U} \mathbf{\Sigma} \mathbf{V}^\star,
\end{equation}
where $\mathbf{U}$ is an $N_\mathrm{obs} \times N_\mathrm{obs}$ orthogonal matrix with
the left singular vectors in columns
while $\mathbf{V}$ is an $N_\lambda \times N_\lambda$ orthogonal matrix with the 
right singular vectors in columns. $\mathbf{\Sigma}$ is a diagonal
matrix with the singular values on the diagonal. 
The real power of PCA lies in the fact that one can truncate the previous 
decomposition by only leaving $r$ singular values equal to their original value and
setting the rest to zero. This way, one gets $\tilde{\mathbf{O}}$, a 
reconstruction of the original matrix constrained to have $\mathrm{rank}(\tilde{\mathbf{O}})=r$.

Although carrying out the PCA decomposition using the $\mathbf{O}$ matrix
is possible, it is often much more efficient from a computational point of view to compute 
the singular vectors of the correlation or the cross-product matrices. With the use of simple 
algebra for the case of the correlation matrix, $\mathbf{X}=\mathbf{O}^\dag \mathbf{O}$,
one can verify that:
\begin{equation}
    \mathbf{O}^\dag \mathbf{O} = \mathbf{V} \mathbf{\Sigma}^\star \mathbf{U}^\star 
    \mathbf{U} \mathbf{\Sigma} \mathbf{V}^\star = \mathbf{V} \left( \mathbf{\Sigma}^\star 
    \mathbf{\Sigma} \right) \mathbf{V}^\star,
\end{equation}
so that the right singular vectors of the $\mathbf{X}$ matrix  are equal to the right singular
vectors of the observation matrix. Likewise, the singular values of the correlation
matrix are those of the original matrix but squared. The advantage of this approach is that the matrix 
$\mathbf{X}$ has size $N_\lambda \times N_\lambda$, so that the diagonalization
becomes more computationally efficient if $N_\lambda \ll N_\mathrm{obs}$.

On the contrary, in cases in which $N_\lambda \gg N_\mathrm{obs}$, one can use
a similar approach but using the cross-product matrix, $\mathbf{X}'=\mathbf{O} \mathbf{O}^\dag$.
In such case, the left singular vectors of the matrix $\mathbf{O}$ are obtained.
It is important to remark that both descriptions are dual and they are completely 
equivalent, so one should choose the one that provides the most efficient computation.

Once an orthogonal basis is found, one can reconstruct the original matrix
by just computing the projection along each direction and multiplying each
projection by the orthogonal basis:
\begin{equation}
    \mathbf{O} = \left( \mathbf{O} \mathbf{V} \right) \mathbf{V}^\star.
\end{equation}








 


\subsubsection{Denoising}
\label{sec:pca_denoising}
Given that the largest amount of variance of the input dataset is explained
with the first singular vectors, reconstructing the data with only a few 
such vectors lead to a very efficient denoising technique. This can
be technically achieved by reconstructing the original dataset 
with the matrix $\mathbf{V}'$, a submatrix of $\mathbf{V}$ of lower rank
that only contains the columns associated with the largest singular vectors.
A denoised dataset can then be obtained by computing:
\begin{equation}
    \mathbf{O}_\mathrm{denoised} = \left( \mathbf{O} \mathbf{V'} \right) \mathbf{V'}^\star.
\end{equation}
The selection of the number of eigenvectors to keep is often done by
computing the residual $\mathbf{O}_\mathrm{denoised}-\mathbf{O}$ and characterizing
its statistical properties. In the case of Gaussian noise with fixed variance, one can stop adding
eigenvectors once the variance of the residual is similar enough to the noise variance. 
When this is achieved, one can be sure that the
selected eigenvectors are retaining the signal and removing the uncorrelated noise. 
For a more automatic way, see \cite{gavish2014optimal}, which provides a strategy for optimally 
picking the rank of $\mathbf{V'}$.

If the observations are affected by systematic effects, like interferometric fringes 
(in the case of Stokes observations), they will be part of the output. In many cases in which
these systematic effects are strong, one can be lucky and find them isolated
in one or two eigenvectors. If this is the case, it is possible to remove
these systematics from the observations by deleting these eigenvectors
from $\mathbf{V'}$. However, it is often the case that systematic effects
are extracted together with real signals in some eigenvectors and they 
cannot be easily separated. A technique that has recently been proposed
is to carry out a rotation in the subspace described by these 
eigenvectors with the aim of isolating the contribution of systematic
effects and real signal \citep{2019ApJ...872..173C}. Other techniques, with
more control from the user side, are based on the techniques presented
in Sect.~\ref{sec:rvm}.

PCA denoising is now systematically used for removing Gaussian noise
from the observations \citep{2007ApJ...659..829A,2016A&A...596A...5M,2018A&A...619A..60J}.
This denoising is also very helpful in stabilizing spatial
deconvolution methods, like the one developed by \cite{2013A&A...549L...4R}, which
is routinely used to remove the effect of spatial smearing in 
observations \citep{2015A&A...579A...3Q,2016MNRAS.460..956Q,2016MNRAS.460.1476Q,2016A&A...596A..59F,2016A&A...596A...2B,2017A&A...601L...8B}.

\subsubsection{Interpretability}
Somehow surprising, it has been found that, in some specific cases, the leading singular
vectors of the PCA decomposition have a well defined physical
meaning. This was first pointed out by \cite{skumanich02}, who demonstrated 
this for spectropolarimetric observations of a sunspot. 
They showed that the first singular vector of Stokes $I$
is associated with the average spectrum, the second one
gives information about the velocity, and the third one gives
information about magnetic splitting or any other broadening
mechanism. Likewise, for Stokes $V$ they found that the first singular 
vector correlates with the longitudinal component of the field, 
the second one correlates with velocities in the magnetic component 
and the third one correlates with broadening mechanisms. This is
not surprising if one realizes that PCA is akin to a 
the Taylor expansion of the Stokes profiles \citep{skumanich02}.
Despite the results discussed so far, PCA often does not extract
interpretable physical information from the observations. The reason has to be
found on the fact that PCA focuses on global properties of the observations to maximize the amount of variance
explained. However, many of the interpretable features of data are local (i.e., the
position of the core of the line, the presence of several velocity or magnetic components in the
line, etc). When looking for interpretability, other techniques like
t-SNE \citep[Student-t Stochastic Neighbor Embedding;][]{Hinton_Roweis_2003} can 
be more useful (see Sect.~\ref{sec:nonlinear_unsupervised}).

\subsubsection{Inversion with lookup tables}
The compression capabilities of PCA have been also exploited
for accelerating the inversion of Stokes profiles. The process of inverting
Stokes profiles consist of inferring the physical properties
that produce a given observation. This inversion is usually solved
using a maximum likelihood approach in which a merit function (often
the $\chi^2$ as a consequence of the assumption of Gaussian noise), that
measures the difference between the observations and synthetic Stokes
profiles is minimized. This minimization can be done using several
techniques. However, the idea when using PCA is
to use one of the simplest methods of inversion one can think of:
generate a large database with Stokes profiles synthesized in
model atmospheres parameterized with $N_\mathrm{par}$
parameters and pick up the model providing the best fit. 

This inversion method, first suggested
by \cite{rees_PCA00}, requires some specialized methods for the
construction of the database. The reason is that a trivial
method in which every parameter of the model is sampled at $n$ 
values requires a database of size $n^N_\mathrm{par}$, which 
quickly becomes impractical. Moreover, because of the curse of 
dimensionality, an exponentially large amount of sampled models will lie
in the borders of the space and can become useless for the inversion process.
As a consequence, this inversion method only works for very simple models. 

With these problems in mind, \cite{rees_PCA00} developed a Monte Carlo approach for
populating the database. They started from a model chosen at random.
A new model is randomly proposed and the resulting Stokes profiles are
compared with the existing ones. If they both lie in a small Euclidean ball
of radius $\epsilon$, they are assumed to be coming from very similar
atmospheres and only one of them is kept. This procedure is iterated until
a sufficiently large database is obtained or when the fraction of accepted
new models becomes impractically low. This can be understood as an indication that the
space of models is densely sampled. PCA compresses the database
by only storing the projections along a few relevant singular vectors. This
can lead to compression factors of an order of magnitude, which also
accelerates the database search.

After the first pioneering work of \cite{rees_PCA00}, more
works followed. They were especially centered on the interpretation of
scattering polarization signals and the Hanle effect in lines of
He \textsc{i}. The fundamental reason for this is that the solution of
the forward problem for these lines is very time consuming, so one better
spends the time building a database that can later be used to carry out
very fast inversions. This is in contraposition with what happens when a classical
iterative algorithm is used for fitting the observations. 
\cite{2002ApJ...575..529L} proposed using PCA to compress a 
database of synthetic profiles of the He \textsc{i} D$_3$ multiplet at
5876 \AA\ using the optically thin approximation. They used the database
to invert observations of prominences carried out with the THEMIS telescope, showing that
this technique is promising. They obtained magnetic fields that are almost
parallel to the solar surface and with strengths around 40 G. The same code
was applied to prominence data from the High Altitude Observatory Stokes II polarimeter
\citep{1985SoPh...96..277Q} by \cite{2003ApJ...582L..51L}. For computational
reasons, the database building process was specifically tailored for the observations, by 
restricting the ranges of some of the model parameters. Again, the method
yields strongly inclined magnetic fields, almost parallel to the solar
surface, with strengths as large as 50 G.

The availability of the PCA-based inversion code opened the
possibility of quickly inverting 2D maps. For this reason, 
\cite{2003ApJ...598L..67C} observed a prominence with the Dunn
Solar Telescope (DST) of the National Solar Observatory (NSO)
with a spatial resolution close to 1". A database of 2$\times$10$^5$
was built for inverting the physical properties of prominences. The resulting maps show 
magnetic fields with an average of $\sim$20 G, but with blobs
displaying strengths above 50 G. Again, the fields are almost
parallel to the solar surface. Some possible limitations of the model used
for the inversion were discussed in \cite{2005ApJ...622.1265C}. The same approach
of using PCA-compressed databases were also used by \cite{2005A&A...436..325L}
to deal with the inversion of He \textsc{i} D$_3$ profiles
in spicules. This demonstrates that the generation of a look-up
table is a suitable inversion procedure for any observation once
the database is built with the appropriate ranges of the model
parameters. Their conclusion is that the magnetic field vectors are aligned
with the visible structure of the spicule, finding fields above 30 G
in some cases. 

Later on, databases for the simultaneous inversion of the He \textsc{i}
D$_3$ and 10830 \AA\ multiplets were developed. Lines that are sensitive to scattering
polarization and the Hanle effect (as is the case in the mentioned He \textsc{i} lines)
suffer from more ambiguities than
those whose polarization is controlled only by the Zeeman effect. A
careful interpretation of the polarization signal of several lines can
potentially help in solving these ambiguities. However, more care
needs to be taken when constructing the database precisely to deal
with these ambiguities. \cite{2009ApJ...703..114C} built a database with
2.5$\times$10$^5$ models and used it to invert simultaneous observations
of He \textsc{i} D$_3$ and 10830 \AA, resulting in an improved determination
of the magnetic field.

Despite its success, the look-up method has some drawbacks, some of them
a consequence of the curse of dimensionality:
\begin{enumerate}
    \item The procedure followed to fill the database has difficulties 
    dealing with ambiguous and quasi-ambiguous solutions. The Zeeman effect
    is subject to the well-known 180$^\circ$ ambiguity in the azimuth in the
    reference system of the line-of-sight. In other words, fields whose
    azimuth on the plane of the sky for 180$^\circ$ produce exactly the
    same Stokes profiles. When scattering polarization and the Hanle effect
    dominate, possible additional 90$^\circ$ ambiguities (Hanle ambiguities) appear. 
    Rejecting synthetic profiles that lie inside the $\epsilon$-ball of other preexisting profiles
    in the database disfavor the representation of physical properties that are subject to
    ambiguities. This is of almost no importance for profiles controlled by the
    Zeeman effect but can turn out to be important for those cases dominated by
    scattering and the Hanle effect.
    \item Current observations of Stokes profiles produce noise standard deviations 
    of the noise that reach 10$^{-4}$ in units of the continuum intensity. This means
    that the $\epsilon$-balls have to be really tiny so that the number of 
    profiles needed to fill a database with such precision quickly becomes 
    unmanageable. Most existing databases have been constructed with larger
    $\epsilon$-balls, so that we are at the risk of confusing cases in which 
    different (ambiguous) physical configurations but produce similar Stokes profiles.
    Ideally, one would like to push the limit on the $\epsilon$-balls to very small
    values, even smaller than the noise level, but this is clearly unfeasible.
    \item Filling up the database using the Monte Carlo approach can take
    a very long time. The first proposed profiles will always be accepted but
    the fraction of acceptance drops substantially when a few hundred thousand
    profiles are already part of the database. Additionally, every time one
    checks for the addition of a new profile, it must be tested against
    all profiles already present in the database. The number of comparisons
    to carry out is $N_\mathrm{tot}(N_\mathrm{tot}+1)/2$ to fill $N_\mathrm{tot}$ profiles in the database.
    Even though each comparison
    is very fast, the number of them one needs to carry out 
    rapidly makes this approach difficult to use. To partially compensate for this
    problem, other approaches based on the Latin hypercube sampling have also been
    used \citep{mckay79}.
    \item When used in evaluation mode, the 
    inversion requires the comparison of the Stokes profiles of interest with 
    all the profiles in the database. This requires the calculation of $N_\mathrm{tot}$
    comparisons for each observed Stokes profile. Recently, \cite{casini13} has
    devised an indexing method that accelerates the search. It is based on the
    use of a binary search tree built using the signs of the first $n$ PCA
    coefficients of each profile. This can potentially lead to an acceleration
    of a factor $2^{4n}$ in computing time.
\end{enumerate}

\subsection{Fuzzy clustering}
The solar corona is the outermost layer of the solar atmosphere and can be observed in 
various wavelengths \citep{2021PhRvL.127y5101K}. In the corona, magnetic pressure dominates over plasma 
pressure \citep{2001SoPh..203...71G}, and 
closed magnetic field lines confine plasma, appearing as bright coronal loops in extreme 
ultraviolet (EUV) wavelengths. Coronal holes are observed as dark areas \citep{2009LRSP....6....3C} and are regions with 
plasma depletion and lower temperature and density due to the continuous outflow of plasma 
along "open" magnetic field lines. Proper identification of coronal hole boundaries is crucial 
as they are a major source of the solar wind (SW), which can affect the Earth's environment \citep{2006JGRA..111.7S01T}, 
especially during the declining phase of a solar cycle \citep{2006JGRA..111.7S01T}. Accurate detection of coronal holes 
is challenging due to their varying boundaries with wavelength and resolution \cite{2021Ervin}.

Developing connectivity models between the Sun and the Earth requires observational constraints 
from the Sun, and a good evaluation of coronal hole boundaries can help improve these models. 
Machine learning methods have become increasingly popular for identifying coronal hole boundaries
\citep[e.g.,][]{2008AdSpR..42..917B, 2009A&A...505..361B},
replacing laborious and experienced observer-based methods. Accurate determination of coronal hole 
boundaries could also help solve new solar coronal questions such as the ``open flux'' problem, where 
the magnetic field in the Earth is two orders of magnitude higher than estimated from the Sun \citep{2017ApJ...848...70L}.

Before machine learning techniques were used, researchers used different techniques for the
identification of the CH boundaries. Previously, the identification and mapping
of CH were performed based on the helium spectroheliograms and photospheric
magnetograms by iterative visual inspection \citep{2005ASPC..346..261H}, a
laborious process requiring experienced observers. Automatic detection of the
CH was realized initially using spectroheliogram images in He \textsc{i} 1083 nm
wavelength, or spectral line properties of He \textsc{i} 1083 nm multiplet and
other multi-wavelength analysis \cite[and the references therin]{2005ASPC..346..261H}. 
In the last decades, we are witnessing a strong increase in the application of
ML methods for the identification of the CH boundaries. 

One of the earlier identification methods is the spatial 
possibilistic clustering algorithm\footnote{The latest version
can be found in \url{https://github.com/bmampaey/SPoCA}.}(SPoCA) which is
implemented as part of JHelioviewer
\footnote{See \cite{2017A&A...606A..10M} and find it on \url{http://swhv.oma.be/user_manual}.} 
\citep{2008AdSpR..42..917B, 2009A&A...505..361B,2014A&A...561A..29V}. Other
approaches are based on segmentation techniques together with ML
algorithms \citep{2015JSWSC...5A..23R}, or on the fuzzy
\citep{2013SoPh..283..143C} and k-means clustering \citep{2022ApJ...930..118I}.
The SPoCA method, based on an unsupervised fuzzy clustering method \citep{2008AdSpR..42..917B},
is a generalization of the k-means clustering discussed
in Sect.~\ref{sec:kmeans}. SPoCA implements three types of fuzzy clustering
algorithms considered to be appropriate for the EUV solar images: the Fuzzy
C-means (FCM); a regularized version of FCM known as Possibilistic C-means (PCM), 
and a Spatial Possibilistic Clustering Algorithm (SPoCA) that
integrates neighbouring intensity values. The SPoCA algorithm was described and
implemented by \cite{2008AdSpR..42..917B, 2009A&A...505..361B} for the automatic
identification of the CH, AR and quiet sun (QS) in EUV images. The
reason for using fuzzy clustering for the EUV images lies in the 
inherent uncertainty when categorizing visible structures. The SPoCA algorithm works by 
optimizing the following objective function:
\begin{equation}
\label{eq:SPoCA}
J_\mathrm{SPoCA}(B,U,X) = \sum_{i=1}^{C} \left( \sum_{j=1}^{N}  u_{ij}^m \sum_{k \in \mathcal{N}_j} \beta_k d(\mathbf{x}_k,\mathbf{b}_i)  + \tau_i  \sum_{j=1}^N (1-u_{ij})^m \right),
\end{equation}
where $C = 3$ is the number of clusters (\{CH, AR, QS\} in this case), $N$ is the number
of pixels of the image, $B = \{\mathbf{b}_1,...,\mathbf{b}_C\}$ are the cluster centers,
$X = \{\mathbf{x}_j, j=1,\ldots,N\}$ are the feature vectors of dimension $p$ 
that describe the Sun at each location, $U$ is a fuzzy partition matrix that
encodes the membership of feature vector $\mathbf{x}_j$ to class $i$, $m \geq 1$
is a parameter that controls the degree of fuzzification (a value of $m = 1$ means no fuzziness),
$\beta_k = 1$ if $k=j$, and  $\beta_k = (\text{Card}(\mathcal{N}_j)-1)^{-1}$, for any other $k$, with
$\text{Card}(\mathcal{N}_j)$ being the number of elements in the neighborhood of pixel $j$, 
$d$ is a distance function in the space of features, and $\tau_i$ is the intraclass mean fuzzy distance.

\cite{2014A&A...561A..29V} build upon the SPoCA software to extract,
characterize and track CH and AR from EUV images. They used an FCM to initialize
a PCM, which is considered more robust to noise and outliers. For the map
segmentation, they used different decision rules. \cite{2014A&A...561A..29V} looked mostly at CH and AR
and performed a parametric study to determine optimal configurations
of the algorithm. The dataset was built based on different EUV imagers.
The data used for the study was between 1997 and 2011 and obtained from the
EIT/SOHO in 171 and 195 \AA. 
The output of the program is a mask that overlays onto the original image.
The solution provides also the location of the AR or CH barycenter. From the
results, it was concluded that the FCM yields the best output for extracting CH.
The SPoCA detection method is regularly used for feature identification within
JHelioviewer and also as a training set for other methods.

One of the challenging tasks in the determination of the CH
boundaries is how to discriminate them from filament channels. This is
a consequence of the fact that, sometimes, filaments (prominences seen
in the solar visible disk) can be mistaken with CH. One of the early
attempts in tackling this issue was made by \cite{2015JSWSC...5A..23R}, who
used image segmentation methods together with supervised ML techniques for 
distinguishing between filaments and CH using AIA/SDO images in the channel at 193 \AA. 
The data is preprocessed by applying intensity-based thresholding and then the 
CH is identified using SPoCA. After the feature
extraction, CH and filaments were manually labeled based on simultaneous H$\alpha$ images.
Additionally, the line-of-sight magnetic field is obtained from HMI/SDO. 
They analyzed support vector machines (SVM) for classification \citep{svm95},
decision trees, and random forests as classifiers. They found that the SVM provided
the best result, especially when using the magnetic field information.

\subsection{k-means}
\label{sec:kmeans}
The k-means algorithm \citep{macqueen1967} has been widely applied in solar physics,
fundamentally in the field of spectropolarimetry, to classify the 
observed Stokes profiles in 2D maps\footnote{The k-means algorithm is available, for instance,
on the \texttt{scikit-learn} Python package.}. k-means tries to cluster
the $n$ observed $M$-dimensional data points $(\mathbf{x}_1,\mathbf{x}_2,\ldots,\mathbf{x}_n)$
into $k$ sets $\mathbf{S}=(S_1,S_2,\ldots,S_k)$, defined by their respective
cluster centers $(\boldsymbol{\mu}_1,\boldsymbol{\mu}_2,\ldots,\boldsymbol{\mu}_k)$.
This is done by obtaining the cluster centers that minimize the intracluster distance for
all the points in the dataset:
\begin{equation}
    \argmin_{\boldsymbol{\mu}} \sum_{i=1}^k \sum_{j=1}^n
    \mathbf{1}_{S_i}(\mathbf{x}_i) \Vert \mathbf{x}_i - \boldsymbol{\mu}_j \Vert^2,
\end{equation}
where $\mathbf{1}_{S}(x)$ is the indicator function, which takes the value 1 if
the elements belongs to class $S$ and zero otherwise:
\begin{align}
    \mathbf{1}_S(x) = 
\begin{cases} 
      1 & \textrm{if }  x \in S \\
      0 & \textrm{if } x \notin S.
   \end{cases}
\end{align}
Formally, this results into an $M$-dimensional Voronoi diagram\footnote{A Voronoi diagram is 
a partition of a hypervolume into regions close to each of a given set of objects.}, which has
linear decision boundaries.
From a practical point of view, this loss function is optimized 
iteratively as follows:
\begin{enumerate}
    \item Define a set of cluster centers.
    \item Compute the distance between all the observations and the cluster centers.
    \item Associate each observation to its closest cluster.
    \item Recompute cluster centers and repeat from step 2.
\end{enumerate}
The distance metric used can be tuned for 
the problem at hand but it is often simply the Euclidean distance (e.g. $\Vert \mathbf{x}_i - \boldsymbol{\mu}_j \Vert^2$ in the previous equation). Other metrics 
like the Mahalanobis distance\footnote{The Mahalanobis distance measures the distance between
two points taking into account the covariance structure of the underlying distribution.}
can be used to account for the covariance in the clusters. All of
them produce an $M$-dimensional Voronoi diagram but the decision boundaries 
depend on the specific distance metric.

k-means suffers from two fundamental problems. The 
first one is the inability of the algorithm to infer the number of clusters (i.e., it is a hyperparameter). To 
automatically extract the number of clusters one needs to resort to more advanced
methods. 
The second problem is that the final positions of the clusters
depend on the initialization.
The simplest solution to the first problem is to carry out k-means with different values of $k$
and deciding the optimal number by minimizing approximate estimations of the Bayesian
evidence like the Bayesian information criterion \citep[BIC][]{schwarz_bic78}.
A better option is to use ``density-based spatial clustering of applications 
with noise'' \citep[DBSCAN, see][]{Ester96adensity-based}, which can infer the number of
cluster centers by finding core samples of high density and expanding clusters from 
them\footnote{DBSCAN is available on the \texttt{scikit-learn} Python package.}.
Arguably the most robust option is to build a fully hierarchical Bayesian model \citep{teh_jordan_2010}
in which the number of clusters is considered a random variable\footnote{In this case, 
Dirichlet processes are often used as priors. A Dirichlet process is a probability distribution whose 
range is itself a set of probability distributions. It is used in Bayesian inference to describe 
the prior knowledge about the distribution of random variables.}. Concerning the second problem,
it is often the case that k-means need to be carried out several times to check 
for proper convergence.

\subsubsection{Spectral clustering}
k-means was used by \cite{2011A&A...530A..14V} to analyze the circular polarization
profiles in the quiet Sun as observed with Hinode/SP \citep{2013SoPh..283..579L}. They ended up inferring 
that the optimal number of
classes is $\sim$35 and that they can be grouped in six families
according to their general shape. One of the 
most prominent outcomes
is that a large fraction of the observed circular polarization profiles
are asymmetric. This means that the inversion of these profiles has to
be done with atmospheric models with gradients along the line of sight
\citep[see ][for an explanation regarding the physical origin of asymmetric profiles]{GrossmannDoerth:1988}. Later,
\cite{2015ApJ...806....9K} used k-means in the analysis of filament eruption
that produced an X-class flare. The strong variability of the 
observed profiles of the Ca \textsc{ii} 8542 \AA\ line (showing emission, absorption, asymmetric
, and also flat profiles) made it difficult to estimate the velocity from their Doppler
shift. By using k-means to cluster all the profiles into classes, the authors
were able to better define a model for each class of profiles to robustly
estimate the Doppler shift. 

Along this very same line, \cite{2018ApJ...861...62P} used
k-means to analyze observations of the Mg \textsc{ii} h and k spectral lines in flares with
the Interface Region Imaging Spectrograph \citep[IRIS;][]{2014SoPh..289.2733D}. Their
conclusion, by studying hundreds of thousands of profiles from several tens of flares,
is that profiles in flares show a single peak, instead of the double peak typical
of the quiet Sun. Additionally, these profiles also show enhanced broadenings and
blueshifted central reversals. 

Recently, \cite{2019ApJ...875L..18S} applied
k-means for the fast inversion of IRIS profiles. The idea is to cluster
the Mg \textsc{ii} h and k profiles from a large selection of observations, leading to
what they call Representative Profiles (RP). A detailed inversion of this representative profiles
with inversion codes like the STockholm Inversion Code \citep[STiC;][]{2019A&A...623A..74D}
can then be done, with the necessary care and the large computing time that these
inversions require (they sometimes require of the order of 2 CPU hours per profile). The result
is a one-to-one relation between RP and Representative Model Atmospheres (RMA). Afterward, 
the inversion of maps is carried out by comparing each observed pixel with the list of 
RP and setting the associated RMA as the solution. When the code is working in evaluation, 
one finds acceleration factors of 5-6 orders of magnitude in computing time.

The use of RP leads to a huge gain in computing time at the expense of
precision in the results. k-means will always select the RP with 
the smallest distance to the observed profile at every pixel. Although
this distance is the smallest, nothing avoids this distance to be
large in absolute units, when none of the RP produces a good fit to the observed profile. 
This would happen, for instance, in pixels with rare spectra, sufficiently
rare that it was not statistically present on the training set. Therefore, one should always be
cautious and avoid overinterpreting the results. It is always a good practice to
visualize the distance (in suitable units) between the observations and the
selected RP, paying special attention to those pixels in which the distance is
large and its specific reason.

\subsubsection{Segmentation of coronal holes}
Recently, k-means has been implemented for the
identification of CH by \cite{2022ApJ...930..118I}. Three of the AIA/SDO
wavelengths (171, 193 and 211 \AA) were used in different combinations,
individual channels, 2-channels (2CC) and 3-channels (3CC) composites, for
building the data sets. By computing the within-group sum of square distances as
a function of the number of clusters, they manage to give an optimal
number of clusters by locating the elbow of the plot (also known as the scree-plot method). 
The results obtained by applying the k-means method on each of the data sets were
compared among themselves to identify the best-performing data set. The results
were also compared with CH identified with other methods such as CATCH and HEK \citep[Heliophysics Event
Knowledge;][]{2012SoPh..275...67H}. They concluded that k-means has a good overlap 
with the CHs obtained with CATCH, especially when using the AIA 193 \AA\ channel.

\section{Linear models: supervised}
Linear regression is arguably the simplest model used in statistics and
machine learning and it has become the workhorse of these two disciplines,
also in solar physics. Its main assumption is that the signal of
interest can be developed as the weighted sum of basis functions:
\begin{equation}
    I(x) = \sum_{i=1}^M w_i K_j(x),
    \label{eq:linear_model_expansion}
\end{equation}
where $w_i$ are the weights associated with the $M$ basis functions $K_j(x)$.
The flexibility in the selection of the basis functions is one of the reasons 
for the power and flexibility of linear models. Additionally, the linear 
character of the model simplifies the calculations, in many cases
allowing to carry out analytical calculations.

\subsection{Hermite functions}
While PCA provides a purely empirical orthogonal basis set to
represent the Stokes profiles, other more classical approaches
have been tried in the literature. One that is particularly relevant
is the use of Hermite functions, developed by \cite{hermite_deltoro03}.
They realized that these functions when defined as
\begin{equation}
    h_n(\lambda) = \left( 2^n n! \sqrt{\pi} \right)^{-1/2} \exp\left[-\lambda^2/2 \right] H_n(\lambda),
\end{equation}
where $H_n(\lambda)$ are the Hermite polynomials as a function of the wavelength $\lambda$, look very similar to the Stokes
profiles when displayed in wavelength units normalized to the width of the
spectral line. $h_0(\lambda)$ is a Gaussian function, very similar to Stokes $I$,
$h_1(\lambda)$ looks similar to Stokes $V$ when dominated by the Zeeman effect, while
$h_2(\lambda)$ looks similar to Stokes $Q$ and $U$ in the same regime. Although interesting
from a mathematical point of view, the Hermite expansion has not been used
in real situations because they work well only when all the profiles have
a definite width. When Stokes profiles of different widths are present in the
field of view, empirical decompositions like PCA are definitely much more efficient.

\subsection{Relevance vector machines}
\label{sec:rvm}
Very powerful methods have been proposed and used in solar physics for regression based on non-parametric models. 
Non-parametric regression relies on the application
of a sufficiently general function that only depends on observed
quantities and that is used to approximate the observations. A very
flexible and efficient non-parametric regression method is that of the relevance
Vector Machines \citep[RVM;][]{tipping00}, a Bayesian update of the 
support vector machine learning technique of \cite{vapnik95}\footnote{\url{https://github.com/aasensio/rvm}}. 
In this
case, the general function is just the linear combination of user-defined kernels
of Eq. (\ref{eq:linear_model_expansion}) with $x=\lambda$.
The $K_j(\lambda)$ functions are arbitrary and defined in advance,
and $w_i$ is the weight associated to the $i$-th kernel function. 
The parameters 
we infer from the data appear linearly in the model once
the kernel functions are fixed. For instance, if the kernel functions 
are chosen to be polynomials, one ends up with a standard
polynomial regression.

The main advantage of non-parametric regression is that the
model automatically adapts to the observations. For this adaptation
to occur, the basis functions should ideally capture all possible
ways in which the signal can behave. The number of
basis functions one can include in the linear regression can
be arbitrarily large, 
making Eq. (\ref{eq:linear_model_expansion}) a very powerful model for any unknown signal.
As an example, one can use a combination of polynomials of many different
orders, sinusoidal of many frequencies, and Gaussians at different positions and with different 
widths to approximate a very general spectral line.

Obviously, this makes the regression problem ill-defined and the solution
se\-ve\-rely overfits the data provided that $M$ is large enough. For this reason, \cite{tipping00}
proposed to circumvent overfitting by pursuing a hierarchical Bayesian approach. In this
case, a Gaussian prior is imposed for each one of the $w_i$. This prior is
made dependent on a set of hyperparameters $\alpha_i$, which are learned from
the data. If a Jeffreys' prior is imposed on $\alpha_i$, i.e., $p(\alpha_i) \propto \alpha_i^{-1}$,
the resulting prior for $w_i$ is $p(w_i) \propto |w_i|^{-1}$. 
In essence, with the specific priors, in the limiting case
that $\alpha_i$ tends to infinity, the marginal prior for $w_i$ is so peaked at
zero that it is compatible with a Dirac delta. This means that this
specific $w_i$ does not contribute to the model and can
be dropped without impact. This regularization
proposed by \cite{tipping00} leads to a sparse $\mathbf{w}$ vector, so an 
automatic relevance determination is implemented in the method.

This method was applied for the first time for denoising purposes
by \cite{asensiomanso12} in solar physics. For this purpose, one selects Gaussian functions
of different widths centered at each one of the spectral points observed. This
obviously constitutes an overdetermined non-orthogonal dictionary\footnote{A dictionary is
a set of potentially non-orthogonal functions that are used to represent
a signal as a linear expansion.} but the
regularizing properties of RVM help in keeping only a few active 
Gaussians, which explain the observations. The remaining
signal is considered to be noise. \cite{ariste_2014} proposed it as a very
efficient method for fringe removal from data. Fringes appear in observed spectra
because of the internal reflection in thin plates in the optical path.
As a consequence, the observed spectrum can be understood as a combination of
quasi-periodic fringes plus the original spectrum. \cite{ariste_2014} proposed
Gaussian functions, $G_j(\lambda)$ of different widths for explaining the spectral lines and
a combination of sines and cosines, $P_j(\lambda)$ for explaining the fringes:
\begin{equation}
    I(\lambda) = \sum_{i=1}^M p_i P_j(\lambda) + \sum_{i=1}^M w_i G_j(\lambda).
\end{equation}
The sparsity regularization that is part of RVM produces that spectral lines are
not efficiently developed with periodic functions. One would need lots
of sines and cosines to do that and this is penalized by the model. Likewise, fringes are not
efficiently developed with Gaussians for precisely the same reason. Once
the regression is done, defringing is done by removing the quasi-periodic 
component and computing:
\begin{equation}
    I_\mathrm{defringed}(\lambda) \approx \sum_{i=1}^M w_i G_j(\lambda).
\end{equation}

\subsection{Compressed sensing and sparsity regularization}
\label{sec:regularization}
The theory of compressed sensing has emerged recently to solve
strongly undetermined problems. One case of that is the recovery
of signals from measurements. It is a well-known fact that 
band-limited signals need to be sampled according to the
Nyquist-Shannon theorem. If not, the latent function cannot be properly
recovered from the samples. During the last few years, the emerging theory of compressed 
sensing \citep[CS;][]{candes06,donoho06} has shown
that this sampling is indeed too restrictive when some details of
the signal structure are known in advance. Although this might
sound counterintuitive, it is indeed true that, in many instances, 
natural signals have a structure that is known in advance, in many
cases motivated by physical arguments. For instance, stellar
oscillations can be represented by sinusoidal functions of 
different frequencies, images can be represented in a multiresolution 
analysis using wavelets, etc. The key point is that, typically,
only a few elements of the basis set in which we develop the
signal are necessary for an accurate description of the important
physical information. The innovative character of CS is that this
compressibility of the observed signals is inherently taken into
account in the measurement step, and not only in the post-analysis, 
thus leading to efficient measurement protocols.
Instead of measuring the full signal (wavelength variation of
the Stokes profiles in our case), under the CS framework one
measures a few linear projections of the signal along some 
vectors are known in advance and reconstruct the signal solving a
nonlinear problem. For a more in-depth description, we refer the reader 
to recent references \citep[e.g.,][and references therein]{baraniuk07,candes08}.

The usage of compressive sensing techniques for the measurement 
of a signal, represented as a vector $\mathbf{x}'$ of length $M$, is
based on the following two key ideas:

\begin{enumerate}
    \item Instead of measuring the signal itself, one measures the
    scalar product of the signal with carefully\footnote{The precise meaning
    of ``carefully'' can be found in \cite{candes06}.} selected vectors:
    \begin{equation}
        \mathbf{y} = \boldsymbol{\Phi} \mathbf{x}' + \mathbf{e},
        \label{eq:cs_measurement}
    \end{equation}
    where $\mathbf{y}$ is the vector of measurements of dimension $N$, $\boldsymbol{\Phi}$ is
    an $N \times M$ sensing matrix and $\mathbf{e}$ is a vector of dimension $N$
    that characterizes the noise on the measurement process.
    Note that the previous equation describes the most general
    linear multiplexing scheme in which the number of measurements 
    $M$ and the length of the signal $N$ may differ. In
    the standard multiplexing case, the number of measured scalar products 
    equals the dimension of the signal ($N = M$).
    Consequently, it is possible to recover the vector $\mathbf{x}'$ provided
    that $\textrm{rank}(\boldsymbol{\Phi}) = N$, so that the problem is not ill-conditioned.
    In other words, one has to verify that every row of the $\boldsymbol{\Phi}$
    matrix is orthogonal with respect to every other row.
    \item The assumption that the
    signal of interest is sparse in a certain basis set (or can be
    efficiently compressed in this basis set). Any compressible
    signal can be written, in general, as:
    \begin{equation}
        \mathbf{x}' = \mathbf{W}^T \mathbf{x},    
    \end{equation}
    where $\mathbf{x}$ is a $K$-sparse vector (if only $K$ elements
    of the vector are different from zero) of size $M$ and $\mathbf{W}^T$
    is the transpose of an $M \times M$ transformation matrix associated with the
    basis set in which the signal is sparse. For instance, $\mathbf{W}$ can be
    the Fourier matrix if the signal $\mathbf{x}$ is the combination of a few
    sinusoidal components. Other transformations of interest are
    the wavelet matrices or even empirical transformation matrices like those 
    found using principal component analysis.
\end{enumerate}
The combination of those ingredients leads to the multiplexing scheme:
\begin{equation}
    \mathbf{y} = \boldsymbol{\Phi} \mathbf{W}^T \mathbf{x} + \mathbf{e},
    \label{eq:cs}
\end{equation}
with the hypothesis that $\mathbf{x}$ is sparse, which renders CS feasible.
It has been demonstrated by \cite{candes06} that, even
if $\mathrm{rank}(\boldsymbol{\Phi} \mathbf{W}^T) < N$ (we have 
fewer equations than unknowns), the signal $\mathbf{x}$ can be recovered with 
overwhelming probability when using appropriately chosen sensing matrices $\boldsymbol{\Phi}$.
When the number of equations is less than the number of unknowns, 
it is usual to solve Eq. (\ref{eq:cs}) using least-squares methods
that try to minimize the $\ell_2$ norm\footnote{The $\ell_q$ norm of 
a vector is given by $\Vert \mathbf{x} \Vert_q=\left( \sum x_i^q \right)^{1/q}$ when $q\geq 1$.} of the residual. This is usually 
accomplished using techniques based on the singular value
decomposition \citep[see, e.g.,][]{numerical_recipes86}. However, such 
minimization is known to return non-sparse results \citep[e.g.,][]{romberg08}. 
A more appropriate solution is to look for the vector with the smallest $\ell_0$ pseudo-norm
(the number of non-zero elements of the vector) that fulfills the
equation:
\begin{equation}
\argmin_x \Vert \mathbf{x} \Vert_0 \quad \text{subject to} \quad 
\Vert \mathbf{y} -
\boldsymbol{\Phi} \mathbf{W}^T \mathbf{x} \Vert_2 < \epsilon,
\end{equation}
where $\epsilon$ is an appropriately small quantity. The solution to the previous problem is, in general, not computationally feasible.
However, \cite{candes06,candes06_2} demonstrated that, under certain
conditions for the matrix $\boldsymbol{\Phi} \mathbf{W}^T$ \citep{candes06}, 
the problem reduces to:
\begin{equation}
\argmin_x \Vert \mathbf{x} \Vert_1 \quad \text{subject to} \quad 
\Vert \mathbf{y} -
\boldsymbol{\Phi} \mathbf{W}^T \mathbf{x} \Vert_2 < \epsilon,
\label{eq:l1_optimization}
\end{equation}
The advantage lies in the fact that very efficient numerical methods exist 
for the solution to such a problem\footnote{Some CS problems can be
solved using the \texttt{scikit-learn} Python package.}.

The theory of CS has extensively been used in solar physics after
\cite{asensio_lopez_cs10} introduced it into the field of spectropolarimetry.
Given that the theory relies on the compressibility of signals, these authors
tested whether this is indeed practically the case for the Stokes profiles. 
They showed that polarimetric signals in many spectral lines can be 
efficiently compressed using PCA and also non-empirical basis sets like
different families of wavelets. Once this is verified, they proposed
several potential applications of the CS theory to the measurement of
Stokes profiles. The first one is the conceptual idea of a multiplexing spectro-imager.
This is an extension of the classical double pass subtractive spectrographs which
work as follows: i) the slit of the standard spectrograph is removed; ii) a 
coded narrow slit following a Hadamard orthogonal sequence
is located in the focal plane of the spectrograph, together with a
device to return the light through the spectrograph in subtractive mode.
As a consequence, an image is formed at the entrance of the spectrograph
where each column corresponds to a different wavelength.
This idea later became real with the development of the
Tunable Universal Narrowband Imaging Spectrographs \citep[TUNIS;][]{2010AN....331..658L,2011CoSka..41...99L}. 
The inverse problem to recover the original
monochromatic images is solved using CS and a sparsity constraint in the
spectral direction. 

\cite{asensio_lopez_cs10} also proposed a sub-Nyquist
spectrograph, in which the pixel size is several times larger than the spectral sampling
of the spectrograph. The original spectral resolution of the spectrograph
is obtained by solving again a CS problem. The authors demonstrated that,
under certain conditions, the original resolution can be recovered. This
might be of relevance for very high resolution spectrographs, which also
require cameras with a large number of pixels to cover a sufficiently large
spectral range.

Another idea suggested by \cite{2010AN....331..652A} was
the application of the CS theory to Fabry-Perot etalons (FPE). Almost all successful FPE consists of three optical elements: a relatively narrow filter and two etalons of different free spectral ranges. An etalon is
a thin plate that works as a periodic frequency filter with well-defined
transmission peaks of high transparency. When
the two etalons are appropriately aligned and tuned, the transmission profile
of the combination has a very high transmission peak.
The secondary transmission peaks are strongly reduced, although this spectral
structure is again periodic with a much larger period. The narrow
filter serves to isolate only one of the transmission peaks. Tuning 
the etalons is a very difficult task and, for this reason, only
a few such instruments exist. \cite{2010AN....331..652A} suggested
that one can use the CS theory to remove one of the etalons and 
still recover the original signal. The numerical experiments using
PCA as a sparsity-inducing basis set was successful. However, no
instrument is still based on this idea. Probably one of the reasons
is that one needs to precompute the basis set, and for this, a 
normal spectrograph is needed. Recently, \cite{molnar20} demonstrated
that the application of neural networks for the solution of CS problems can overcome
this difficulty and recover a large fraction of the spectral resolution
lost during the observation with the instrument.

The idea of a sub-Nyquist polarimeter was put forward by 
\cite{asensio_ao_16} using the CS theory. Such a polarimeter modulates
the polarimetric properties of the incoming light
at very high frequencies (roughly at kHz rates) to freeze the 
variations of the refraction index
of the Earth atmosphere, but measures at a much slower rate (of only a few
hundred Hz). The camera then integrates the modifications to the
Stokes parameters produced by seeing variations. Consequently, one ends
up solving a linear recovery problem like that of Eq. (\ref{eq:l1_optimization}),
under the assumption that the seeing variations are compressible in the 
Fourier basis. This is indeed approximately the case given that the
power spectrum of the seeing roughly follows a $1/f^2$ law. The simulations
carried out by \cite{asensio_ao_16} demonstrated that it is possible
to recover the seeing variations at kHz frequencies from integrations
one order of magnitude slower, with a very robust behavior with noise.

Another very fruitful field of application of compressed sensing is in the thermal
diagnostics of the corona. The multiband observations capabilities of the
Atmospheric Imaging Assembly instrument \citep[AIA;][]{lemen_aia12} onboard the
Solar Dynamics Observatory \citep[SDO;][]{sdo2012} can be potentially used
to constrain the temperature and densities of the optically thin plasma in the
solar corona. This is done via the solution of the linear Differential Emission Measure (DEM)
problem, which can be posed as a linear system once the problem is 
discretized. The DEM problem is severely undetermined and its solution must
be regularized to find a reliable result. \cite{2015ApJ...807..143C} applied a sparsity 
constraint by posing the DEM inversion problem in the form of Eq. (\ref{eq:l1_optimization}), with 
the additional constraint that each component of the solution vector be non-negative (since the 
EM is proportional to the square of the free electron density, it must be non-negative to be 
physically meaningful). The method works by proposing an overcomplete and non-orthogonal dictionary composed of
Dirac-delta and Gaussian functions that cover the expected range of temperatures
in a logarithmic scale. The solution method imposes sparsity on the coefficients
associated with the elements of the dictionary to find the final combination
that explains the observations. The method was validated in a large variety of synthetic cases, from
simple ones to thermodynamic models obtained from a fully compressible, 3D
magneto-hydrodynamic (MHD) simulation of an active region. 
Later, \cite{2018ApJ...856L..17S} pointed out that the selection of
widths of the Gaussians that are part of the dictionary proposed by default
by \cite{2015ApJ...807..143C} could lead to some problems in flaring regions. 
They decreased the default width of some of the Gaussians and also increased
the $\log T$ gridding to allow for more thermal structure. The thermal structure
inferred from AIA data alone is, consequently, more consistent with thermal X-ray observations.

Compressed sensing has also been proposed by \cite{2019ApJ...882...13C} for the
analysis of current and future multi-slit spectroscopic instruments, like the Multi-slit Solar
Explorer \citep[MUSE;][]{DePontieu:MUSE}. These multi-slit instruments observe different regions of the solar
surface. The dispersive element used for the analysis of the spectrum produces, at the detector, a
superposition of spectra originating from all slits. Disentangling this mixture is again
done by solving a linear problem like that of Eq. (\ref{eq:cs_measurement}), where the
mixture matrix depends on the specifics of the instrument. \cite{2019ApJ...882...13C} proposed
that an $\ell_1$ constraint can be used to successfully solve the problem. This method has also been adapted for unfolding overlapping EUV spectra in slitless imaging spectrometer data, e.g., for the COronal Spectroscopic Imager in the EUV \citep[COSIE][]{Winebarger:COSIE,Golub:COSIE}.

Sparsity constraints can also be applied to the solution of nonlinear
problems, like the inversion of Stokes profiles. In this case, the 
problem to be solved is:
\begin{equation}
\argmin_x \Vert \mathbf{x} \Vert_1 \quad \text{subject to} \quad 
\Vert \mathbf{y} -
f(\mathbf{W}^T \mathbf{x}) \Vert_2 < \epsilon,
\label{eq:l1_optimization_nonlinear}
\end{equation}
where $\mathbf{y}_\mathrm{syn}=f(\mathbf{p})$ are the synthetic Stokes profiles. These
are obtained by solving the radiative transfer equation on a model
atmosphere parameterized by the vector of physical properties
$\mathbf{p}$. Using this approach, \cite{2015A&A...577A.140A} 
developed a new 2D inversion code under the Milne-Eddington approximation \cite[see][]{landi_landolfi04}.
The solution is regularized by assuming that the maps of physical properties
are sparse in a wavelet basis. The sparsity constraint effectively reduces
the number of free parameters of the problem and produces much cleaner inverted
maps. This approach has also been exploited by \cite{2016A&A...590A..87A} to
invert Stokes profiles that can be affected by systematic effects that are not
part of the line formation model (e.g., fringes, blends, etc.).

Hybrids of unsupervised and supervised models including sparsity have also been built
for solar flare prediction \citep{2018ApJ...853...90B}. In this case, a sparsity
constrained linear model is used to extract relevant features from the
observations, while a variant of k-means is used to cluster the
resulting features. The results show that the synergy between the
supervised and unsupervised methods performs classification better than
previous approaches.

\section{Deep neural networks}
\label{sec:ann}
Arguably the most successful machine learning methods nowadays are based on deep
nonlinear artificial neural networks (ANN), especially deep neural networks (DNN). 
For this reason, we focus this section on the 
description of ANNs which we consider to be models with great potential in
the field.

\begin{figure}
\centering
\includegraphics[width=0.6\textwidth]{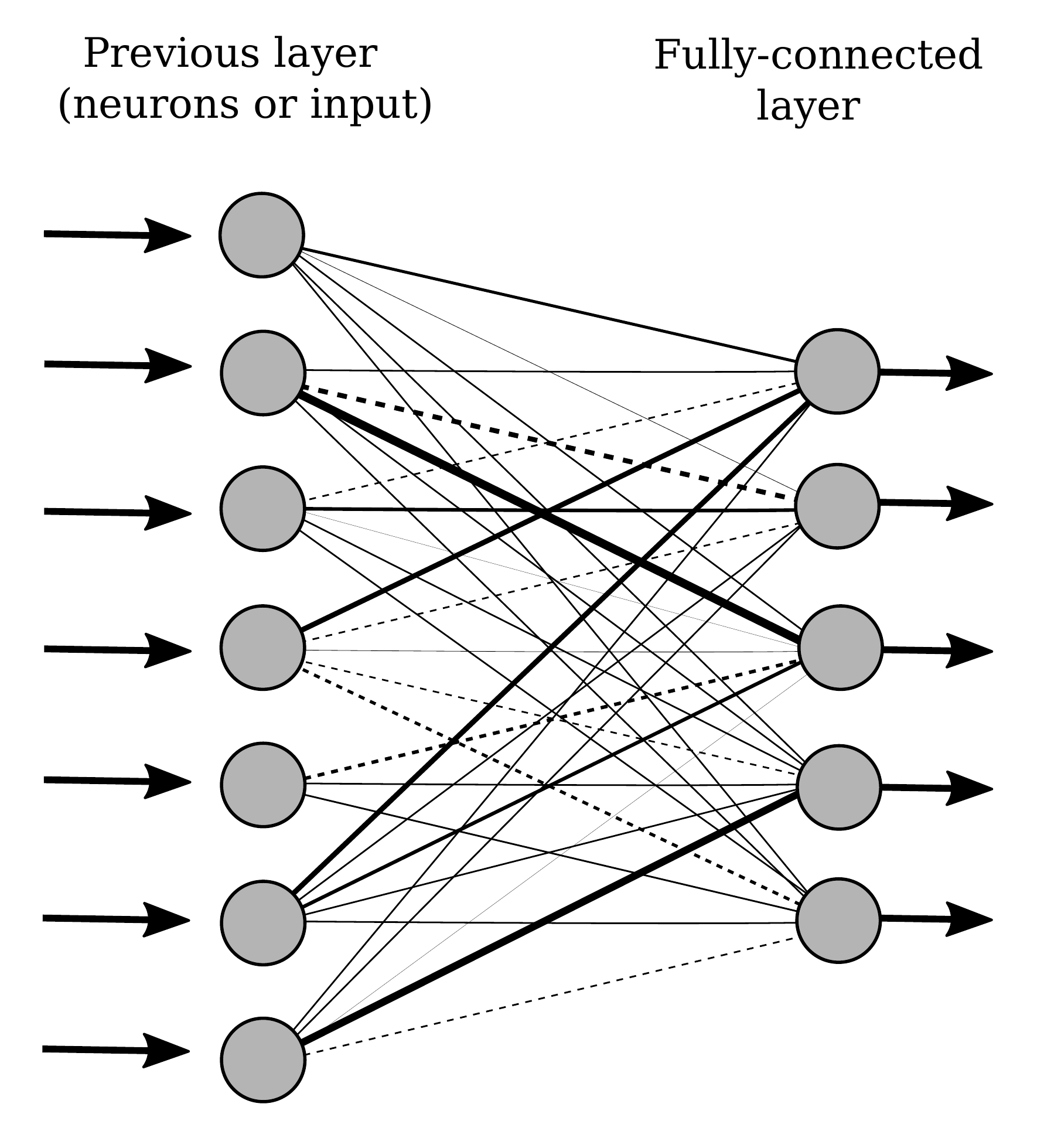}
\caption{Building block of a fully-connected neural network. Each input of the previous 
layer is connected to each neuron of the output. Each connection is represented by different 
lines where the width is proportional to the absolute value of the weight. Solid
lines represent positive weights while dashed lines refer to negative weights.}
\label{fig:networks}
\end{figure}

ANNs are well-known computing systems based on connectionism
that can be considered to be universal approximants \citep{B96} to arbitrary functions 
\citep[a theorem demonstrated by][]{cybenko88}. They are inspired by the
connectivity of animal brains and their origin can be traced back to the 1940s. At that time,
some ideas of how to carry out computations based on mimicking animal brains appeared
\citep{mcculloch_pitts43}.
After some theoretical advances, \cite{Rosenblatt58theperceptron} built the Mark I Perceptron machine, 
the first implementation of a perceptron, a supervised algorithm for 
binary classification. ANNs slowly evolved over the decades but never emerged as
the method of choice for machine learning. The fundamental reason for this was, as demonstrated
in recent years with the success of deep learning, fundamentally wrong. Their training is based on the optimization of 
a scalar loss function that is non-convex in the parameters. Consequently, 
locating the global minima is a daunting task. In fact, it may not exist at all and the loss
landscape is made of a plethora of local minima. For this reason, the machine learning
community preferred to use methods based on convex loss functions, like the ones presented in the previous
section. Only in recent years, researchers are starting to understand the loss landscape
and realize that non-convexity is in fact the property that has opened up
the current revolution in machine learning.

The building block of an artificial neural network is shown in Fig.~\ref{fig:networks}. The most basic constituent of a neural network is the neuron (inspired by biological
neurons but not strictly equivalent), shown as grey circles in the 
figure. From a mathematical point of view, a neuron can be understood as a simple storage of a real number, which is then
used in some predefined operations when this neuron is connected with other neurons. These
connections can be massive and this connectivity is precisely the one that gives
enormous representation power to neural networks. The state of each neuron $i$ is computed
by a very basic operation on the input vector: it multiplies all the input values $x_j$ by some weights $w_j$, 
adds some bias $b_i$ and finally returns the value of a certain user-defined
nonlinear activation function $f(x)$. In mathematical notation, a neuron computes:
\begin{equation}
y_i = f \left( \sum_j x_j\cdot w_j + b_i \right),
\end{equation}
which is a generalization of the simple model for a neuron of \cite{mcculloch_pitts43}.
The output $y_i$ is then input in another neuron that does a similar operation. Therefore,
neural networks can be considered to be a complex composition of very simple nonlinear functions.
Each layer $k$ is parameterized by a set of parameters $\boldsymbol{\theta}^{(k)}$. After passing
through the $L$ layers the output can be written as:
\begin{equation}
    \mathbf{y} = f(\mathbf{x};\boldsymbol{\theta}) = f^{(L)}_{\boldsymbol{\theta}^{(L)}}(\cdots f^{(2)}_{\boldsymbol{\theta}^{(2)}}
    (f^{(1)}_{\boldsymbol{\theta}^{(1)}}(\mathbf{x}))).
\end{equation}
It is sometimes useful to make explicit all intermediate features of the neural network:
\begin{eqnarray}
    \mathbf{y}^{(1)} &=& f^{(1)}_{\boldsymbol{\theta}^{(1)}}(\mathbf{x}) \nonumber \\
    \mathbf{y}^{(2)} &=& f^{(2)}_{\boldsymbol{\theta}^{(2)}}(\mathbf{y}^{(1)}) \nonumber \\
    &\cdots& \nonumber \\
    \mathbf{y}^{(L-1)} &=& f^{(L-1)}_{\boldsymbol{\theta}^{(L-1)}}(\mathbf{y}^{(L-2)}) \nonumber \\
    \mathbf{y} &=& f^{(L)}_{\boldsymbol{\theta}^{(L)}}(\mathbf{y}^{(L-1)})
\end{eqnarray}
Using the standard notation for function composition ($\circ$), a neural network then
provides the following output:
\begin{equation}
    \mathbf{y} = (f^{(L)}_{\boldsymbol{\theta}^{(L)}} \circ f^{(L-1)}_{\boldsymbol{\theta}^{(L-1)}} \circ \cdots \circ f^{(2)}_{\boldsymbol{\theta}^{(2)}} \circ f^{(1)}_{\boldsymbol{\theta}^{(1)}})(\mathbf{x}).
    \label{eq:nn_composition}
\end{equation}
Precisely this composition character is the one that allows graphical models like the
one depicted in Fig.~\ref{fig:networks} to be useful. The compositional character opens
up the possibility to split very complex models as the combination of smaller submodels. This 
abstract \emph{block} representation gives neural networks an enormous potential because they can be engineered quite easily
to the solution of a very broad class of problems. 

In many cases, an ANN can be understood as a pipeline where the information goes from the input to the output, 
where each neuron makes a transformation like the one described above. 
Each transformation deforms the topology of the input space \citep{Naitzat:2020} with the
hope that the final prediction turns out to happen in a much simpler space.
Neurons are usually grouped in layers and the number of connected layers defines
the depth of the network. For reasons that will become clear in Sect.~\ref{sec:backpropagation}, 
very deep neural networks are hard to train, and only in the last decade, we have been able to
do that. Currently, some of the most successful neural networks contain millions or billions of neurons organized in
several tens or hundreds of layers \citep{veryDeep2014}. 

One may ask: How do we know which weights and biases to use to get an optimal result for our supervised learning problem? 
The optimal values for the weights and biases are unknown before training. They are parameters of the ANN, typically 
initialized by sampling from a random distribution (e.g., normal distribution). The task of supervised training 
is to provide samples of input and targets (rows of $\mathbf{X}$ and $\mathbf{Y}$) so that the loss function 
can be evaluated, and gradient descent be used to update the parameters of the ANN. See 
Sect.~\ref{sec:training} for a discussion of how ANNs are efficiently trained.

\subsection{Architectures}
\subsubsection{Multi-layer fully connected neural networks}
The most used type of neural network from the 1980s to the 2000s
is the fully connected network \cite[FCN; see][for an overview]{Overview2014}, 
in which every input of all considered layers is connected to every neuron of the following layer. Likewise, the output 
transformation becomes the input of the following layer (see left panel of Fig.~\ref{fig:networks}).
This kind of architecture succeeded to solve problems that were considered to 
be not easily solvable, such as the recognition of handwritten characters \citep{B96}. 

\subsubsection{Convolutional neural networks}
Despite the relative success of neural networks, their application to 
high-dimensional objects like images or videos turned out to be
an obstacle. The fundamental reason was that the number of
weights in a fully connected network increases extremely fast with the 
complexity of the network (defined by the number of neurons) and the computation quickly becomes unfeasible.
As each neuron of a given layer is connected to every neuron of the previous one, adding a new neuron to a layer
also implies adding a large number of weights, equal to the number of neurons in the layer. 
The number of weights of a deep fully connected neural network is, then:
\begin{equation}
    N = \sum_\mathrm{i \in layers} N_i N_{i-1},
\end{equation}
where $N_i$ is the number of neurons of layer $i$.
A larger number of neurons implies then a huge increase in the number of connections.
This became an apparently insurmountable handicap, which was only solved with the
appearance of convolution neural networks \cite[CNN or ConvNets;][]{LeCun1998}.
The idea brought forward by \cite{LeCun1998} was motivated by biological processes
and exploit the fact of sharing weights across the input. From a mathematical
point of view, CNNs define a set of kernels of small size that are then used
as convolution kernels. The input is then convolved with them, providing as output 
that is known as \emph{feature map}. The fundamental advantage of CNNs is that sharing
the weights across the whole input drastically reduces the number of unknowns. As a side effect,
convolutions also make CNN's shift invariant (features can be detected in an image irrespectively of
where they are located), a very powerful inductive bias. 

For a two-dimensional input $X$ of size $N \times N$ with $C$ 
channels\footnote{The term channels is inherited from
the those of a color image (e.g., RGB channels). However, the term has a much more general
scope and can be used for arbitrary quantities \cite[see][for an application]{Asensio2017}.} 
(a cube or tensor of size $C \times N \times N$), each output feature 
map $O_i$ (with size $1 \times N \times N$) of a convolutional layer is computed as:
\begin{equation}
O_i=K_i * X + b_i, \qquad i=1,\ldots,M,\label{eq:conv}
\end{equation}
where $K_i$ is the $C \times K \times K$ kernel tensor associated with the output feature map $i$, 
$b_i$ is a bias value ($1 \times 1 \times 1$) and the symbol $*$ is used to refer to the the convolution 
operation {\footnote{In most ML 
framework implementations of convolutional layers, the $*$ operator is actually the cross-correlation 
instead of convolution, as is usually defined in the mathematical literature. The difference between
the two operations are irrelevant because kernels will be learned during training, but the correlation is
more computationally efficient.}}. 
Once the convolution with $M$ different kernels is carried out and stacked together, the output 
$O$ will have size $N \times N \times M$. All convolutions are here indeed intrinsically three-dimensional, 
but one could see them as the total of $M \times C$ two-dimensional convolutions plus the 
bias.

Like the weights of a fully-connected network, the optimal weights of a convolutional kernel are unknown. Instead 
they are initialized (often with values sampled from random distributions) and updated during training time. 
Regardless of the value of the kernels, the individual application of the convolutional operator 
(as given in Eq. (\ref{eq:conv})) and the serial composition of such operations remain linear 
operations. To introduce nonlinearities and increase the expressivity, convolutional layers are often 
succeeded by activation functions (see Sect.~\ref{sec:activation}). Pooling layers are 
also used to improve the spatial connectivity of CNNs and to reduce the dimensionality of the input. 
For example, the~\emph{maxpool} operation returns the maximum value in 
non-overlapping windows of size $N_{\rm sub}\times N_{\rm sub}$ pixels. 
Often, applications of the convolutional 
layer and/or maxpool layers are strided. For a stride of one, a convolution displaces
the kernel on the input with a step of one pixel. When the stride is larger than one, the
convolution kernel is displaced in larger steps. This practice reduces the nominal 
dimensionality (i.e., the number of components of the output, not the intrinsic rank) of the output. 
Repeated application of this type of downscaling reduces the number of trainable 
parameters, and thus the computational effort needed to train the network.

Like fully connected layers, CNNs are typically composed of several layers. This 
layer-wise architecture exploits the 
property that many natural signals are generated by a hierarchical composition of 
patterns. For instance, faces are composed of eyes, while eyes contain a similar internal structure. 
This way, one can devise specific kernels that extract this information from 
the input. CNNs work on the idea that each convolution layer extracts information about certain patterns, 
which is done during the training by iteratively adapting the set of convolutional
kernels to the specific features to locate. This obviously leads to a much more optimal solution as
compared with hand-crafted kernels. Despite the exponentially smaller 
number of free parameters as compared with a fully-connected ANN, CNNs often produce much better 
results. 

It is interesting to note that, since a convolutional layer just computes 
sums and multiplications of the inputs, the same operation could be done with
a multi-layer FCN. However, training such a neural network would require 
huge amounts of training data to learn the natural inductive biases of locality and
shift invariance of CNNs \citep{Peyrard15}.

Although a convolutional layer significantly decreases 
the number of free parameters as compared with a fully-connected layer, it 
introduces some hyperparameters (global characteristics of the network) to be set in 
advance: the number of kernels
to be used (number of feature maps to extract from the input), the size of 
each kernel with its corresponding padding (to deal with the borders of the image)
and the stride (step to be used during the convolution
operation) and the number of convolutional layers and specific architecture to 
use in the network. As a general rule, the deeper the CNN, the better the result, 
at the expense of a more difficult and computationally intensive training. 

\subsubsection{Recurrent neural networks}
\label{subsubsection:RNN}
The efficient description of sequences of data requires neural networks
with a different architecture. In this case, it turns out to be important to 
have feedback connections to keep track of long-term dependencies in the
input sequences. Recurrent neural networks \citep[RNNs;][]{1986Natur.323..533R}
can keep track of these dependencies by unrolling the 
network for all the elements of the sequence and connecting the output of
each neuron in the sequence to the input of the next one. 
RNNs are designed to learn sequential or time varying patterns \citep{Medsker21}, 
like for example the solar cycle variation. RNNs started to be used initially for 
solving character recognition problems, but they were also implemented in many other fields, like 
financial predictions, the verification of the water quality, etc. The architecture 
can be built on fully or partially-connected layers, including multilayer feedforward 
networks and specific learning algorithms were developed for the RNNs \citep{Medsker21}. 
The RNNs were initially difficult to train because they suffer from the vanishing gradient
problem (see Sect.~\ref{sec:backpropagation}). Different architectures were
proposed to cure this problem, and the long short-term 
memory \citep[LSTM;][]{LSTM} is arguably the most
successful.
In solar physics, the LSTM was intensively applied for the prediction of the current 
solar cycle (see Sect.~\ref{subsection:SCpred}). 

\subsubsection{Attention and Transformers}
The attention mechanism, which is a variety of algorithms that compute the
output by weighting the importance of different features of the data, has
become important thanks to the Transformer model \citep{Transformer17}.
Transformers can translate a sequence of arbitrary length into a sequence 
of the same length of features of arbitrary dimensionality using self-attention.
Given an input $\mathbf{X}$, self-attention works by building matrices of values ($\mathbf{V}$), queries 
($\mathbf{Q}$) and keys ($\mathbf{K}$) by using trainable weight matrices:
\begin{equation}
    \mathbf{V} = \mathbf{W}_V \mathbf{X}, \quad
    \mathbf{Q} = \mathbf{W}_Q \mathbf{X}, \quad
    \mathbf{K} = \mathbf{W}_K \mathbf{X}
\end{equation}
and computing:
\begin{equation}
    \mathrm{Att}(\mathbf{Q},\mathbf{K},\mathbf{V}) = 
    \mathrm{softmax} \left( \frac{\mathbf{Q} \mathbf{K}^T}{\sqrt{d_k}} \right) \mathbf{V},
\end{equation}
where $d_k$ is the dimensionality of the queries and keys.
The product of the query and key matrices is a score matrix that
defines the amount of attention that each element of the output pays 
to every element of the input sequence. This score matrix is then scaled down to
allow for more stable gradients, and a softmax is applied to trans-
form the scores into probabilities. Finally, these attention weights
are applied to the values. Transformers have not been yet
applied in solar physics (see Sect.~\ref{sec:farside_imaging} for an
usage of attention). However, given their success in other fields, 
we anticipate that this kind of attention model will eventually 
emerge as suitable ones in the analysis of
images or sequences.

\subsubsection{Graph neural networks}
Neural networks can also be defined in graphs, which is sometimes appropriate
for specific problems. These problems are still hard to find in solar physics
but at least one application already exists (see Sect.~\ref{sec:inversion_stokes}).
A connected graph $G=(V, E)$ is defined by the set of grid points $V$
(also known as nodes or vertices) and the set of edges connecting the grid points, $E$. 
Each node can encode relevant properties $\mathbf{p}_i$. Each edge $\mathbf{e}_{ij}$ 
connects the two nodes $i$ (sender) and $j$ (receiver), and describes relevant inter-node properties.
The computation inside the graph is based on the so-called \emph{processor}, 
made of $N$ consecutive message passing processes. Message passing is fundamental to connect 
the information in very distant nodes in the graph, given that all updates are local, as
shown in the following. Each message
passing consists of updating the latent information contained in all edges and then in all nodes, 
as follows:
\begin{eqnarray}
    \textbf{e}^{t+1}_{ij} &=& \textbf{e}^t_{ij} + f_E^{t+1}(\textbf{e}^t_{ij}, \textbf{v}^t_i, \textbf{v}^t_j), \nonumber \\
    \bar{\textbf{e}}^{t+1}_{j} &=&\sum_k f_A^{t+1}\Big( \textbf{v}^t_k, \textbf{e}^{t+1}_{kj} \Big), \nonumber \\
    \textbf{v}^{t+1}_{j} &=& f_V^{t+1}\Big( \textbf{v}_j, \bar{\textbf{e}}^{t+1}_{j} \Big),
\end{eqnarray}
where $f_E$, $f_A$ and $f_V$ are neural networks. After a predefined number of message
passing steps, one ends up with updated information in the nodes and in the edges. In
a supervised training setup, this updated information is then compared with that of the
training set and the weights of the neural networks are updated until convergence.

\begin{figure}
\centering
\includegraphics[width=\textwidth]{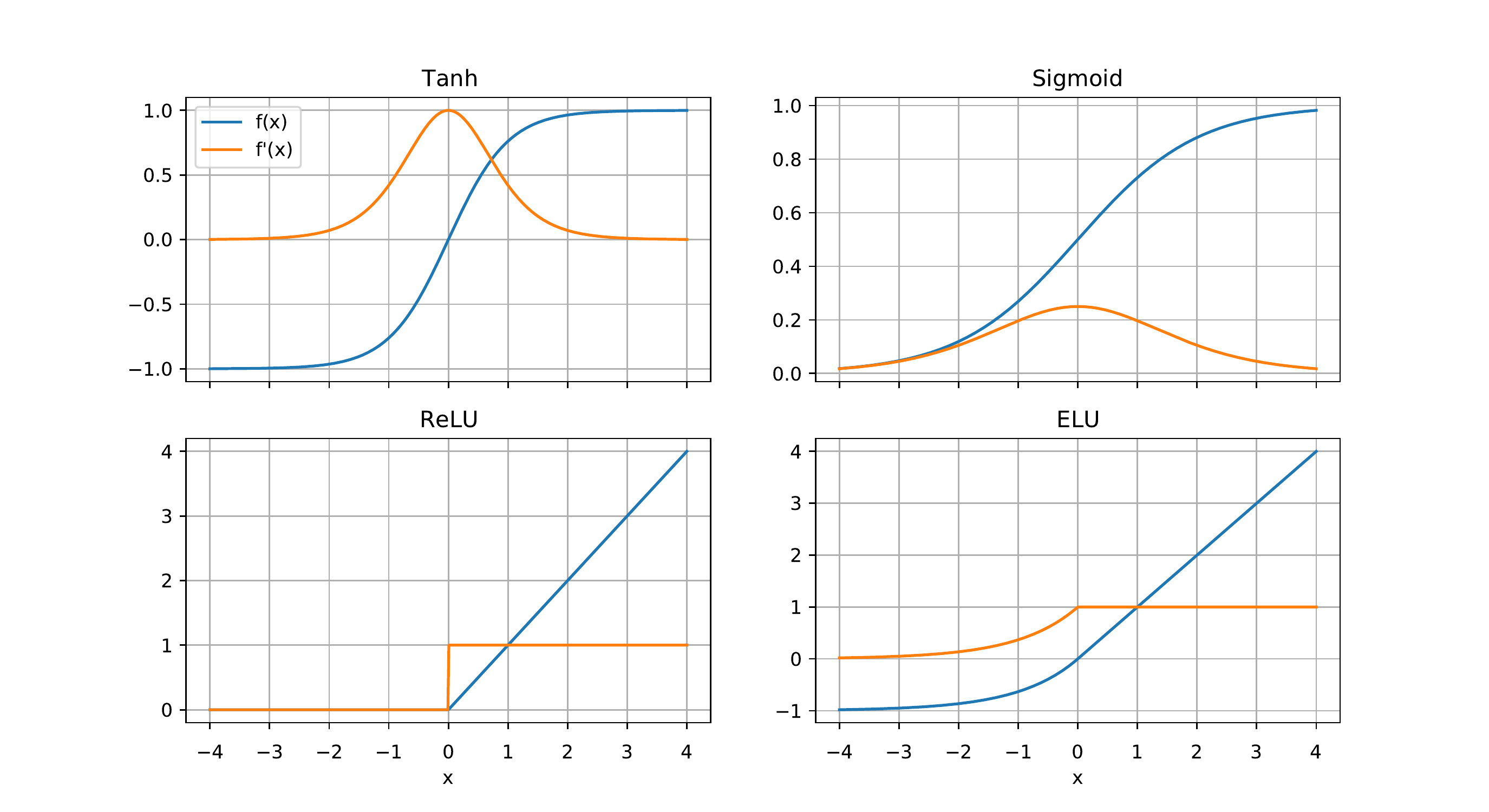}
\caption{Some activation functions often used in ANNs: hyperbolic tangent (Tanh), 
sigmoid, rectified linear unit (ReLU), and exponential linear unit (ELU).}
\label{fig:activation}
\end{figure}

\subsection{Activation layers}
\label{sec:activation}
The output of a linear layer of a neural network is often passed through a nonlinear function,
known as the \emph{activation function}. This function
introduces the non-linear character into the neural networks, which is the
source of its strength. Although hyperbolic
tangent, $f(x)=\tanh(x)$, or sigmoidal, $f(x)=[1+\exp(-x)]^{-1}$, activation units were originally 
used in ANNs (see Fig.~\ref{fig:activation}), nowadays a panoply of more convenient nonlinearities are used. 
Probably the most common activation function is 
the Rectified Linear Unit \cite[ReLU;][]{relu10} or slight variations of it, like
the exponential linear unit \citep[ELU;][]{elu2015}. The ReLU
replaces all negative values in the input by zero and keeps the rest untouched:
\begin{equation}
    \mathrm{ReLU}(x) = \max(0,x).
\end{equation}
This activation has the desirable property of having a constant
derivative for positive arguments, which greatly accelerates the training and
reduces the vanishing gradient problem.
Examples of a few activations functions are displayed in Fig.~\ref{fig:activation}.

\subsection{Training}
\label{sec:training}
Neural networks (either deep or shallow) can be seen as a very flexible parametric function that produces
an output, $\mathbf{y}$, from an input, $\mathbf{x}$, with the aid of some internal parameters,
$\boldsymbol{\theta}$. These parameters are the weights and biases of all layers, 
together with any possible learnable parameter
of the activation layers. Training is performed by iteratively modifying the vector of 
parameters $\boldsymbol{\theta}$ until a loss function is minimized. 
This can be seen as a standard maximum-likelihood optimization when the loss function
is given by the likelihood function.

\subsubsection{Loss function}
\label{sec:loss_function}
In general, a loss function is a differentiable scalar function that depends on the inputs and
outputs, as well as any parameter or internal feature of the neural network. 
Using the definition of the neural network functional form of Eq. (\ref{eq:nn_composition}), 
the most general loss function is represented by the following scalar:
\begin{equation}
    L = g(\mathbf{x},\mathbf{y},\{\boldsymbol{\theta}^{(L)},\ldots,
    \boldsymbol{\theta}^{(1)}\},\{\mathbf{y}^{(L-1)},\ldots,\mathbf{y}^{(1)}\}),
\end{equation}
where the dependence on $\boldsymbol{\theta}^{(i)}$ shows 
the contribution of the weights of all intermediate layers in the neural network, while
the dependence on $\mathbf{y}^{(i)}$ shows the dependence on all intermediate 
features.

\subsubsection{Gradient descent}
In general, and irrespective of the specific loss function, the optimization
is routinely solved using simple first-order gradient descent algorithms 
\cite[GD; see][]{Rumelhart1988}, which modifies the weights using
the gradient of the loss function with respect to the model parameters. 

In practice, procedures based on the so-called stochastic gradient descent (SGD) are used, in which
only a few examples from the training set (a batch) are used
during each iteration to compute a noisy estimation of the gradient and adjust the weights 
accordingly. A training set is then divided into $n$ batches, each one containing $B$
training examples. Although the calculated gradient in a batch is a noisy estimation of the one 
calculated with the whole training set, the training is often faster and more reliable. 
To formalize SGD, let us consider the loss function as the addition of losses over all the
$n$ batches of the training set, so that:
\begin{equation}
    L(\boldsymbol{\theta})= \sum_{j=1}^n \sum_{k=1}^B L_{kj}(\boldsymbol{\theta}),
\end{equation}
where $L_{kj}$ is the loss function for the $k$-th element of the $j$-th batch.
The standard gradient descent algorithm optimizes the loss function by updating
the parameters of the neural network using:
\begin{equation}
\boldsymbol{\theta}_{i+1} = \boldsymbol{\theta}_i -\eta\nabla L(\boldsymbol{\theta}_i) = 
\boldsymbol{\theta}_i - \eta \sum_j^n  \nabla L_j(\boldsymbol{\theta}_i),
\end{equation}
where $\eta$ is the learning rate. The SGD method updates the parameters following the same idea but
calculating the gradient using only a single batch:
\begin{equation}
\boldsymbol{\theta}_{i+1} \approx \boldsymbol{\theta}_i -\eta\nabla L_j(\boldsymbol{\theta}_i)
= \boldsymbol{\theta}_i -\eta \sum_{k=1}^B \nabla L_{jk}(\boldsymbol{\theta}_i), \qquad j=1,\ldots,n.
\end{equation}

The learning rate is used to tune the step size defined by the
gradient, which is often not optimal unless one is very far
from the optimal solution. The learning rate can be kept fixed or it can be changed 
according to our requirements. It is usually tuned to find a compromise between the 
accuracy of the network and the speed of convergence. If $\eta$ is too large, the steps 
will be too large and the solution could 
potentially overshoot the minimum. On the contrary, if it is too small it will take too 
many iterations to reach the minimum. In recent years, adaptive methods like Adam \citep{adam14} 
or RMSProp \citep{rmsprop12} have been developed to automatically tune individual learning rates
for each variable. These are still first-order algorithms in which some second-order
information from the Hessian is estimated using consecutive iterations.

\subsubsection{Backpropagation}
\label{sec:backpropagation}
The gradient of the loss function with respect to the 
free parameters of the neural network needed during training is obtained via the
backpropagation algorithm \citep{LeCun1998b}. The composite character of 
neural networks makes the calculation of these gradients easier than
for a general nonlinear function because one can recursively apply the
chain rule. To demonstrate this, it is advisable to start 
with the simple case of two layers:
\begin{align}
    L &= g\left(\mathbf{x}, \mathbf{u} \right) \\
    \mathbf{u} &= f^{(2)}_{\boldsymbol{\theta}^{(2)}} (\mathbf{v}) \\
    \mathbf{v} &= f^{(1)}_{\boldsymbol{\theta}^{(1)}} (\mathbf{x}),
    \label{eq:backprop_simple}
\end{align}
where $\mathbf{u}$ and $\mathbf{v}$ are used as intermediate results of 
hidden layers.
The gradient of the loss function with respect to both sets of $\boldsymbol{\theta}$ parameters
are given by:
\begin{align}
    \frac{\partial L}{\partial \boldsymbol{\theta}^{(1)}} &= 
    \frac{\partial L}{\partial \mathbf{u}} \frac{\partial \mathbf{u}}{\partial \mathbf{v}} 
    \frac{\partial \mathbf{v}}{\partial \boldsymbol{\theta}^{(1)}} \\
    \frac{\partial L}{\partial \boldsymbol{\theta}^{(2)}} &= 
    \frac{\partial L}{\partial \mathbf{u}} \frac{\partial \mathbf{u}}{\partial \boldsymbol{\theta}^{(2)}}.
\end{align}
The case of three layers is similarly given by:
\begin{align}
    L &= g\left(\mathbf{x}, \mathbf{u} \right) \\
    \mathbf{u} &= f^{(3)}_{\boldsymbol{\theta}^{(3)}} (\mathbf{v}) \\
    \mathbf{v} &= f^{(2)}_{\boldsymbol{\theta}^{(2)}} (\mathbf{w}) \\
    \mathbf{w} &= f^{(1)}_{\boldsymbol{\theta}^{(1)}} (\mathbf{x})
    \label{eq:backprop_simple2}
\end{align}
The gradients are given by:
\begin{align}
    \frac{\partial L}{\partial \boldsymbol{\theta}^{(1)}} &= 
    \frac{\partial L}{\partial \mathbf{u}} \frac{\partial \mathbf{u}}{\partial \mathbf{v}} 
    \frac{\partial \mathbf{v}}{\partial \mathbf{w}}
    \frac{\partial \mathbf{w}}{\partial \boldsymbol{\theta}^{(1)}} \\
    \frac{\partial L}{\partial \boldsymbol{\theta}^{(2)}} &= 
    \frac{\partial L}{\partial \mathbf{u}} 
    \frac{\partial \mathbf{u}}{\partial \mathbf{v}} 
    \frac{\partial \mathbf{v}}{\partial \boldsymbol{\theta}^{(2)}} \\
    \frac{\partial L}{\partial \boldsymbol{\theta}^{(3)}} &= 
    \frac{\partial L}{\partial \mathbf{u}} 
    \frac{\partial \mathbf{u}}{\partial \boldsymbol{\theta}^{(3)}}
\end{align}

In general, except for the first element in both previous equations, the rest
of the terms of the shape $\partial \mathbf{u}/\partial \mathbf{v}$ are the Jacobian matrices. Therefore,
the backpropagation can be understood as the multiplication of Jacobian matrices of the effect
of each individual layer. The algorithm can be implemented with relative 
simplicity by just multiplying Jacobian matrices when traversing the neural
network in the backward direction (that is precisely the reason for the name
of the algorithm). Note that this calculation is also very efficient because
one can store precomputed products of Jacobian matrices and use them afterward.

To efficiently calculate all gradients, one starts by computing 
the gradient $\partial L/\partial \mathbf{u}$ with
the loss function and the last layer of the network. Then one goes to the previous
layer and computes the Jacobians $\partial \mathbf{u}/\partial \mathbf{v}$ and
$\partial \mathbf{u}/\partial \boldsymbol{\theta}^{(3)}$. Both Jacobians
are used to update the gradients with respect to the variables $\boldsymbol{\theta}^{(2)}$
and $\boldsymbol{\theta}^{(3)}$ respectively. The procedure is iterated until the first
layer is found. In practice, this process is currently done with automatic differentiation 
techniques, implemented in packages like PyTorch\footnote{\url{https://pytorch.org/}} \citep{pytorch19}, 
Tensorflow \footnote{\url{https://www.tensorflow.org/}} \citep{tensorflow2015-whitepaper}
or JAX\footnote{\url{https://github.com/google/jax}} \citep{jax2018github}.
Because these tools deal 
with the product of Jacobians in the
neural network graph, they allow the user to easily define flexible neural network
architectures tailored to specific needs.

\subsubsection{Vanishing gradient problem}
The way neural networks are trained suffers from a problem known as 
the vanishing gradient problem \citep[e.g.,][]{Kolen:2001}. This was the reason why the field 
of artificial neural networks was somehow stalled during the years
before the first decade of the 21st century. If one considers
typical nonlinear activation functions like the $\tanh(x)$, their
derivative becomes very close to zero if the input is relatively far from 
zero. Consequently, the Jacobian of this activation function becomes very 
small and the gradient is not propagated backwards to the previous layers.
As an effect, the gradient of the loss function with respect to the first 
layers of the neural networks using $tanh(x)$-like activation function rapidly 
becomes zero. As a result, the stochastic gradient descent
cannot produce any correction on their weights. As commented before,
new activation functions like $\mathrm{ReLU}(x)$ largely solve this problem
because their derivative does not saturate.

\subsection{Bag-of-tricks as of 2023}
\subsubsection{Initialization}
Tuning the initial value of the weights and biases of all the connections turned out to
be crucial for the success of deep learning. The aim of the initialization is to
avoid the explosion or vanishing of the layer activations so that gradients can
seamlessly be backpropagated and producing changes in all the layers of the model.
If symmetric activation functions like $tanh$ are used, \cite{glorot10} noticed
that good results are found when initializing weights with a uniform distribution bounded in 
the interval $[-\sqrt{6}/\sqrt{n_\mathrm{in}+n_\mathrm{out}},\sqrt{6}/\sqrt{n_\mathrm{in}+n_\mathrm{out}}]$, where
$n_\mathrm{in}$ and $n_\mathrm{out}$ are the number of input and output connections at a given layer, respectively. 
This is currently known as \emph{Xavier} initialization. For asymmetric activation
functions, \cite{kaiming15} checked that initializing weights from a normal distribution
with zero mean and variance $2/n_\mathrm{in}$ can be efficiently used to train
very deep neural networks. This is currently known as the \emph{Kaiming} initialization.

\subsubsection{Augmentation}
The supervised training of deep neural networks often requires a large number
of examples in the training set. Many times, especially in science, building 
such large databases is
unfeasible simply because of the lack more training examples. In such a 
case, one can apply augmentation techniques as a remedy to artificially 
increase the training set. Rotations, reflections, changes in contrast, and many
other such transformations can produce new training cases that produce a more
stable result and better generalization after training.

\subsubsection{Regularization and overfitting}
Because of the large number of free parameters, especially in very deep CNNs, overfitting can be 
a problem. One would like the network to generalize well and avoid any type of ``memorization'' of
the training set. 
There is increasingly stronger empirical and theoretical evidence showing that
stochastic gradient descent methods, specific neural architectures, and the
overparameterization of very large models leads to
flat minima in the loss function that automatically produce good generalization 
\citep[e.g.,][]{hochreiter97,2020arXiv200911162B}. In other words, the
non-convex optimization problem is plagued with local minima but all of
them are equally good in their generalization properties.

One could argue that deep neural networks seem to be self-regularizing.
But, in those cases in which overfitting is found, there
are a few ways to introduce extra regularization during training. Many of them
can be understood as an addition of a prior term in the loss function so that
one optimizes for the maximum a posteriori solution instead of the maximum 
likelihood. The most used ones are weight decay and dropout. Weight decay
consists of forcing the weights of the neural network to be
small. Large weights tend to produce neural networks that are very specialized to
the training data and do not generalize well. For this reason, one typically
adds an $\ell_2$ (also known as Tikhonov) regularization term like the following:
\begin{equation}
    L_\mathrm{regularized} = L + \lambda |\boldsymbol{\theta}|^2.
\end{equation}
The strength of the regularization is controlled by the hyperparameter $\lambda$.
Dropout consists of randomly removing connections among neurons in the
neural network with probability $p$. This makes the training noisier but
introduces a certain regularization by sparsifying the weights. In essence,
neural networks learn how to solve the problem at hand even with
random perturbations to the architecture. 

\subsubsection{Normalization}
Several techniques have been described in the literature to accelerate
the training of CNNs and also to improve generalization. Batch normalization \citep{batch_normalization15}
is a very convenient and easy-to-use technique that consistently produces large accelerations in the
training. It works by normalizing every batch to have 
zero mean and unit variance. Mathematically, the input is normalized so that:
\begin{align}
y_i = \gamma \hat{x_i} + \beta \nonumber \\
\hat{x_i} = \frac{x_i - \mu}{\sqrt{\sigma^2 + \epsilon}},
\end{align}
where $\mu$ and $\sigma$ are the mean and standard deviation of the inputs on the batch and
$\epsilon=10^{-3}$ is a small number to avoid underflow. The parameters $\gamma$ and $\beta$
are learnable parameters that are modified during the training. Although batch normalization
can stabilize and accelerate training, it is true that it requires the usage of relatively
large batches so that the statistics $\mu$ and $\sigma$ are not too noisy.
Other variants of normalization have also been developed: layer normalization, instance
normalization, group 
normalization, \ldots\footnote{See \url{https://bit.ly/3XleCff}}.

We caution against the liberal use of batch normalization in physics 
applications (especially for regression) without 
careful testing. If a feedforward network $F: \mathbf{x} \longrightarrow \mathbf{y} $ is 
thought of as a mapping between two 
dimensional quantities, the use of batch norm essentially means the units in which 
the inputs are provided 
changes for every batch. 

\subsubsection{Residual blocks and skip connections}
Very deep networks usually
saturate during training, producing higher errors than shallow networks because
of difficulties during training (fundamentally produced by the vanishing
gradient problem). Residual networks \cite{residualnetwork16}
came to the rescue by obtaining state-of-the-art results with
exceptionally deep networks without adding any extra parameters and with
practically the same computational complexity. It is based 
on the idea that if $y=F(x)$ represents the desired effect of the block on the
input $x$, it is much simpler for a network to learn the deviations from the input.
This residual mapping works then by rewriting $y=x+R(x)$, with $R(x)$ a 
new neural network that describes the residual. Skip connections are specific
types of residual connections in which intermediate features of the neural
network are added or concatenated in later stages of the network (see
the U-Net architecture of Fig.~\ref{fig:unet}). They also help in
propagating gradients to the initial layers of the neural network.

\subsubsection{Specialized hardware}
The embarrassingly parallel character of the operations to be carried out in
a layer of a neural network (for instance, convolutions with different kernels can be carried out 
simultaneously without any dependence) has opened up the possibility of
using specific hardware to accelerate the calculations. GPUs were 
traditionally architected for parallel graphics rendering (using fragment shaders). 
They are optimized for Single Instruction Multiple Data (SIMD) processing. This type of parallel 
programming paradigm is suited for application to large-scale scientific datasets, and 
for dense matrix multiplication. This means GPUs are ideal for accelerating deep neural networks,
giving increases in the computation power
of more than an order of magnitude with respect to general purpose CPUs. Tensor
Processing Units (TPU) are even more specialized hardware that are, in essence,
very fast matrix multipliers. Recently, even optics-based computation hardware
has been proposed, with the promise to accelerate some computations by orders
of magnitude at very reduced power consumption \citep{Miscuglio:2020}.

It has also been verified that deep neural networks are especially tolerant to
floating point errors so that they can be easily (and routinely) trained in single-precision.
Even half-precision can be used, provided one does the backpropagation in single-precision.
Specialized GPUs and TPUs can accelerate half-precision calculations by a large factor
when compared with single-precision.

\section{Unsupervised deep learning}
\label{sec:nonlinear_unsupervised}
One of the weakest points of all linear methods described in the previous sections is
that they rely only on the information provided by second-order
statistics (correlation). Therefore, they cannot efficiently describe a dataset which
is lying in a nonlinear manifold of the original high-dimensional space. 
We expect this to be true in general, so
relying on nonlinear models has become a necessity.
Several unsupervised nonlinear models were developed in the first years of the century: 
locally linear embedding \citep[LLE;][]{lle00}, Isomap \citep{isomap00}, 
a kernelized version of PCA \citep{kpca98}, self-organizing maps \citep[SOM;][]{kohonen_SOM01}, 
autoassociative neural networks \citep{Bourlard1988} and t-SNE \citep{Hinton_Roweis_2003}. Only the last three methods 
have been used in solar physics but without much continuity. However, the
landscape in recent years has changed completely thanks to the 
deep learning revolution. It is now possible to train excellent generative models that 
capture the statistical properties of a training set 
and we should expect this line of research to produce very
interesting applications in solar physics.

\subsection{Self-organizing maps}
A self-organizing map\footnote{An implementation can be found in \url{https://github.com/bougui505/quicksom}.} is a specific type of neural network that is
trained unsupervisedly. A SOM is a way to project a high-dimensional dataset into a two-dimensional
space by keeping, as much as possible, the topological information present in the 
original space. It consists of a predefined set of $N\times N$ neurons that are
connected locally. The training is done by competitive learning starting from a random 
initial distribution of weights. Weights are updated after each observation is
used by computing the neuron that is closer (typically in Euclidean distance) 
to the observation. The information of the neuron is then propagated to the neurons
around within a predefined distance. One of the problems of this training
is that it results in an unpredictable distribution of classes along the whole map. 
However, we point out that reproducibility can easily be solved by fixing the random seed used for training.
It was used by \cite{asensio_mn07} to classify profiles of the
Mn \textsc{i} line whose Stokes $I$ profile is especially sensitive to the 
magnetic field strength. 
SOMs were later used by \cite{2012ASPC..463..215A} to classify profiles
in IMaX \citep{imax11} observations and they also proposed them as a poor's man 
inversion method with reduced precision because it is fundamentally a classification-based
inversion. Although self-organizing maps look promising for classification purposes, the
lack of control of the output reduces their attractiveness.

\subsection{t-SNE}
Student-t Stochastic Neighbor Embedding (t-SNE) is a nonlinear 
dimensionality reduction method\footnote{t-SNE is available 
on the \texttt{scikit-learn} Python package.} that has had some success in
recent years. The idea is to embed high-dimensional data for visualization in a 
low-dimensional space of two or three dimensions, which are especially suited
for human understanding. It models each high-dimensional object by a two- or three-dimensional 
point in such a way that similar objects are modeled by nearby points and 
dissimilar objects are modeled by distant points with high probability. Thanks to
the perplexity hyperparameter, one can make the mapping focus more on global or
local properties of the observations. This multi-scale characteristics makes t-SNE
a good candidate for exploring purposes. However, contrary to
PCA, t-SNE does not set up a basis. Mapping new observations requires training
the algorithm from scratch. It has been used by \cite{2020ApJ...891...17P} for the
classification of Mg \textsc{ii} line profiles. t-SNE can reliably distinguish between
profiles associated with flaring regions and non-flaring regions. Additionally, it has
been used by \citep{2021ApJ...907...54V} for classifying H$\alpha$ profiles and
identifying those that are suitable for a simple inversion method based on the
cloud model. Both works demonstrate that t-SNE is promising for understanding
the general picture of large observations. However, as any unsupervised method,
this interpretation can only be done a posteriori.

\subsection{Mutual information}
\cite{2021ApJ...912..121P} explored the use of neural networks to compute
the mutual information between pairs of spectral lines observed with the
IRIS satellite. Mutual information can be seen as a generalization of
correlation\footnote{An implementation can be found in \url{https://github.com/gtegner/mine-pytorch}.}. For two random variables $X$ and $Y$, the mutual information measures
the difference between the joint distribution $p(x,y)$ and the product
of their marginal distributions $p(x)p(y)$. \cite{2021ApJ...912..121P} showed
that an encoder-type neural network can be trained to measure the mutual information.
This training proceeds by using the same neural network to encode two spectral
lines observed at the same pixel, which are samples from the joint distribution. The same neural 
network is used to encode two spectral lines from different pixels, which are seen
as samples from the marginal distributions. By maximizing the distance between
both encodings, the neural network learns how to approximate the mutual information.
After training such an architecture with millions of IRIS profiles, they found that 
lines are weakly correlated in quiet conditions. The coupling strongly increases in
flaring conditions, with Mg \textsc{ii} and C \textsc{ii} having the strongest
coupling. \cite{2021ApJ...915...77P} used this tool to analyze in detail the
full atmospheric response during flares. This tool is very promising for the study
of multispectral data.

\subsection{Autoencoders}
\label{sec:autoencoders}
Perhaps the most promising nonlinear dimensionality reduction 
are autoencoders\footnote{An implementation can be found in \url{https://github.com/dariocazzani/pytorch-AE}.} (AE), also known in the past as autoassociative neural networks \citep[AANNs;][]{socas_navarro05}.
They were not very often used because of their inherent computational
burden, given that one has to train a neural network for every
new type of observation that one needs to analyze. This is currently not a problem
because of the availability of libraries for training neural networks
and the powerful hardware to which we have access. However, even during the first
decade of the 2000s, training these neural networks was still problematic.
AANNs are a special case of an encoder-decoder fully connected neural network. 
For the case of analyzing Stokes profiles, the input Stokes profiles are 
encoded by decreasing the size of the
layers until a bottleneck layer of only $d$ neurons is found. $d$ is the
expected intrinsic dimensionality of the Stokes profiles. They
are again expanded in the decoder part until recovering the original
size of the Stokes profiles. They are trained by forcing the network
to output exactly the same profiles used as input. This way, the neural
network has to compress the relevant information for each Stokes profile
into only $d$ numbers. \cite{socas_navarro05} showed a comparison
of AANNs and PCA. Given the nonlinear character of AANNs, they are able to much
better reconstruct a set of Stokes profiles using a lower dimensionality.

We anticipate that, in the current era of deep learning, AE will find a central role in many fields of solar physics, especially those
related with spectroscopy and spectropolarimetry, although imaging
could certainly obtain gains. The projection of 
the observations into a latent space of reduced dimensionality 
introduces a strong regularization, that can be efficiently
exploited by many inversion methods. The first applications of modern
AEs are very recent. \cite{DBLP:conf/cbmi/SadykovKDOKI21} used them
to show that the spectroscopic data of the Mg \textsc{ii} line observed
withe the NASA's IRIS satellite can be compressed by a factor of 27 without
any relevant impact on the line profiles. Additionally, the authors find
that the features found by the AE are interpretable. More recently, \cite{2022A&A...659A.165D} use an
AE to compress Stokes $I$ profiles to facilitate the computation of uncertainties
during the inference process using a Bayesian framework (see Sect.~\ref{sec:uncertainty}
for more details).

\begin{figure}
    \centering
    \includegraphics[width=\textwidth]{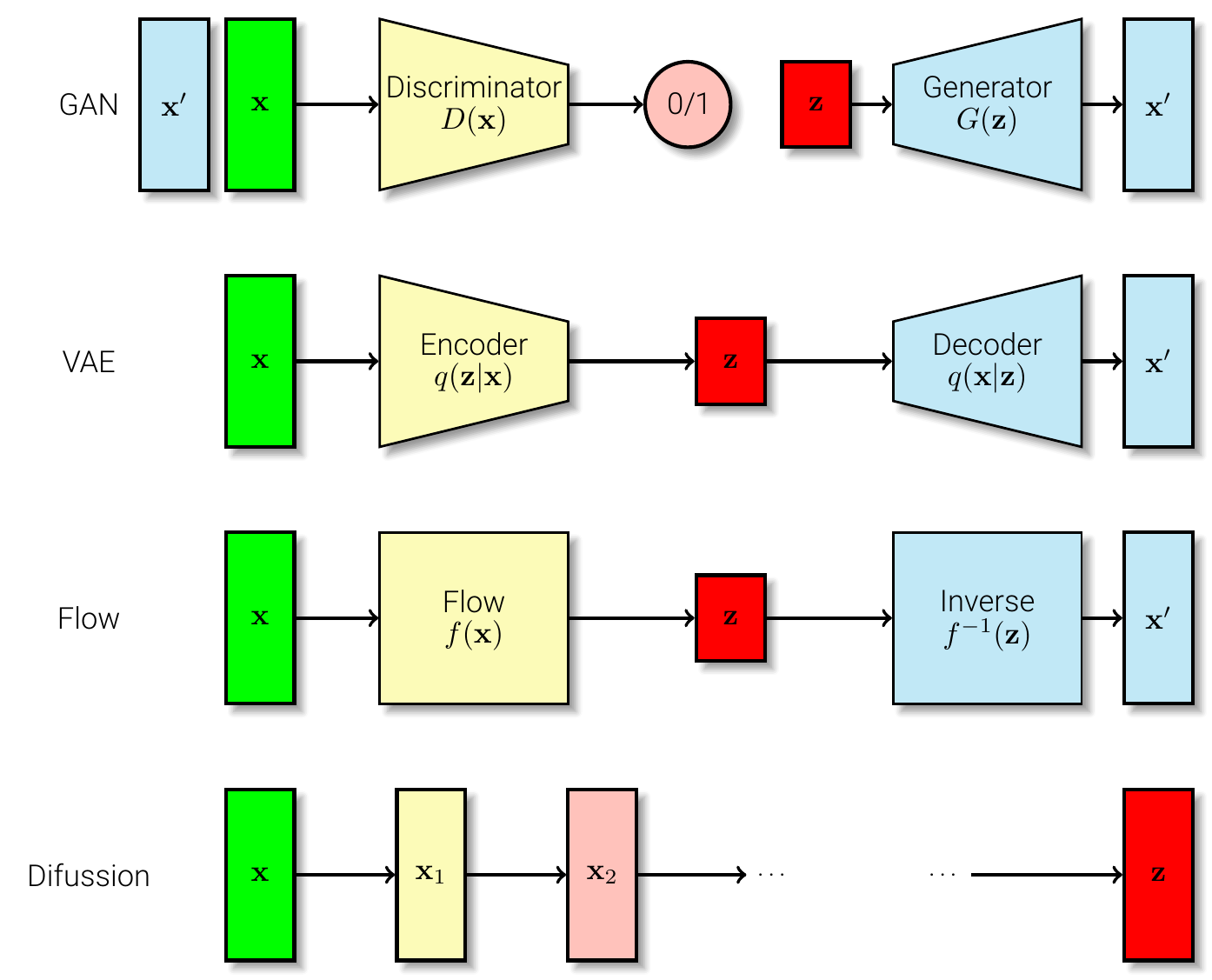}
    \caption{Overview of the most successful nonlinear generative models for
    signals in high dimensions (from \url{https://lilianweng.github.io/posts/2021-07-11-diffusion-models}).}
    \label{fig:generative}
\end{figure}

\subsection{Generative models}
\label{sec:generative_model}
Generative models are probabilistic models, $p(\mathbf{X})$, that can approximate the distribution
of objects of interest, $\mathbf{X}$, given a sufficiently large training set
while being accompanied with an efficient way of sampling from $p(\mathbf{X})$. As such, they can be
used as priors for $\mathbf{X}$ in any subsequent inference process. Modern
generative models, especially for objects of large dimensionality like images, 
are either based on variational autoencoders \citep[VAE;][]{Kingma2014}, 
generative adversarial networks \citep[GAN;][]{Goodfellow2016},
normalizing flows \citep[NF;][]{Dinh2014} or denoising diffusion probabilistic
models \citep[DDPM;][]{NEURIPS2020_4c5bcfec}. A diagram with the specific architecture
of each generative model is shown in Fig.~\ref{fig:generative}. All of them can be seen
as an instance of a latent-variable model (displayed as red blocks). In these models, we assume the
existence of a hidden latent variable with a dimensionality that can be equal to or smaller than 
that of the signal of interest. This latent variable is often extracted from
very simple probability distributions (Gaussian noise in many cases) and 
transformed, thanks to the action of a neural network, into samples
from the distribution of interest.

\subsubsection{Generative adversarial networks}
Generative adversarial networks\footnote{An implementation can be found in \url{https://github.com/eriklindernoren/PyTorch-GAN}.} (GAN) have had a huge impact on image generation, arts, language, and on
some fields of research. They are based on two networks (see the upper row
of Fig.~\ref{fig:generative}): a generator $G(\mathbf{z})$,
that maps the latent variable into the signals of interest, and a discriminator $D(\mathbf{x})$
that tells whether a sample $\mathbf{x}$ is coming from the distribution of
interest or not. Both neural networks are trained simultaneously using adversarial
training \citep{Goodfellow2016}.

Despite the huge impact in many fields, the impact in solar physics has been somewhat
reduced. \cite{kim19} proposed conditional GANs for the generation of artificial
magnetograms from STEREO data. The interest of such an approach is that once trained,
GANs can generate artificial magnetograms on the far side of the Sun. They can be compared
with current observations carried out with the Polarimetric and Helioseismic Imager (PHI) on Solar Orbiter 
\citep{2020A&A...642A..11S}. The quality of farside magnetograms is still
reduced, even after the improvements provided by \cite{felipe19} and \cite{2022A&A...667A.132B}. \cite{kim19} trained the 
generator by using extreme UV data from AIA and magnetograms from HMI, both observed on
the near side. The generated magnetograms in active regions look very similar to the target ones while
also providing very strong correlations in the total unsigned magnetic flux. More
daunting is the task of correctly generating the polarity structure of the active regions, whose
information is absent or barely present in the EUV images. According to \cite{kim19}, their
GAN is able to correctly produce Hale's law, purely learned from the data.

The reverse process, to produce EUV images from magnetograms, was approached by 
\cite{park19} with some success using GANs. Trained again with SDO data, the model is
able to produce brightenings in all AIA filters in active regions, which compares
well with the real data. In filters like 171 \AA\ which show conspicuous loops, the
GAN has a hard time reproducing them probably because the connectivity information is
not present in the magnetograms.

\cite{shin20} developed a model that generates artificial magnetograms from Ca \textsc{ii} K images.
They improved over previous works by using a training scheme that takes into account
both large-scale and small-scale properties of the images simultaneously, as proposed by
\cite{wang17}. This allows them to generate high-resolution magnetograms with sizes up
to 1024$\times$1024 pixels. Again, the polarity structure
of very active regions is correctly captured by the model even though this information
is probably absent from the Ca \textsc{ii} images. The only sensible explanation
for this is that this information is extracted from the statistical properties
of the training set. The authors also point out that
the model does a bad job on the quiet regions of the Sun.

An obvious question that arises for the image-generation models that 
we have discussed is what is their final purpose. 
It seems obvious that simply generating the images might have limited applicability, except
perhaps homogenizing very long baseline datasets. On the
contrary, having an efficient generative model for such complex processes will surely become key
for future research. Generative models map a latent vector $\mathbf{z}$ of reduced dimensionality
onto a complex and large image $I$. Consequently, introducing a pretrained generative model in an elaborate
inference scheme is a very good prior and can strongly inform the output and lead to very efficient
inference methods directly from images. For instance, one can think of data
assimilation methods in which a physical simulation is set up to explain a specific
observation. In this case, the physical model is very efficiently related to the observation
via the latent space, which automatically avoids outliers.

\subsubsection{Variational autoencoders}
Standard autoencoders, as shown in Sect.~\ref{sec:autoencoders}, are not generative models because there
is no way of sampling from the distribution. A variational autoencoder\footnote{An implementation
can be found in \url{https://github.com/dariocazzani/pytorch-AE}}
\citep[VAE;][]{Kingma2014} is a modification of a standard autoencoder that
works as a generative model (see the second row of Fig.~\ref{fig:generative}). 
To this end, the latent space is forced to have a fixed 
probability distribution during training. Once trained, sampling from this
fixed distribution (often a Gaussian distribution) and passing the samples
through the decoder, produces samples of the variable of interest according to the
prior. A VAE was used by \cite{2021ApJ...912..121P} as a means of compressing Mg \textsc{ii}
profiles. When the VAE is trained with line profiles from the so-called quiet Sun, it 
represents a very efficient outlier detector. Out-of-distribution profiles
(i.e., flaring profiles) cannot be efficiently reproduced by the VAE. Therefore, if the
difference between the reconstructed profile and the original profile is large, one can
safely say that the profile is not coming from the inactive Sun.

\subsubsection{Normalizing flows}
Another powerful way of producing samples from the posterior distribution
is via normalizing flows\footnote{\url{https://github.com/bayesiains/nflows}} (NF), which are a very flexible, tractable, and easy-to-sample family 
of generative models, that can
approximate complex distributions. Simply put, an NF is a transformation of a simple
probability distribution (often a multivariate standard normal distribution, with zero
mean and unit covariance) into the desired probability distribution (see the
third row of Fig.~\ref{fig:generative}). Normalizing flows accomplish this by
the application of a sequence of invertible and differentiable variable transformations. Let us
assume that $\mathbf{Z}$ is a $d$-dimensional random variable with a simple and tractable probability
distribution $q_\mathbf{Z}(\mathbf{z})$, with the condition that it is fairly straightforward to sample. 
Let $\mathbf{X}=f(\mathbf{Z})$
be a transformed variable, with a function $f$ that is invertible. If this condition holds, then
$\mathbf{Z}=g(\mathbf{X})$, where $g=f^{-1}$. The change of variables formula
states that the probability distribution of the transformed variable is given by:
\begin{equation}
    q_\mathbf{X}(\mathbf{x}) = q_\mathbf{Z}(g(\mathbf{x})) 
    \left| \mathrm{det} \left( \frac{\partial g(\mathbf{x})}{\partial \mathbf{x}} \right) \right|.
\end{equation}
The term $\partial g(\mathbf{x}) / \partial \mathbf{x}$ is the Jacobian matrix and takes into
account the change of probability volume during the transformation. Its role is to
force the resulting distribution to be a proper probability distribution with unit
integrated probability. Since the transformation is invertible, the equality 
$\partial g(\mathbf{x}) / \partial \mathbf{x}=(\partial f(\mathbf{z}) / \partial \mathbf{z})^{-1}$ holds, so that
one can rewrite the previous expression as:
\begin{equation}
    q_\mathbf{X}(\mathbf{x}) = q_\mathbf{Z}(\mathbf{z}) 
    \left| \mathrm{det} \left( \frac{\partial f(\mathbf{z})}{\partial \mathbf{z}} \right) \right|^{-1}.
\end{equation}

Designing an invertible transformation that can be trained to produce generative models over 
complex datasets is difficult. For this reason, normalizing flows make use of the fact that
the composition of invertible transformations is also invertible. Then, if $f=f_M \circ f_{M-1} \circ \cdots \circ f_1$, the
transformed distribution is
\begin{equation}
    q_\mathbf{X}(\mathbf{x}) = q_\mathbf{Z}(\mathbf{z}) 
    \prod_{i=1}^M \left| \mathrm{det} \left( \frac{\partial f_i(\mathbf{y_i})}{\partial \mathbf{y_i}} \right) \right|^{-1},
    \label{eq:flow}
\end{equation}
where $\mathbf{y}_i=f_{i-1} \circ \cdots \circ f_1(\mathbf{z})$ and $\mathbf{y}_1=\mathbf{z}$.
Compositional invertible transformations have made it possible to define
very flexible normalizing flows through the use of deep neural networks. 


Despite their potential as a flexible probabilistic generative
model, they have not been used in solar physics for this purpose yet. We refer the
reader to Sect.~\ref{sec:uncertainty} for a discussion on how NFs have been applied for
the acceleration of Bayesian inference from spectropolarimetric observations by 
directly fitting the posterior distribution.

\subsubsection{Denoising diffusion probabilistic models}
Denoising diffusion models\footnote{\url{https://github.com/lucidrains/denoising-diffusion-pytorch}} 
\citep[DDPM;][]{NEURIPS2020_4c5bcfec} are based on two chains of processes (see Fig.~\ref{fig:generative}). The first one adds a small 
amount of noise to a certain sample from the variable of interest. When this noise addition is repeated
many times, the final result cannot be distinguished from pure noise and is assumed
to be the latent variable. This process is, obviously, easy to simulate. The inverse 
process takes the latent variable and proposes a neural network that ``cleans''
the noise, trying to undo what the first process did to the signal. This generative
model is at the base of the most recent image generative models, of enormous
success when coupled with powerful language models. We still need to see applications
of DDPMs as prior for solar data.

\section{Applications of supervised deep learning}
The vast majority of applications of nonlinear models in supervised training are 
based on CNNs. The models have been increasing in complexity in the last
few years, motivated by the success of CNNs in learning directly from 
the data. In the following, we describe relevant applications to different
subfields of solar physics.

\subsection{Segmentation of solar images}
\begin{figure}
\includegraphics[width=\textwidth]{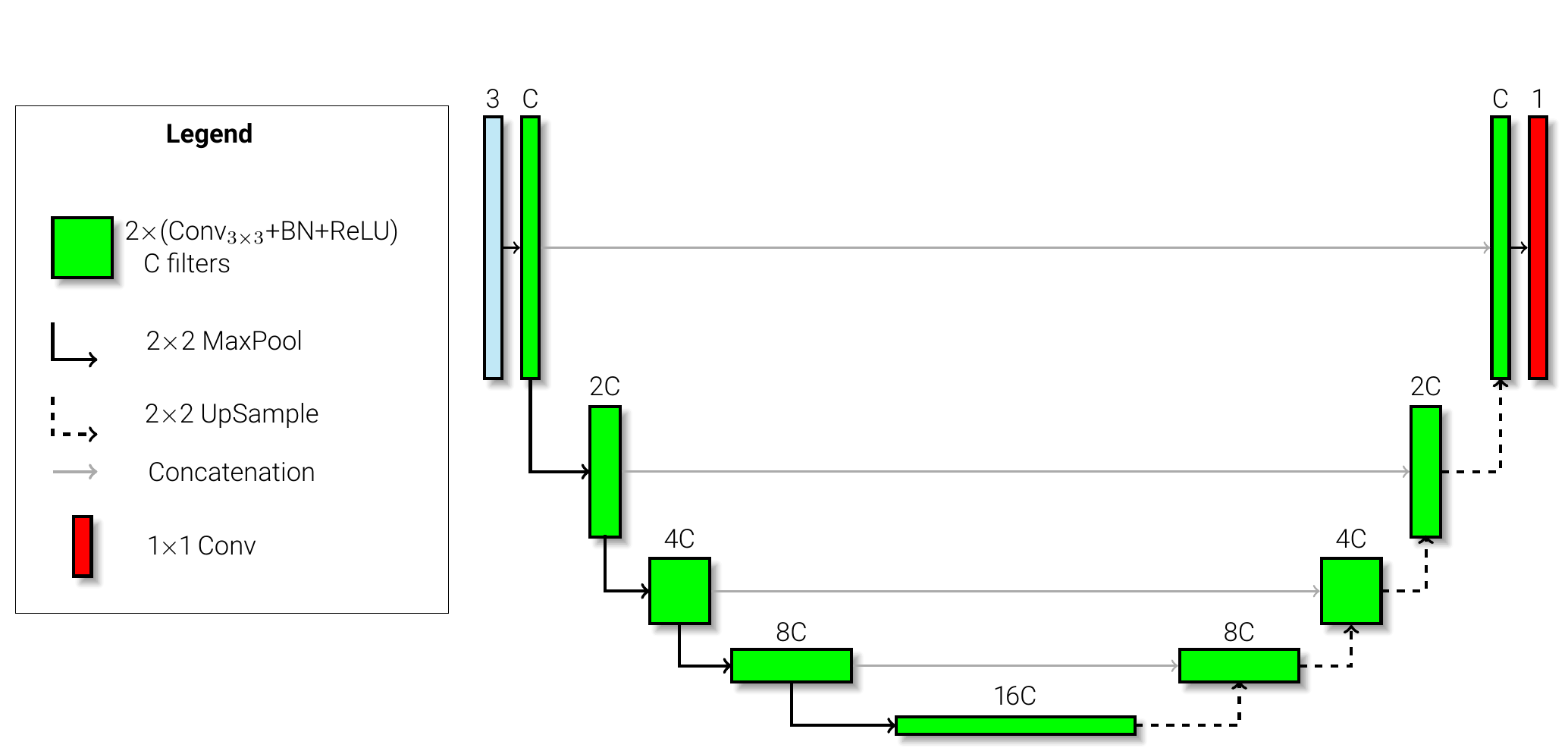}
\caption{Schematic drawing of the encoder-decoder U-Net architecture. In this case, the input
has 3 channels, while the output contains only one channel. The number of channels after
the first convolutional layer is C. Green blocks summarize the application of a convolutional
layer with a $3\times 3$ kernel, followed by batch normalization and a ReLU. These operations
are repeated twice. Solid arrows correspond to the MaxPool operation, while dashed arrows
refer to bilinear upsampling of the feature images. Grey arrows refer to skip connections that
are simply concatenated in the decoder.}
\label{fig:unet}
\end{figure}

Deep learning has produced a huge advance in the dense (per pixel) classification of solar 
images, of special relevance due to the large amount of synoptic solar observations
that we currently have. Automatic detection and segmentation of solar structures 
in images could allow us to build databases for an enormous amount of images. Detecting
sunspots, flares, coronal holes, and other structures are potential candidates for such
applications. 
CNNs have recently been used for the identification of CHs. \cite{2018MNRAS.481.5014I}
proposed a U-Net architecture\footnote{\url{https://lmb.informatik.uni-freiburg.de/people/ronneber/u-net}}
as proposed by \cite{2015arXiv150504597R} (see Fig.~\ref{fig:unet}) to identify CHs on solar AIA/SDO images obtained
in the 193 \AA\ wavelength. \cite{2018MNRAS.481.5014I} trained the model with 2385 binary maps from the
Kislovodsk Mountain Astronomical Station. 
The output of the U-Net is a binary image that tells whether
the pixel belongs to a coronal hole or not. The training is carried out using the binary
cross-entropy (BCE) as a loss function:
\begin{equation}
    L = -\sum_i y_i \log \hat{y}_i + (1-y_i) \log(1-\hat{y}_i),
\end{equation}
where $y_i$ is the target label for the $i$-th pixel and $\hat{y}_i$ is the
prediction of the network. The results of
the CH identification were compared with feature maps of other methods, such as
CHIMERA \citep[Coronal hole identification via multi-thermal emission
recognition algorithm;][]{2018JSWSC...8A...2G} and SPoCA, from January 2017 to
July 2018. One of their conclusions is that the U-Net architecture
produces segmentation maps that are more consistent than those of SPoCA. By comparing
the area variation of the CH, they observe that CHIMERA and U-Net show similar
results and the two methods have a correlation coefficient of 0.76. In a
follow-up study, \cite{2020ApJ...903..115I} extended the identification of the
CH for synoptic maps and they constructed a catalogue for 2010-2020 based on
the AIA/SDO 193 \AA\ data. 
The Solar Corona Structures Segmentation Network (SCSS-Net) was also developed by
\cite{2021MNRAS.508.3111M}, again inspired by the U-Net
architecture, for the dense segmentation of solar images and 
the localization of CH and AR.
U-Nets were also used by \cite{2020ApJS..250....5J} to identify and track solar magnetic
flux elements observed in magnetograms. This will largely facilitate tracking of small-scale
magnetic elements, something that is currently done with ad-hoc techniques and large
human intervention \citep[e.g.,][]{2014ApJ...797...49G}. Since tracking involves some
degree of time coherence on the labeling of the elements, we anticipate that taking into
account the time evolution could produce a large improvement over single-frame
segmentation \citep[e.g.,][]{Ventura_2019_CVPR}.

\cite{2021A&A...652A..13J} used a CNN to identify the boundaries of
CH using the seven extreme ultraviolet (EUV) channels of AIA/SDO as input, together
with the line-of-sight magnetograms provided by the HMI/SDO. Their
identification method is termed Coronal Hole RecOgnition Neural Network Over multi-Spectral-data (CHRONNOS).
Their analysis of the CNN concludes that the CNN has the
ability to learn directly from multi-dimensional data and can identify CH
and distinguish them from prominence channels. To this end, the 
CNN takes advantage of the shape, structural appearance, global context information,
and the multi-wavelength representation. 


Later, using three different and independent wavelengths, \cite{2021SoPh..296..160B}
implemented the Single Shot MultiBox Detector
(SSD) and the Faster Region-based Convolutional Neural Network (R-CNN), for the
detection of the CH, prominences and sunspots. The training set is based on
full-disk data from AIA/SDO and HMI/SDO between 2010 and 2019. The data cadence is 12 hr for sunspots and
CH and four hours for the prominences. The events in each observed image
was manually labelled, including the bounding boxes. The total number
of images with coronal holes was 5085 (from the AIA 193 \AA\ channel), those
with sunspots was 4383 (from the intensity images of HMI/SDO) and those
with prominences were 2926 (from the AIA/SDO 304 \AA\ channel). Once trained, 
they checked that the models do a good job in locating the CH after a 
direct comparison with the HEK database.

\begin{figure}[]
\centering
\includegraphics[width=0.4\textheight, trim = 0 0 0 10, clip]{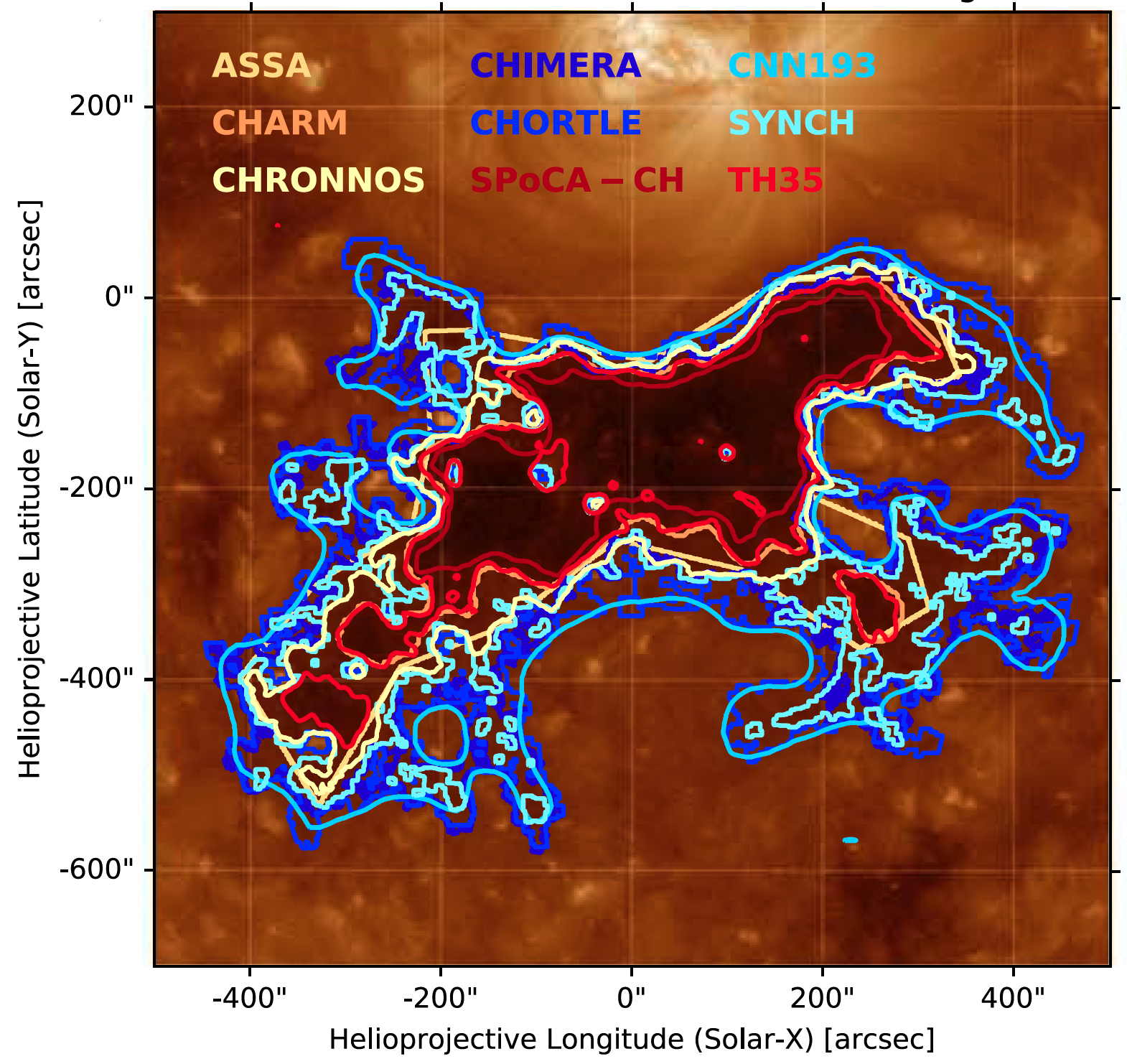}
 \caption{A comparison of the estimated coronal hole maps from nine different 
 automated detection schemes overlaid on the AIA 193 \AA\  \citep[from][]{2021ApJ...913...28R}.}
 \label{fig:uncertainty}
 \end{figure}

Although several methods (some of them based on CNNs) have been developed
in the recent years, we are still missing an estimation of the 
uncertainties in the segmentation of solar images. In a recent paper, \cite{2021ApJ...913...28R}
analyzes these uncertainties in the detection of CH boundaries. 
Nine automatic methods are compared using a CH from the southern hemisphere,
close to the sun center, and observed for a couple of solar rotations. Multiple
EUV wavelengths and measurements of the radial component of the photospheric
magnetic field from the SDO spacecraft were used as preparation of the data to
be used by different methods. The compared methods are ASSA-CH, CHIMERA
\citep{2018JSWSC...8A...2G}, CHORTLE, CNN193 \citep{2018MNRAS.481.5014I},
CHRONNOS \citep{2021A&A...652A..13J}, SPoCA-CH \citep{2014A&A...561A..29V}, and
SYNCH. They also evaluated the mean CH intensity in AIA 193A, the mean signed and
unsigned line-of-sight magnetic field component (B$_\mathrm{LOS}$), the degree of
unipolarity, and the net open magnetic flux (sum of B$_\mathrm{LOS}$ over the CH area).
They found that different methods produce significantly
different outcomes. The differences are small in the center of the CH and
they start to be larger when approaching the boundary of the CH. Differences
in the shape of the CH and its physical properties are also found 
(see Fig.~\ref{fig:uncertainty}).

As a consequence, the choice of the method has
a non-negligible impact on the predicted solar wind.
As a final conclusion, one of the fundamental problems to characterize the uncertainties is the
absence of a well-agreed definition of a CH (or, by extension, of any feature on the solar surface). We can only compare automatic methods with a segmentation made by eye by the observer. This manual segmentation can also depend on the wavelength used for its evaluation. We urgently need a community effort toward defining an agreed training set.

Dense segmentation of photospheric images has been pursued recently by \cite{2022FrASS...9.6632D},
with the aim of classifying granular structures. The access to high-resolution images
has shown the overwhelming complexity of granulation. One can only hope to 
understand the physical mechanisms by first applying a semantic segmentation of the
images and isolating interesting phenomena (intergranular lanes, exploding granules, \ldots). 
Although it is still work in progress, \cite{2022FrASS...9.6632D} demonstrated that a U-Net 
is able to learn this segmentation problem and then apply it to large datasets to analyze
the statistical properties of magnetoconvection.

\begin{figure}
    \centering
    \includegraphics[width=0.8\textwidth]{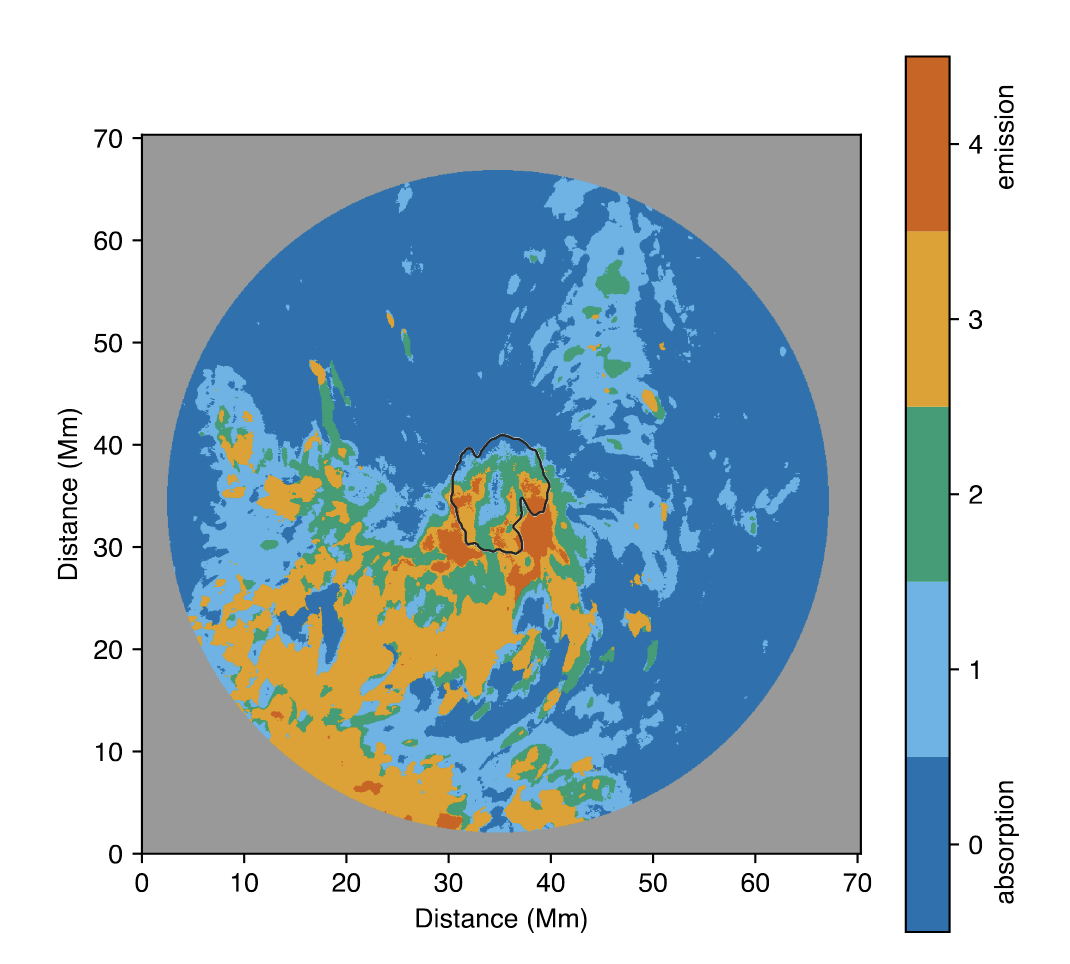}
    \caption{SPectral classifications of an IBIS observation, where the color bar relates to the spectral shape classified, with `0' and `4' representing pure absorption and emission profiles,
respectively. The umbra/penumbra boundary is highlighted using a black contour \citep[from][]{MacBride_2020}}
    \label{fig:MacBride_2020}
\end{figure}

\subsection{Classification of solar images}
Arguably one of the most used ways of understanding physical phenomena in the 
solar surface and the heliosphere has been by painfully classifying all events into
different classes. This allows researchers to pinpoint typical properties of 
similar events and associate them with their physical properties. In the era of
photographic or video images, this classification could be done by hand. However,
the amount of data that we are currently generating, as well as the expected increase
in data rates in the near future, is so large that we need the help of machines to classify
all events. \cite{armstrong19} used a deep neural network to automatically classify 
solar events in different classes: quiet Sun, prominence, filaments, sunspots, and flare ribbons.
The trained neural network achieves an extremely high performance (close to 99.9\%). They also
demonstrate that transfer learning can be used in solar physics. Transfer learning is the idea
of reusing complete, or parts of a neural network that have been previously trained and
applying them to another problem. Transfer learning is part of many successful applications of
deep learning in general and is based on the idea that the initial layers of a convolutional
neural network are able to extract features from the images that are used by the
last layers of the neural network to carry out the classification. \cite{armstrong19} successfully
demonstrated that the first layers of a CNN trained with solar structures in one wavelength are able to
extract features that can be used with images in a different wavelength.

Along the same line raised in the previous section, \cite{armstrong19} also advocated for 
the generation of a huge and curated database of
classified solar images. This is in parallel to similar efforts in the machine learning
community like ImageNet \citep{imagenet_cvpr09}, in which more than 14 million images have
been labeled by hand using more than 20 thousand words describing the images. The
solar ImageNet would be a large database of multiwavelength multi-instrument images
ideally labeled by hand.

Apart from the pure classification of images, one of the potential applications of these neural
networks is to automatically detect ``interesting'' events that are far from the typical cases.
In the case of a classifier, a flag can be raised by the machine when it finds an event that is
associated with similar probability to several classes. This means that the class is not known with certainty so it does not look like any of the examples in the training set.
Another option to develop an outlier detector is by using a generative model (see Sect.~\ref{sec:generative_model})
that learns how to produce images like those in the training set. If the properly trained generative model
is able to correctly reproduce the input (some generative models are even
able to output an estimation of the likelihood), the input is compatible with the trained data and
cannot be considered an outlier. If it is not well reproduced (or the likelihood is very small), then
it can be tagged as an outlier and a flag can be raised for later analysis.

Although not a whole-image classification scheme, \cite{MacBride_2020} used 
a FCN classifier to identify and classify spectral line profiles. The aim was to
rapidly cluster the profiles into different categories depending on the number of 
components in the emission peak and absorption dip present in the line.
The model is able to determine the underlying properties of complex profiles, not 
only to identify the peak and dips in the profiles but also to classify sub-classes 
that will then be used to constrain better the fitting of a single or multiple velocities
components inside the pixel.
They tested the method using Ca \textsc{ii} 8542 \AA\ line profiles observed by the 
Interferometric BIdimensional Spectrometer (IBIS) using an uneven spectral sampling, with a 
higher density in the line core, as proof of concept of the model, and also as a  benchmark 
for two-component atmospheric profiles studies that are commonly present in sunspot chromospheres
(see Fig.~\ref{fig:MacBride_2020}).

\subsection{Prediction of flares}

Due to the consequences of an impact of a major solar flare on terrestrial space weather, 
there has been an increasing interest in applying statistical learning techniques 
for the prediction of flares (typically M-class flares and greater) and coronal mass 
ejections (CMEs). Since not all flares have associated CMEs (and vice 
versa)\footnote{\cite{Sheeley:1983} reported that every GOES X-ray flare (lasting six 
hours or longer) had an associated CME. Their data set comprised events observed between 1979 and 1981.}, 
forecasting of the two are considered related, but different problems.

In either case, the problem is usually posed as a classification problem. Given input 
parameters $\mathbf{x}$ sampled at time $t_0$, does a flare occur in the time 
period $t\in (t_0,t_0 + \Delta t]$, with $\Delta t$ on the order of hours to days? 
Variations on this problem statement can include multiclass classification  
(e.g., whether there is an M- or X-class flare), or regression to predict the maximum 
soft x-ray flux (e.g., as measured by GOES XRS) in the time period of interest.

Early attempts at flare/CME prediction focused on the use of input features inspired 
by physics models (or heuristics) of how solar flares are thought to operate. Ample 
theoretical considerations, observational evidence, and numerical simulations support the commonly accepted picture that solar
flares are powered by abrupt reconfigurations of the solar coronal magnetic field \citep[see reviews by][]{PriestForbes:2002,ShibataMagara:2011} which results in the coronal magnetic field entering a 
lower energy state. The lowest energy state of the coronal magnetic field above an active region is
the potential field configuration, which has zero current density, since 
$\mathbf{j} = \nabla \times \mathbf{B} = \nabla \times [-\nabla \Phi] = 0$ \citep[e.g., see][]{Altschuler:1969}. 
Without  available \emph{free energy} (i.e., the magnetic energy in excess of the energy stored in a potential 
field configuration), an active region should not be flare-productive. This physical argument suggests
the photospheric (since this is the layer for which magnetograms are most easily acquired) current density 
should have predictive power for flare prediction. A related quantity of interest is the twist parameter, which
is the current density normalized by the magnetic field strength. 

\cite{Falconer:2001} performed a pilot study of the association between CME-productivity 
and an AR's perceived non-potentiality with two quantities derived from vector magnetograms. 
The latter was visually assessed from the morphology of loops in Yohkoh X-ray images 
(e.g., the presence or absence of a sigmoid). The paper suggests that the length of the main 
polarity inversion line (PIL) and the net current (measured through one polarity) are quantitative 
indicators of CME productivity. However, only eight vector magnetograms covering three distinct 
ARs were available for this study. \cite{Falconer:2002} extended the work using 17 vector 
magnetograms covering 12 ARs. In addition to the PIL length and net current, they examined a 
dimensionless twist parameter and reported all three are correlated with the flux content of the 
AR, and are correlated with CME productivity. To address the limitations imposed by the 
lack of regular vector magnetogram coverage available at the time, \cite{Falconer:2003} developed 
a proxy for the main PIL length parameter using line-of-sight magnetograms from the Michelson Doppler 
Imager \citep{Scherrer:MDI} onboard the ESA/NASA Solar \& Heliospheric 
Observatory \citep*[SOHO; ][]{SOHO} mission. This work opened up the possibility 
to use MDI full-disk magnetograms (available at 90 min cadence) for assessing the CME productivity of ARs. 

In a series of papers \citep{LekaBarnes:2003a,LekaBarnes:2003b,BarnesLeka:2006,LekaBarnes:2007}, Leka \& Barnes performed 
systematic analyses of how vector magnetogram-derived parameters such as current and twist are different 
between flaring and non-flare active regions. Of particular relevance is \citet{LekaBarnes:2003a}, in which 
they performed discriminant analysis on flaring and non-flaring regions. This is a linear model for binary 
classification. Suppose $\mathbf{x}$ is the feature vector (consisting of vector magnetogram-derived quantities)
and $\overline{\mathbf{x}}^0$ and $\overline{\mathbf{x}}^1$ denotes the mean of $\mathbf{x}$ over the two separate 
populations (in this case, flaring and non-flaring active regions). 
The sign of the linear functional 
\begin{equation}
f(\mathbf{x}) = \mathbf{x}\mathbf{C}^{-1}(\overline{\mathbf{x}}^0 - \overline{\mathbf{x}}^1) + \frac{1}{2} (\overline{\mathbf{x}}^0 - \overline{\mathbf{x}}^1) \mathbf{C}^{-1} (\overline{\mathbf{x}}^0 + \overline{\mathbf{x}}^1),
\end{equation}
was used to classify whether an active region with feature vector $\mathbf{x}$ is in the flaring or 
non-flaring population. They computed discriminate functions for single variate as well as multivariate 
feature vectors. However,  the data set available only included 24 blocks of roughly 1-hour long observations 
(spanning over 7 active regions and 10 C, M and X flares) from the  University of Hawaii Imaging Vector Magnetograph. 
The data set was enough to establish that a small number of input parameters is insufficient to distinguish 
the two flare-active and flare-quiet active region populations with low error rates. With six input features 
(standard deviation of the horizontal magnetic vector, skew of vertical current density $J_z$, kurtosis of $J_z$, 
area of pixels with a shear angle greater than 80 deg, time rate of the change of the best-fit linear force-free 
parameter and the time rate of change of the mean unsigned normal flux density), they were able to construct a function $f(\mathbf{x})$ which linearly separates the two populations.

\subsubsection{HMI era}
The \citet{LekaBarnes:2003a} study was limited by the quantity of data available and it is not 
clear how generalizable the results are when applied to other active regions. However, within 
the data set studied, it appears variables measuring the distribution of magnetic twist helps 
with discerning whether an active region is flare-quiet or productive.

\begin{figure}
    \centering
    \includegraphics[width=0.8\textwidth]{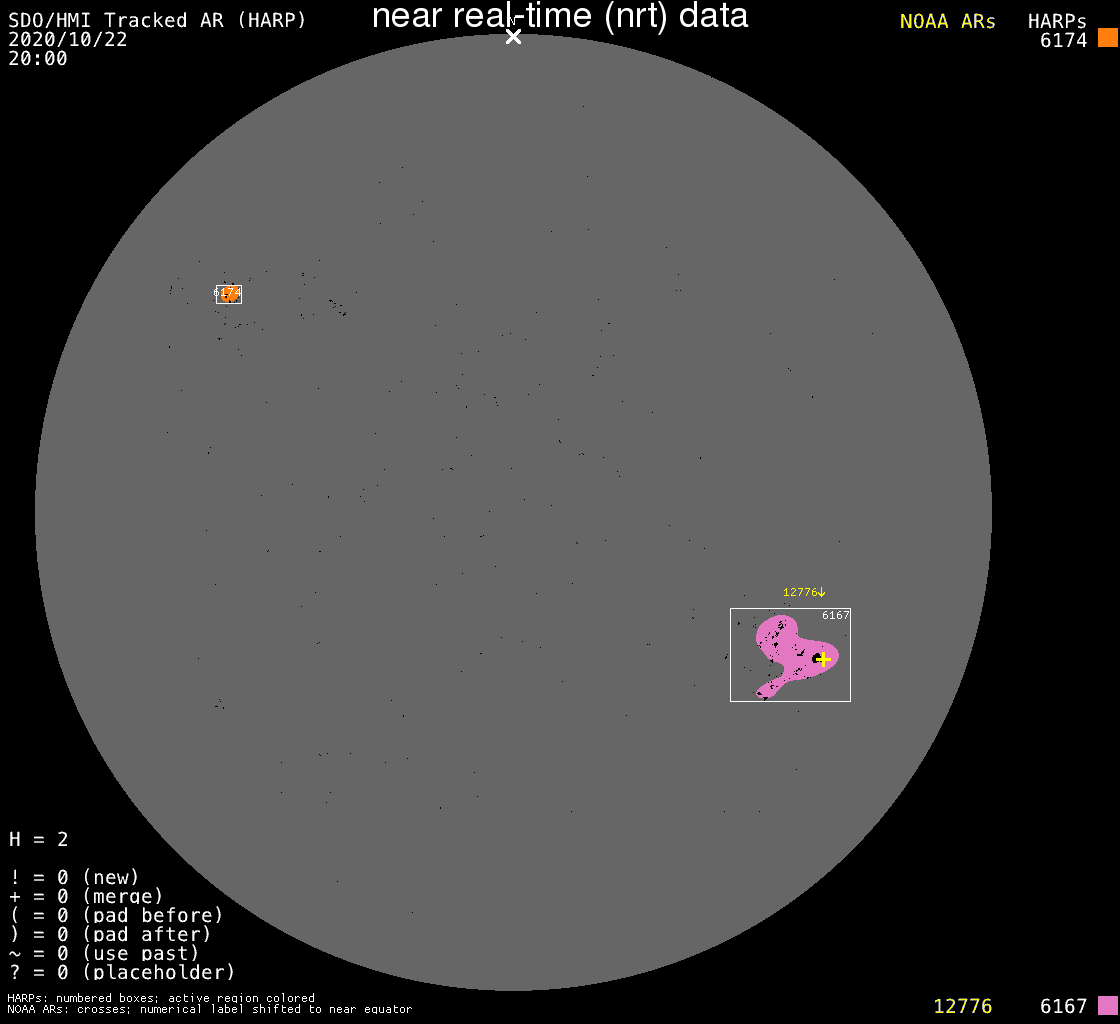}
    \caption{Space Weather HMI Active Region Patches (SHARPs) identified by a computer tracking algorithm. In this image, two SHARPs have been identified and are marked by rectangular bounding boxes.}
    \label{fig:sharps}
\end{figure}
The availability of regular vector magnetograms at 12 minute cadence from SDO/HMI was a game changer. 
It enabled the curation of flare prediction datasets with hundreds and thousands of samples. 
The Space Weather HMI Active Region Patches (SHARPs) data product provides vector magnetograms 
in rectangular patches (on the  CCD or on a longitude-latitude grid; see Fig.~\ref{fig:sharps}) following 
active regions as they emerge and rotate across the solar disk. The metadata for this data product includes 
quantities that summarize the size (in terms of flux content and area spanned), and magnetic twist 
(e.g., area-averaged current density) for each magnetogram in the temporal 
sequence \citep[see][for the complete list]{Bobra:2014}. When used for problems like flare prediction, the SHARP 
parameters can be considered `hand-engineered' features extracted
by experts.

The paper by \citet{BobraCouvidat:2015} ushered in a new era for solar flare prediction. Their contributions 
to the field of flare prediction research are manyfold. This study was the first to use the 
HMI SHARP data set to demonstrate the potential utility of continuous vector magnetogram data 
for flare prediction. The paper introduced many standard data science practices to the solar physics 
community. This includes the practice of $n-$fold cross-validation (CV). Specifically, they evenly 
split the dataset into $n$ tranches, picked the union of $n-1$ tranches to train a model, and then 
used the remaining tranche as a test set. They then rotated through the tranches, each time using a 
different tranche to be the test set. This allowed the model to be trained and tested $n$ times. 
The spread of the evaluation metrics over the so-called $n-$folds provides a measure of the reliability of the metrics.

\subsubsection{Evaluation metrics}
As more research groups began to tackle the flare prediction problem, there emerged a need to standardize how 
flare prediction models are evaluated. Consider any binary classification problem. The aim is for 
the model to classify whether an element is in class A. For any binary classification problem 
(including flare prediction), consider the contingency Table~\ref{tab:contingency}. The 
contingency table completely specifies the joint probability density function (JPDF) regarding 
whether an event has occurred, and whether it was forecast to occur.

Metrics regarding the performance of the forecasting method are functions of the JPDF 
(see Table~\ref{tab:binary_metrics}). For example, the recall (false alarm rate) is the conditional 
probability of a positive forecast, \emph{given} the event did (not) take place. In contrast, the precision is the conditional probability that the event occurred, given a positive forecast was issued.

Which metric is the correct one to use? It depends on the goal and there is no single correct answer. 
For instance, the stakeholder of an operational space weather forecast may prioritize the need to 
minimize false positives, because false positives will trigger protocols (e.g., shutting down the 
power supply). For other stakeholders, avoidance of a false negative (i.e., a reliable 
\emph{all-clear"} forecast) may be the priority \citep[][]{Barnes:2016}. If the stakeholders 
were solar physicists with research interests in flares and must submit their observing plans 
to instruments on a space-borne observatory (e.g. Hinode or IRIS) a day in advance, the cost for 
false alarms may be comparatively small. 

\begin{table}[]
    \centering
    \begin{tabular}{|r|c|c|}
        \hline 
         & Event: Yes & Event: No \\
        \hline
        Forecast: Yes & \TP & \FP \\
        \hline
        Forecast: No & \FN & \TN \\
        \hline
    \end{tabular}
    \caption{Contingency table for binary classification. We denote \TP, \TN, \FP~and \FN~as the number of true positives, true negatives, false positives and false negatives, respectively.}
    \label{tab:contingency}
\end{table}

\begin{table}[!h]
    \centering
    \begin{tabular}{|r|c|c|c|}
    \hline
    \textbf{Metric} & \textbf{Definition} & \textbf{Meaning} & \textbf{Range} \\
    \hline 
    Recall &$\frac{\TP}{\TP + \FN}$ & $P(\rm{Forecast:Yes} | \rm{Event: Yes} ) $ & $[0,1]$\\
    \hline 
    Precision &$\frac{\TP}{\TP + \FP}$ & $P(\rm{Event:Yes} | \rm{Forecast: Yes} ) $ & $[0,1]$\\
    \hline
    Specificity &$\frac{\TN}{\TN + \FP}$ & $P(\rm{Forecast:No} | \rm{Event: No} ) $ & $[0,1]$\\
    \hline
    False Alarm Rate &$\frac{\FP}{\TN + \FP}$ & $P(\rm{Forecast:Yes} | \rm{Event: No} ) $ & $[0,1]$\\
    \hline
    Accuracy & $\frac{\TP}{\TP + \FN + \TN + \FP}$ & $P(\rm{Forecast: Yes}~\&~\rm{Event:Yes})$ & $[0,1]$\\
    \hline
    Rate Correct & $\frac{\TP + \TN}{\TP + \FN + \TN + \FP}$ & $P(\rm{Forecast == Event})$ & $[0,1]$\\
    \hline    
    Critical Success Index &$\frac{\TP}{\TP + \FP + \FN}$ &  - & $[0,1]$\\
    \hline
    Gilbert Skill Score &$\frac{\TP -\rm{CH}}{\TP + \FP + \FN-\rm{CH}}$ & CSI excluding chance hits  & $[0,1]$\\
    \hline
    Heidke Skill Score (v1) & $\frac{\TP}{\TP+\FN} - \frac{\FP}{\TP+\FN}$ & Recall$\times (2-\rm{ Precision}^{-1}) $ & $(-\infty ,1]$\\
    \hline
    Heidke Skill Score (v2) & $\frac{\TP + \TN - \rm{E}}{\TP+\FN + \TN +\FP - \rm{E}}$ & - & $[0 ,1]$\\
    \hline
    True Skill Statistic & $\frac{\TP}{\TP+\FN} - \frac{\FP}{\TN +\FP}$ & Recall - False Alarm Rate & $[-1 ,1]$\\
    \hline
    \end{tabular}
    \caption{Evaluation metrics for binary classification. Refer to Table~\ref{tab:contingency} for the definition of classes. In the definition for the Gilbert Skill Score, $\rm{CH}$ (chance hits) is the Accuracy for a random forecast model. The probability that a random forecast outputs a positive is uncorrelated with the underlying probability of the event. Hence, the joint probability for the Accuracy can be factored, giving $\rm{CH} = P(\rm{Forecast: Yes}) \times P(\rm{Event:Yes}) = \frac{(\TP+\FP)}{n} \frac{(\TP+\FN)}{n}$, where $n = \TP+\FP+\FN+\TN$. In the definition for the Heidke Skill Score \citep[v2;][]{SWPC:Forecast}, E refers to the Rate Correct for a random forecast model: $\rm{E} = \rm{CH} + \frac{(\FP+\TN)(\FN+\TN)}{n^2}$.}\label{tab:binary_metrics}
\end{table}


When comparing different flare prediction models, one would ideally wish to compare models using 
identical test data sets. In practice, it is not possible without the coordination of research groups 
tackling the flare prediction problem due to differences in the choice of flares they include in the 
data sets, the choice of train/test set partitioning, etc. These choices impact the class imbalance 
between flaring and non-flaring events in the data sets used. Let the class imbalance be the ratio of 
non-events ($\TN+\FP$) to actual events ($\TP + \FN$). Evaluation metrics comprising sums or 
products of terms, each of which only depends on blue or red quantities (e.g., Recall, Specificity, False 
Alarm Rate, True Skill Statistic) do not depend on the underlying class imbalance.

\cite{Bloomfield_2012} and \cite{BobraCouvidat:2015} offer extensive discussions of the benefits 
of using the True Skill Statistic over the Heidke Skill Score \citep[e.g.,][]{BarnesLeka:2008} to measure 
the performance of flare prediction models. The primary reason is that the latter is sensitive to the class imbalance. Choosing metrics that are insensitive to the class imbalance is especially important 
when data augmentation or sampling strategies are used in the attempt to improve model performance. Since 
the number of X-class (M-class) flares in a solar cycle is in the dozens (hundreds) in a solar cycle, the 
underlying population has a high class imbalance. In order to help models train better, ML practitioners may 
use resampling strategies that mitigate the imbalance. Relying on metrics that are sensitive to the imbalance 
ratio makes it difficult to compare metrics evaluated on the training set, testing set (which may reflect 
the population imbalance), and across studies \citep[see discussions by][]{BobraCouvidat:2015,Barnes:2016}.

\begin{figure}
\centering
\includegraphics[width=0.6\textwidth]{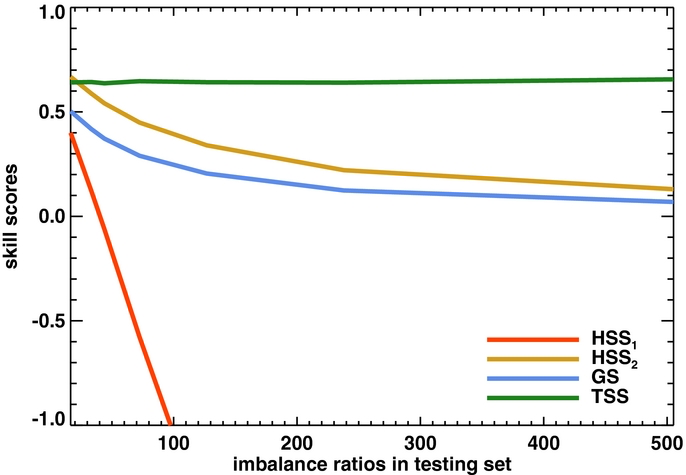}
\caption{Dependence of various binary classification metrics (Heidke Skill Scores, Gilbert Skill Score and True Skill Statistic; see Table~\ref{tab:binary_metrics}) on the underlying class imbalance ratio $\frac{\TN+\FP}{\TP + \FN}$. Reproduced from \citet{BobraCouvidat:2015}.}
\end{figure}

\subsubsection{Baseline models}
When evaluating flare prediction models, whether they are physics-based or purely data-driven, it 
is important to compare their metrics with respect to baseline models. A useful baseline model 
tends to be one that is interpretable, conceptually simple, and computationally inexpensive 
(relative to the models being evaluated). The purpose of a baseline model is to serve as a 
reference point. By evaluating a metric over predictions from the baseline model, and doing 
the same for a more sophisticated model, one can measure the marginal utility of the extra effort.
For flare prediction, two basic baseline models are the random forecast model, and the climatological model.

\noindent \textbf{Random forecast model} This model provides a positive forecast with a set probability $p = \frac{\TP+\FP}{n}$, regardless of the input data. Since the forecast is uncorrelated with the actual event occurrence (and ignores any input features), the Accuracy and Rate Correct of this model are
\begin{eqnarray}
    \rm{Accuracy} & = & P(\rm{Forecast: Yes}) \times P(\rm{Event:Yes})\nonumber \\
    & = & p \frac{(\TP+\FN)}{n}.\\
    \rm{Rate~Correct} & = & P(\rm{Forecast: Yes}) \times P(\rm{Event:Yes}) \nonumber \\ 
    & & +P(\rm{Forecast: No}) \times P(\rm{Event: No}) \nonumber \\
    & = & p \frac{(\TP+\FN)}{n} + (1-p)\frac{(\TN+\FP)}{n}. 
\end{eqnarray}
\noindent The Accuracy and Rate Correct metrics depend on the chosen $p$, and the underlying event class imbalance.
\noindent In contrast, the True Skill Statistic (TSS) for a random forecast model is 
\begin{eqnarray}
\rm{TSS} &=& \rm{Recall} - \rm{False~Alarm~Rate}\nonumber \\
& = & P(\rm{Forecast:Yes} | \rm{Event: Yes} ) - P(\rm{Forecast:Yes} | \rm{Event: No} ) \nonumber\\
& = & p - p = 0,
\end{eqnarray}
\noindent which is true irrespective of the class imbalance and the forecast probability (p) chosen.

\noindent \textbf{Climatological forecast model} This is a special case of the random forecast model, 
with $p=P(\rm{Forecast: Yes}) = P(\rm{Event:Yes})$. Note the forecast probability used here is the 
event rate evaluated over the population (hence the name climatological). For example, the 0.01\% 
of calendar days has at least one X-flare, and the prediction task is to predict whether at least 
one X-flare occurred on a calendar day, $P(\rm{Event:Yes}) = 0.01$.  Suppose $p=0.01$ for the 
climatological model. If the testing set used for evaluating metrics has the same class imbalance 
as the population, $\mathrm{Accuracy} = p^2 = 0.0001$ and $\mathrm{Rate~Correct} = p^2+(1-p)^2 = 0.9802$. 
So in terms of the former, the climatological model appears dismal, and in terms of the latter, 
it performs spectacularly well. In contrast, $\rm{TSS}=0$, which illustrates why this is an unbiased metric. 

The Gilbert Skill Score and the Heidke Skill Score (v2; see Table~\ref{tab:binary_metrics}) are both metrics that are defined relative to the
random forecast model. They partially address the desire that we want to measure the marginal utility of a model against a baseline model. Nevertheless, they still suffer from dependence on class imbalance.

Our recommendation is to decouple metrics from models (as opposed to the Gilbert and Heidke Skill Scores). If possible (and desirable, depending on the stakeholder's needs), choose unbiased metrics like TSS. Then evaluate metrics for baseline and ML models alike to evaluate marginal utility (improved performance, if any). We caveat this recommendation by reiterating that the most relevant metric(s) always depends on the context and the stakeholder(s). 

The choice of appropriate baseline models depends on the application. For flare prediction, the climatological model is used as a reference point by NOAA \citep[e.g., see][]{Barnes:2016}. In some contexts where predictive models (where purely ML-baed and/or physics-based) are already in common use, the State-of-the-Art (SOTA) model may be appropriate. 

\subsubsection{Weakly-labeled supervised training}
The analysis and prediction of flares, especially when done with spectra, is an instance of weakly-labeled datasets. The observations of the IRIS satellite are of special relevance for the analysis of flares in recent times. Although each spectral observation constitutes
a fundamental unit of information, the label associated with the flare (flare/no flare)
cannot be put at the level of individual spectra but only at the time series level. 
\cite{HUWYLER2022100668} approached this classification problem by
using multiple instance learning \citep{DIETTERICH199731}, a supervised
learning technique that associates labels not to individual
instances but to bags of instances. They were able to detect the presence
of flaring regions with tens of minutes in advance from observation of the Mg \textsc{ii} window
with IRIS. These weakly-labeled techniques can also be of great help in segmentation
problems.

\subsubsection{Operational flare forecasting models}
\cite{Leka_2019a} provides a comprehensive review of operational flare forecasting models 
and, for the first time, a consistent comparison between flare models deployed at 
various international agencies and research institutions. While most models perform 
better than a no-skill baseline model, there was no single operational model that consistently 
outperformed others over a broad set of metrics and event distributions.

Further detailed analysis of the behaviors of operational flare models by 
\cite{Leka_2019b} and \cite{Park_2020} provides some important conclusions. Firstly, information 
regarding prior flare activity and active region evolution can improve forecasts. Secondly, 
having a human ``forecaster in the loop'' helps. Thirdly, performance degrades when data 
is restricted to near disk-center. Lastly, the use of ``modern data sources" (e.g., SDO/HMI) 
and statistical approaches improves performance. The data used for the comparison is 
available from \citet{DVN/HYP74O_2019}.

\subsubsection{Deep learning for flare prediction}
\label{sec:deep_flare}
The widespread availability of GPUs, deep learning frameworks and open-source computer 
vision codebases has supercharged the adoption of computer vision methods for flare 
forecasting. \cite{Huang_2018} applied CNNs to MDI and HMI line-of-sight magnetograms 
for flare forecasting. They find that the trained CNNs include intermediate spatial 
filters that are sensitive to magnetic polarity inversion lines. In contrast, some deep 
neural network flare prediction models use ``hand-engineered'' feature extraction \citep[e.g.,][]{Nishizuka_2018}. 
LSTMs have also been applied to flare prediction using 25 SHARP parameters, augmented by 
15 flare history parameters \citep{LiuLiu:2019}. Consistent with prior literature, this work 
shows the incorporation of the prior flare productivity improves prediction performance.

Given the success of deep neural models, \cite{2023ApJS..265...34Y} proposed to
train the CNN model proposed by \cite{Yi_2021} using deep reinforcement learning.
This model predicts the presence of a flare as a binary output from line-of-sight 
magnetograms. The results indicate that RL can improve the quality of the
prediction when compared to more standard training schemes, especially when dealing
with rare events.

Given that these deep neural models will eventually be part of operational flare forecasting 
strategies, it turns out important to check their biases. \cite{2022ApJ...941...20L} analyzed 
several deep neural models to look for the influence of the image resolution on
the prediction abilities. They found that the models analyzed are robust to the
specific image resolution. They pay more attention to global features extracted 
from the active regions, and pay less attention to local information in magnetograms. This
points out that these models will become operational soon.

\subsection{Explainable models for flare prediction}
Many of the deep learning models developed for flare prediction are complex.
Consequently, it is difficult to interrogate the models to understand the reasons why a model
predicts the presence of a flare. For this reason, the community has recently
relied on some of the techniques for explainability developed in machine learning 
in recent times \citep{BARREDOARRIETA202082}. 
\cite{Yi_2021} used Gradient-weighted Class Activation Mapping\footnote{\url{https://github.com/jacobgil/pytorch-grad-cam}}
\citep[Grad-CAM;][]{8237336} to localize
the regions of the solar surface that triggered the model to predict a flare. Grad-CAM can be
used with any CNN-based model. It works by computing the derivative of the prediction with 
respect to the final convolutional layer to produce a coarse importance map. This map
highlights the parts of the input images that trigger the detection of a flare in the neural
network. They find
that the model correctly focuses on the polarity inversion line to forecast a flare, a fact
that is well known. Likewise, \cite{2023A&A...671A..73P} used Grad-CAM and the
game-theoretic method expected gradients \citep{gabriel21} to discover features in
the spectra of Mg \textsc{ii} for predictions of flares. They found that
triplet emission, flows, broadening, and highly asymmetric spectra are features
that appear before a flare. Additionally, the regions to which the neural networks
pay more attention for the prediction are strongly associated with the location of the
maximum UV emission of the flare.

\subsection{Heliosphere and space weather}
This section focuses on applications of ML to heliospheric and space weather problems 
using solar data as inputs. For an overview of the role of ML in space weather studies 
and forecasting and a gentle introduction to ML tailored for the space weather audience, 
we refer the reader to \cite{2019arXiv190305192C}. Another recommended review paper 
is \cite{TenWays}, which lists ten broad categories of approaches to  applying ML 
to space science problems. We show some examples of applications, although
we encourage the reader to consult these review papers focused 
on many aspects of heliophysics beyond solar physics. 

\cite{2022SpWea..2002797T} used fully connected neural networks to predict the
solar energetic particles (SEP) above 10 cm$^{-2} s^{-1} sr^{-1}$ 
with energies above 10 MeV to occur from the properties of the coronal
mass ejection (CME). They found that the neural approach provides consistent
results, although they depend on the availability of observations of the CME. This
makes this method not sufficiently reliable, since SEPs can be present even if
no CME is seen. A similar approach was pursued by \cite{2021SoPh..296..107L}, although
they compared the neural approach with many different linear classifiers.

\cite{Upendran:2020} tackled the problem of solar wind speed prediction at Lagrangian 
point 1 (L1) by training a DNN which takes temporal sequences of SDO/AIA EUV images to 
predict the solar wind velocity as available in the OMNI database. This work made use of 
the technique of transfer learning (TL), whereby the frontend of the DNN was imposed as 
a set of pretrained layers from a well-known computer vision package (in this case, GoogleNet). 
This frontend acts as a preprocessor for feature extraction. The output latent vector 
is then passed on to the remaining trainable layers of the DNN. The SDO dataset 
used by \cite{Upendran:2020} comprised AIA images at a daily sampling frequency. 
By using instead a 30 min sample frequency, \cite{Brown:2022} reported significant 
improvements in evaluation metrics (e.g., root mean squared error) for the solar wind speed prediction. 
Another change they made was to use an attention-based mechanism, though the 
improvement of model performance is largely attributed to the much higher data sampling frequency. 

\cite{Bernoux:2022} trained a DNN to use SDO AIA images (193 \AA) as inputs 
to produce a probabilistic forecast of geomagnetic activity (\texttt{Kp} index). 
The model output is probabilistic in the sense that the output consists of a 
mean and standard deviation of \texttt{Kp}. Similar to the aforementioned work 
on solar wind speed prediction, this model uses TL and has a preprocessor feature extractor. 

\subsection{Solar Cycle predictions}
\label{subsection:SCpred}
It is well known that the Sun passes from a low magnetic activity (measured as the number
of visible sunspots on the disk) to a high magnetic activity with a periodicity of 
roughly 11 years. Over time, researchers have tried to find correlations between
the solar cycle and other observables. Among them, we find flares, CMEs, 
geoeffectiveness (Ap index measured at Earth), cosmic ray flux reaching the 
Earth's environment, and many more. The observations showed that the activity 
of the Sun is correlated with the amount of cosmic rays reaching the Earth \citep{2017LRSP...14....3U}. 
The sunspot number (SSN) displays a high correlation with the total solar 
irradiance (TSI)\footnote{TSI is defined as the radiant energy emitted by the Sun at all wavelengths 
crossing a square meter each second outside Earth's atmosphere \citep{2015LRSP...12....4H}}, which 
turns out to be an important parameter for understanding of the Earth's climate.
It also displays correlation with the occurrence of CMEs \citep{2019SSRv..215...39L}. The periods
of large activity in the Sun produce strong magnetic fields in the atmosphere, which are correlated 
with strong eruptive events. They can be hazardous for the Earth's environment.

In quest of a suitable solar cycle forecast method, \cite{2021SoPh..296...54N} analyzed 
77 predictions made by different research groups for cycle 24 and 37 predictions for the 
current cycle 25. Out of the 77 models, only a couple of models managed to properly predict 
the observed peak of the cycle. Interestingly, none of the methods based on machine learning 
models was able to correctly predict the amplitude of the cycle. 


Cycle 25 is not yet at its peak and the aim of some of the most recent methods based on ML is
the prediction of its maximum amplitude and when it will take place. 
\cite{2021RAA....21..184L} reported two methods employing an auto-regressive neural network 
\footnote{Autoregression implies predicting the future of a 
sequence using previously observed values.} method and a recurrent LSTM network. 
Using the same LSTM 
approach, and based on the SSN variation, three other predictions were reported. 
\cite{2022AdSpR..69..798P} predicted an increase of $\sim$20\% with respect to the previous 
cycle, with the peak in August 2023 and a maximum of SSN of 171.9 $\pm$ 3.4. 
\cite{Wang_2021} reported a decrease in the amplitude of cycle 25, with a predicted peak SSN 
of approximately 114 around 2023. Finally, \cite{2022MNRAS.515.5062B} forecasted a decrease 
of the peak amplitude of $\sim$14\% with respect to cycle 24, and a maximum activity 
peak on cycle 25 around mid-2024. 

\cite{2018SpWea..16.1424O} used a hybrid of regression and a neural network to provide
a prediction. The regression method is used to derive characteristics of the solar 
cycle, which was used afterward as input for a neural network. They predicted
a maximum amplitude for cycle 25 of 112.1$\pm$17.2, to happen in January 2025 ($\pm$6 months).
\begin{figure}[t]
\centering
\includegraphics[width=0.8\textwidth]{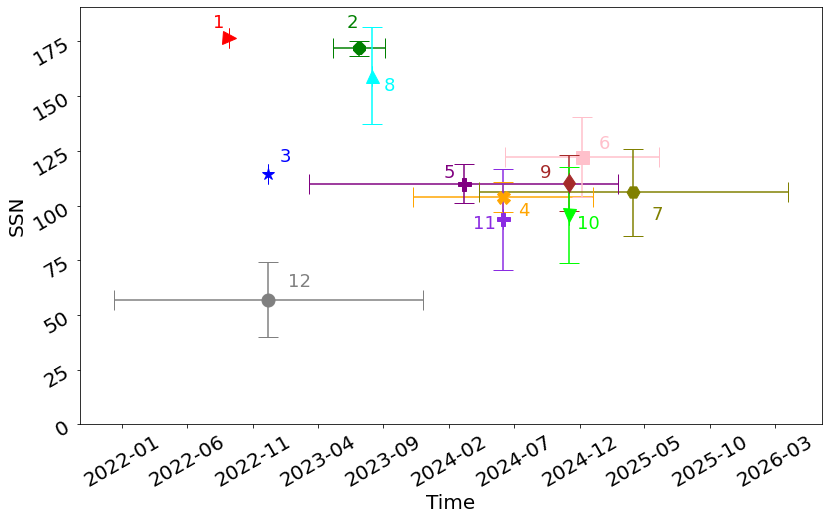}
\caption{Predictions of the solar cycle 25 peak time and SSN for different ML models: mark 1 
with red right-triangle \citep{2021RAA....21..184L}, mark 2 with green octagon \citep{2022AdSpR..69..798P}, 
mark 3 with blue star \citep{Wang_2021}, mark 4 with orange x and mark 5 with purple 
plus \citep{2022MNRAS.515.5062B}, mark 6 with pink square \citep{2018SpWea..16.1424O}, mark 7 
with olive hexagon \citep{2020SoPh..295...65B}, mark 8 with cyan triangle, mark 9 with brown 
diamond, mark 10 with lime triangle, mark 11 with blue-violet 
plus \citep{2019JPhCS1231a2022D}, and mark 12 with grey circle \citep{2019SoPh..294...24C}.}
\label{fig:CyclePred}
\end{figure}
\cite{2020SoPh..295...65B} predicted a weaker cycle 25 when compared with cycle 24, with a 
maximum SSN of 106$\pm$19.75. Their estimations are based on the WaveNet  
\citep{oord2016wavenet} and LSTM architectures. WaveNet is a DNN based on an 
autoregressive generative model. It learns to model the probability 
distribution of a given time-series conditioned on the past. To this end, it uses dilated 
causal convolutional layers \citep[see][for more details]{oord2016wavenet}, which
allows the model to capture time dependencies of very long baselines. They
predicted a peak SSN of 106$\pm$19.75 for cycle 25. Using four machine learning techniques, 
\cite{2019JPhCS1231a2022D} obtained four different predictions for the strength 
of cycle 25. Based on a feed-forward artificial neural network implementation, \cite{2019SoPh..294...24C} predicted the lowest amplitude for cycle 25. A linear regression predicts the maximum to occur in September 
2023 (with an amplitude of 159.4$\pm$22.3). A random forest (RF) and a radial basis function (RBF) 
method predicts the same time for peak, happening in December 2024 but with two different 
amplitudes: 110.2$\pm$12.8 for the RF and 95.5$\pm$21.9 for the RBF. Finally, a SVM method 
predicts a peak around July 2024 with a peak SSN of 93.7$\pm$23.2. All predictions, showing
the dispersion, are summarized in Fig.~\ref{fig:CyclePred}.

All the proposed methods show very good testing and prediction capabilities for the
past solar cycles. However, it is uncertain whether this is true in
future cycles. Predicting a nonlinear process like the solar cycle is a delicate
task, and it remains to be tested that the statistical properties of previous
solar cycles contain enough information to predict the future.

\subsection{Inversion of Stokes profiles}
\label{sec:inversion_stokes}
\subsubsection{Accelerating inversions}
The application of neural networks for the inversion of Stokes profiles goes back
in time to \cite{Carroll2001}, who proved that multi-layer FCN could be used
for estimating model parameters from the observations. 
\cite{Carroll2001} proposed their use for simple Milne-Eddington
inversions and concluded that they were able to obtain physical parameters
without any optimization once the neural networks were trained. As additional
advantages, they showed that the neural networks provided an increase
in speed, noise tolerance, and stability. This
was later verified by other works \citep{socas_nn_03,socas_nn_05}.
Inspired by these advances, \cite{asensio05} also applied neural networks for
the acceleration of the solution of chemical equilibrium. Solving chemical 
equilibrium with a large set of species turns out to be slow and can dominate
the computation time in inversion codes. That is precisely the reason why the inversion
code NICOLE \citep{socas_navarro_2015} has the neural solution as an option.

\cite{Carroll2008} later expanded their original work to use FCNs to
infer the depth stratification in a geometrical height scale of the temperature, velocity, and
magnetic field vector. The network was trained using stratifications and synthetic
Stokes profiles from an MHD simulation of the quiet Sun \citep{vogler05}.
The application of the neural network in a pixel-by-pixel manner allowed them to recover
a tomographic view of the FOV by recombining all individual line-of-sight stratifications. 

After an impasse of more than a decade, the neural inversion of Stokes profiles is
again gaining momentum, driven by modern DNN. \cite{asensio_sicon19}
proposed SICON\footnote{\url{https://github.com/aasensio/sicon}}, a CNN that is trained with MHD
simulations and opens up the possibility of
carrying out extremely fast inversions of 2D maps for observations of the Hinode satellite. 
As an example, a map of 512$\times$512 pixels
can be inverted in an off-the-shelf GPU in merely 200 ms. The authors proposed two
different architectures, both of them displaying consistent results. Apart from the
enormous speed of the inversion, the CNN's have other advantages. One of them is that the inferred
physical properties are not affected by the Hinode PSF, so it essentially deconvolves
the data while inverting. This can only be achieved for space-born observatories because
the PSF is well known and constant with time. Another advantage is that the networks can
provide estimations of quantities that are very difficult to obtain with classical 
inversion methods. This is the case of gas pressure and the Wilson depression. The main 
reason why these CNNs can do this job is because they exploit correlations in the training data.

\cite{Higgins_2021} also used a CNN (a U-Net in this case) to accelerate the production of vector
magnetograms from HMI/SDO. Instead of training with simulations, they trained the
CNN with inversions carried out with the standard pipeline. They also viewed the
inference as a classification problem with a large number of bins for each variable 
of interest, instead of a regression problem. Despite the inherent binning error, this 
allowed them to easily compute uncertainties in the output. \cite{Higgins_2022} expanded
their previous work by training the U-Net with inversions of the same field of view and at
the same time carried out with the Hinode/SOT-SP instrument. They developed SynthIA (Synthetic 
Inversion Approximation), which works under the assumption that the information 
encoded in the Hinode/SOT-SP observations (and the ensuing inversions) is also present in
HMI/SDO (potentially spread over multiplet pixels). This assumption is non-trivial
since HMI/SDO observes only the Fe \textsc{i} spectral line at 617.3 nm at low
spectral resolution, while Hinode/SOT-SP observes the pair of lines at 630 nm at
high spectral resolution. They showed that SynthIA can indeed
extract this information and produce full-disk inversions with a quality similar to
that of Hinode/SOT-SP. 
A similar approach has been pursued by \cite{Jiang:2022} 
to generate vector magnetograms for the Michelson Doppler Imager (MDI)
onboard the Solar and Heliospheric Observatory (SOHO). MDI/SOHO was observing
the Sun between 1996 and 2010, but it was only recording
the longitudinal component of the magnetic field. 
\cite{Jiang:2022} combined this information with
H$\alpha$ observations collected with the Big Bear Solar Observatory (BBSO) to
train a CNN to produce maps of the components of the magnetic field in the plane of the sky. The
trained CNN produces good vector magnetograms, extending the period in which vector
magnetograms are available for the Sun from 1996 to the present day. Despite the
success of these approaches, we caution that using 
data from different instruments should be done with care since small data alignment 
problems might affect the results \citep{Fouhey:2022}.

\cite{milic20} showed that a relatively simple 1D CNN can output temperatures, velocities, and 
magnetic fields at three optical depth heights in the atmosphere
directly from the Stokes profiles. They train the CNN with the aid of
MHD simulations of the quiet Sun. However, in order to more closely mimic standard 
inversion codes, they do not train directly with the data from the simulation. They
first invert the data with SNAPI \citep{snapi18} and use these results as a training
set. The output of the network shows a very good correlation with the original data while
accelerating the inversion by a factor $\sim10^5$.

Along a different line, we find studies of applying DNNs to provide
initial solutions to standard gradient-based inversion codes \citep{2021A&A...651A..31G}.
These methods can greatly accelerate the convergence of inversion codes because the
initialization is close to the expected solution. One of the sub-products of starting close
to the solution is that the Levenberg-Marquardt algorithm often used in
these inversion codes can be made to work close to the Gauss-Newton regime from the 
very beginning, which has an almost quadratic convergence rate.

The inversion of lines affected by departures from local thermodynamic equilibrium (non-LTE)
is computationally demanding. The reason is that one needs to self-consistently solve
the statistical equilibrium equations for the atomic/molecular species producing the observed
spectral lines and the radiative transfer equation \citep[see the review by][]{2017SSRv..210..109D}.
Accelerating this process has been recently tackled with two different approaches. The
first one uses CNNs \citep{2022A&A...658A.182C} to map the populations in LTE to 
the populations in non-LTE (the ratio between the populations in non-LTE and 
those in LTE is known as the departure coefficients) for a hydrogen model atom. Since
populations in LTE can be obtained from the local physical properties, the trained
mapping avoids the solution of the time consuming radiative transfer problem. Another
approach has been recently presented by \cite{2022ApJ...928..101V} based on graph
neural networks and specifically tailored to accelerate inversions of chromospheric
lines of Ca \textsc{ii}. This approach predicts the departure coefficients as a function of the height
in the atmosphere, producing a speedup of a factor $10^3$ without a significant impact
in the synthetic spectral lines. This allows inversions of chromospheric lines, even
those dominated by partial redistribution effects like Ca \textsc{ii} H \& K, to
be carried out as fast as lines formed in LTE.

\subsubsection{Uncertainty characterization}
\label{sec:uncertainty}
Since inversion problems are ill-defined in general, providing a single point estimate
of the physical parameters as output is not optimal. In principle, one should provide
full posterior distributions, which encode the uncertainties and
correlations among all model parameters \citep{2007A&A...476..959A}. A deep learning approach to this has
been pursued by \cite{osborne2019} based on the concept of invertible neural networks 
\cite[INNs;][]{Ardizzone2018}. The idea of INNs is to learn the
forward and inverse mappings simultaneously. The forward mapping, $y=f(x)$ goes from model parameters $x$ 
to observations. The inverse mapping, $x=g(y,z)$ is augmented with a latent
vector $z$ that is assumed to be extracted from a known distribution. This latent
vector takes into account all information lost during the forward
pass, which precisely makes the inverse problem ill-defined. Once the INN is trained, 
an approximation to the posterior distribution can be obtained by sampling the
latent vector. \cite{osborne2019} were able to derive temperatures, electron number
densities and velocities in flaring regions from the interpretation of the
H$\alpha$ and Ca \textsc{ii} 8542 \AA\ line using a RADYN model \citep{carlsson92,carlsson95,carlsson97,radyn15}.

Normalizing flows can also be utilized to characterize uncertainties. If the 
NF is conditioned on the observations, the normalizing flow can be 
trained to return Bayesian posterior probability estimates of the model
parameters for any arbitrary observation. This
amortized posterior estimation is time consuming to train but can then be
applied very fast to observations, opening up the possibility of 
doing Bayesian inference in large fields of view. \cite{2022A&A...659A.165D}
showed how this can be applied to the inversion of Fe \textsc{i} and Ca \textsc{ii} data 
and diagnose the stratification of the solar photosphere and chromosphere. They obtained
the most probable value of the temperature, bulk velocity, and microturbulent velocity in a
very large field of view, together with their uncertainties and correlations.

\subsection{3D reconstruction of the solar corona}
\cite{2023ApJ...948...21R} has recently shown that one can use GANs to build
a mapping from photospheric magnetograms to electron density maps
at different heights in the atmosphere. The model is trained with 
simulations from the Magnetohydrodynamic Algorithm outside a Sphere (MAS) method, that
solves the time-dependent resistive magnetohydrodynamic equations (MHD) in 3D, including 
coronal heating, thermal conduction and radiative losses \citep{Lionello_2008,2015ApJ...802..105R}.

Recent works have demonstrated the potential of using fully connected 
neural networks for the description of continuous fields (scalar, vector,\ldots)
as a function of the position in space 
\cite[e.g.,][]{mildenhall2020nerf}. To this end, neural 
networks, usually termed implicit neural
representations (INR), coordinate-based representations (CBR), or neural fields (NeF), 
are used to map  
coordinates on the space (or space-time) to coordinate-dependent field 
quantities. NeFs have
many desirable properties. They are very efficient in terms of the number of 
free parameters. They 
produce continuous and differentiable fields, which can then be seamlessly part of complex
models. Finally, they have a strong implicit bias, favoring specific signals.
An NeF is given by the following simple, but flexible, fully-connected neural network:
\begin{eqnarray}
     \log N_e(\mathbf{x}) &=& \phi_{n} \circ \phi_{n-1} \cdots \circ \phi_{0}(\mathbf{x}), \nonumber \\
     \phi_i &=& \sigma(\mathbf{W}_i \mathbf{x}_i + \mathbf{b}_i),
     \label{eq:INR}
\end{eqnarray}
where $\mathbf{W}_i$ are weight matrices, $\mathbf{b}_i$ are bias terms, and $\sigma$ 
is an activation function.

NeFs, as defined by Eq. (\ref{eq:INR}), are known to suffer from the so-called 
spectral bias 
\citep{pmlr-v97-rahaman19a,WANG2021113938}, which prevents them from learning 
high-frequency functions.
This problem has been empirically alleviated by first passing the input 
coordinates through a Fourier
feature mapping, which allows the INR to correctly generate high spatial 
frequencies \citep{tancik2020fourier}. Recently,
\cite{sitzmann2020implicit} proposed \texttt{SIREN}s\footnote{\url{https://github.com/vsitzmann/siren}}, which uses periodic 
functions (sines) as activation 
functions so that the electron density can be written as:
\begin{eqnarray}
     \log N_e(\mathbf{x}) = S_\psi(\mathbf{x}) &=& \mathbf{W}_n (\phi_{n-1} \circ \phi_{n-2} \cdots \circ \phi_{0}) 
     + \mathbf{b}_n \nonumber \\
     \phi_{i} &=& \sin \left( \omega_i \left( \mathbf{W}_i \mathbf{x}_i + \mathbf{b}_i \right) \right),
     \label{eq:electron_density}
\end{eqnarray}
where $\psi$ summarize all tunable parameters of the \texttt{SIREN}.
Thanks to a specific initialization procedure, a \texttt{SIREN} can efficiently 
reproduce both low and
high spatial frequencies. In some sense, a \texttt{SIREN} can be seen
as a nonlinear extension of a Fourier series.

NeFs have been recently introduced in solar physics by \cite{PPR:PPR478314}
for the description of the magnetic field in the solar corona. 
\cite{PPR:PPR478314} carry out the extrapolation
of the photospheric magnetic field by describing it with a NeF. They
optimize the neural network by imposing the force-free and the
solenoidal conditions:
\begin{eqnarray}
     L_\mathrm{ff} &=& \frac{\Vert (\nabla \times \mathbf{B}) \times \mathbf{B} \Vert^2}{\Vert \mathbf{B} \Vert^2 + \epsilon} \nonumber \\
     L_\mathrm{div} &=& \Vert \nabla \mathbf{B} \Vert^2.
\end{eqnarray}
The spatial derivatives are easy to compute for a NeF using automatic
differentiation. These losses are optimized by simultaneously fulfilling the
boundary conditions. 

Later, \cite{bintsi2022sunerf} used NeFs to show that 
it is possible to infer the emission properties in the whole 3D corona 
from a set of observations from the ecliptic (with latitudes below
7$^\circ$). To this end, NeFs are used
to describe the local emission properties in the 3D volume. The training requires
accumulating the emission along rays using ray tracing and optimizing
a loss that compares the synthetic images and AIA/SDO observations at 193 \AA.
The simulations demonstrate that 32 observations are enough to obtain
a very accurate description of the corona, even in the polar regions. Extending 
the procedure to infer physical properties like
temperature and electron density (from which the local emission properties
are computed) is just one step ahead.

The force-free extrapolation of magnetic fields in the corona, especially
above photospheric magnetic field concentrations, improves significantly
if one has additional information about the 3D geometrical structure of
coronal loops observed in the UV. A reliable inference of this 3D structure requires the use of triangulation techniques 
with the aid of stereoscopic observations. Since having these observations is not the
case in many occasions, \cite{Chifu_2021} used a CNN to extract the Z component of the loop 
based only on the 2D shape extracted from a EUV single image. The model obtained a very high 
accuracy for short loops with no complex shapes and lower performance in very complex and twisted shapes. 


\subsection{Image deconvolution}
Observing any other astronomical object through the Earth's atmosphere
introduces perturbations that are difficult to correct. The obvious solution
of moving to space is not always possible or feasible. 
Even if adaptive optics systems are working properly, some residual wavefront perturbations are
still present in the images, and the diffraction limit of the telescope is
not reached. A posteriori correction techniques based on phase diversity 
\citep{paxman92,lofdahl_scharmer94,lofdahl98}
and multi-object multi-frame blind deconvolution \citep[MOMFBD;][]{lofdahl02,vannoort05}
have been developed. The main disadvantage of these methods is their
large computational requirements. For this reason, deep learning has been
applied recently by \cite{asensio18} to accelerate the deconvolution process. The method is based 
on a fully convolutional deep neural network that
was trained supervisedly with images previously corrected with the help of
MOMFBD. Once trained, this method can deconvolve bursts of 1k$\times$1k images
containing 7 short-exposure images in $\sim$ 5 ms with an appropriate GPU. This 
opens up the possibility of, for instance, doing the
deconvolution online while analyzing the data\footnote{\url{https://github.com/aasensio/learned_mfbd}}.

Although a step forward in terms of speed, 
the neural approach developed by \cite{asensio18} has two main problems. The
first one is that it is trained with supervision, so one needs to use the
MOMFBD algorithm to build the training set. Though not a major
obstacle, a method that does not need this previous step would be preferable.
The second issue is that it only produces deconvolved images. No estimation of the wavefront in each individual frame
is produced. Estimating the wavefronts can be helpful in checking the performance
of the telescope and instrument and understanding the performance of the
adaptive optics. For this reason, \cite{2021A&A...646A.100A} improved the approach by showing
how the training can be done in a fully unsupervised manner, while also
producing an estimation of the wavefront for each observed frame. Given the
lack of supervision, the method can be generally applied to any type of object, 
once a sufficient amount of training data is available\footnote{\url{https://github.com/aasensio/unsupervisedMFBD}}.


\subsection{Image-to-image models}
Arguably the most powerful property of deep learning models is their ability
to deal with high-dimensionality data like images. Models are powerful enough
to produce very high-resolution natural images. This fact has also been exploited
in solar physics in different applications that we summarize in the following.

\subsubsection{Synthetic generation of solar data}
The multi-wavelength and multi-layer coverage of the solar atmosphere by SDO 
instruments provide opportunities to explore the synthetic generation of solar data. 
In this context, synthetic data generation includes the translation of data from one 
instrument into proxy data for another instrument (or even the same instrument), as 
well as the generation of data that follow the underlying distribution of real observed 
data, but which is not necessarily instantiated on the real Sun. 

One example of data translation is the generation of proxy EUV spectral irradiance 
data. Using the SDOML dataset \citep{Galvez2019}, \cite{szenicer19} trained a CNN 
to translate multi-channel AIA images into disk-integrated line (and band) irradiance 
data observed by EVE. This was done by using data captured by both instruments before mid 
2014, when EVE MEGS-A was still operational. The errors from the CNN model prediction 
are smaller than from a physics model based on differential emission measure (DEM) 
inversions. After the model was trained, a rearrangement of the network components 
allowed for the generation of synthetic line irradiance images. 

\begin{figure}
    \centering
    \includegraphics[width=\textwidth]{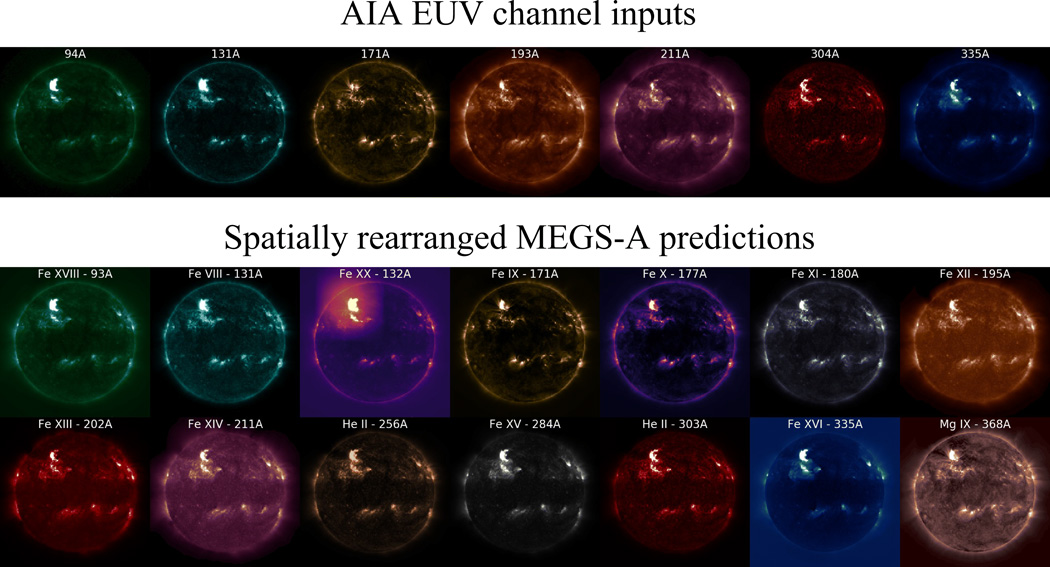}
    \caption{Synthetic EUV line irradiance images generated by a CNN trained to map AIA images to EVE disk-integrated line irradiance data \citep[from][]{szenicer19}.}
    \label{fig:AIA2EVE}
\end{figure}

Conditional Generative Adversarial Networks (cGANs) have been trained to generate 
synthetic magnetograms from EUV/UV images \citep[][]{Kim2019}. Theoretical 
considerations would suggest EUV/UV intensity data would not encode magnetic global 
polarity information \citep{Liu2021}. Specifically, the expectation is that the 
thermodynamic structure of the solar atmosphere is symmetric under the operation 
$\mathbf{B} \to -\mathbf{B}$. However, the synthetic magnetograms from \cite{Kim2019} 
do have bipolar active regions that resemble real active regions with Hale polarity 
rules consistent with solar cycle 24. Detailed comparisons with actual observed active 
regions reveal big differences in the morphology. So these synthetically generated 
magnetograms would not be useful for AR-scale studies. Whether they are suitable for 
use for downstream heliospheric predictions remains to be seen.

Another image translation problem is the synthetic generation of EUV images from other 
EUV images of different wavelengths. This problem was posed by \citet{Salvatelli2019} 
in the context of potentially reducing the number of physical channels needed in future 
EUV telescopes. The approach was to use three AIA input channels to generate another 
AIA channel using U-Nets. This problem was further extensively explored by \cite{Lim2021} using cGANs, 
who considered image translation from single, double, and triple input channels with 
cross-correlation (CC) coefficient between prediction and ground truth as the performance 
metric. \citet{Salvatelli2022} further explored the problem by considering other metrics, 
including commonly used computer vision metrics like structural similarity index 
measures. \citet{Salvatelli2022} showed that the CC metric may not be the ideal 
performance metric, and also showed how various metrics degraded when the trained 
model was applied during flaring conditions. 

\subsubsection{Estimation of velocities}
Motions in the solar photosphere are fundamentally controlled by convection in
a magnetized plasma. Remotely sensing these three-dimensional velocities is important for the
analysis of solar events. The component along the line of sight (LOS) of the velocity
can be extracted from spectroscopic observations thanks to the Doppler
effect. However, the components of the velocity field in the plane perpendicular
to the LOS cannot be diagnosed spectroscopically.
Different algorithms have been used to trace horizontal flows at the solar surface 
by estimating the optical flow from consecutive images. The most widespread
is the method of local correlation tracking \citep[LCT;][]{NovemberSimon_1988}. This method suffers
from problems when dealing with events of a short time duration or with reduced
physical size. To alleviate this, \cite{Asensio2017} developed DeepVel\footnote{\url{https://github.com/aasensio/deepvel}}, an end-to-end deep 
learning approach for the estimation of horizontal 
velocity fields in the solar atmosphere based on a deep fully convolutional neural network. The neural 
network was trained on a set of velocity fields obtained from simulations of 
the quiet Sun. DeepVel is very fast, uses only two consecutive frames 
, and returns the velocity field in every pixel and for every time step. 
DeepVel opened up the possibility of identifying
small-scale vortices in the solar atmosphere that last for a few minutes and with
sizes of the order of a few hundred kilometers, something
impossible with methods based on local correlation tracking

DeepVel was later retrained by \cite{tremblay18} to carry out an exhaustive comparison
with classical local correlation methods and check their ability in extracting transverse plasma
motions at a large scale from SDO/HMI observations. They concluded that DeepVel was
able to beat classical methods by a large margin in small scales, those
of the granulation while being very similar to local correlation methods
in larger scales. It is encouraging that, when applied to simulations, DeepVel is able to nicely
recover the kinetic energy density from the simulation.

A new model, DeepVelU\footnote{\url{https://github.com/tremblaybenoit/DeepVel_DeepVelU}}, based on the U-Net architecture, has been proposed by 
\cite{tremblay20_unet}. This model displays several improvements with respect
to the original DeepVel network. The U-Nets analyze
the inputs in a multiscale fashion, which turns out to be interesting to capture horizontal
velocities at different scales, from granular to the supergranular scales. Additionally, 
\cite{tremblay20_unet} trained DeepVelU in simulations of the quiet Sun and
active regions. They checked that DeepVelU is able to capture the transverse velocities
from simulations with much improved correlation, especially when dealing with
large spatial scales.

\subsubsection{Superresolution}
Instruments are limited by optics to provide a certain spatial resolution on the solar
surface. However, the recent field of research on compressed sensing, which is 
founded on the idea of sparsity and compressibility, has demonstrated that one
can enhance the spatial resolution of images under certain conditions. It is clear
that the presence of spatial correlation in the images of the Sun suggests that one can
enhance current observations to provide a certain degree of superresolution. 
\cite{Enhance18} proposed \texttt{Enhance}\footnote{\url{https://github.com/cdiazbas/enhance}}, 
a deep CNN that provides superresolved
continuum and magnetograms for SDO/HMI. The nominal pixel size of HMI of 0.5$''$ is
transformed into 0.25$''$. These images are compared, as a cross-check, with images obtained
from the Hinode satellite (correctly degraded to provide a resolution of 0.25$''$ per
pixel). The superresolved images provide a very good representation of the small 
scales, enhancing the contrast in the continuum in the quiet Sun by almost a factor 2.
Magnetograms are also properly superresolved although the fact that this is a
signed quantity can produce small artifacts. All-in-all, \texttt{Enhance} is a
very good tool to provide a better picture of the environment around regions
of interest.

More recently, \cite{Dou_2022}\footnote{\url{https://github.com/dfpdl/SPSR}} have used generative 
adversarial network (GAN) to produce high-fidelity and photorealistic super-resolved 
images of Michelson Doppler Imager (MDI) in order to match the Helioseismic and Magnetic 
Imager (HMI) resolution. First, a GAN model is designed to downscale the HMI data to MDI 
resolution to transfer the characterization of the HMI data to the MDI scale. Then a second 
supervised GAN model was developed to produce the superresolved magnetograms based on the MDI data.
We caution the reader to be very critical when using superresolved data for data
analysis since the presence of artifacts and ambiguities can surely affect the
physics inferred from them.

\subsubsection{Denoising}
Although exquisite detail is put on the design of the instruments developed to observe the
Sun, they are always affected by noise. A posteriori methods can be used to denoise the data
by exploting the regularity of the solar structures, helped by the fact that noise
has reduced spatial and temporal correlation. As already discussed in Sect.~\ref{sec:pca_denoising}, 
linear methods have been used with this purpose. However, new, more powerful 
nonlinear methods are appearing in the literature, and they are being used for 
denoising different solar observations. In particular, \cite{diaz_baso19}
got inspiration from the Noise2Noise approach of \cite{lehtinen18}. This is a
method that supervisedly trains a relatively simple denoising neural network by only having pairs of
the same solar structure with two different realizations of the noise. In
contrast, the standard supervised approach needs pairs of noisy and clean images, which are
only possible using synthetic data. It is obvious that obtaining training examples 
for the Noise2Noise approach is much easier than for the standard supervised case. This was
indeed demonstrated by \cite{diaz_baso19}, who used pairs of images taken with the 
CRISP instrument mounted on the Solar Swedish Telescope (SST) at the same wavelength 
but at different times, making sure that the
time separation was small. The denoising results are great, with special relevance
on filterpolarimeters like CRISP, which show conspicuous (preferentially when analyzing
polarimetric signals) systematic artifacts on the observed field, produced either by the
instrument or by the data reduction process.

Later, \cite{park20} also approached the denoising problem of SDO/HMI solar
magnetograms using a deep convolutional conditional GAN, leading to a reduction in the average
noise level of more than a factor 2.5. The GAN is trained so that it maps single 
magnetograms to the average of the 21 magnetograms centered on the one of interest
(including 10 before and 10 after). Potentially, this could lead to a reduction in 
the noise standard deviation of a factor $\sim 4.6$, although it can also lead to
a slight blurriness of the generated images produced by motions in the solar
surface. The generator network is conditioned on the noisy
magnetogram and its purpose is to produce a denoised version of the magnetogram. Following
the standard GAN paradigm, a second discriminator network is in charge of distinguishing the
magnetogram produced by the generator and the real ones from the training set. The equilibrium
of the two networks is produced when the generator produces images indistinguishable from
the training set so that the discriminator is fooled roughly 50\% of the time.

\subsubsection{Image desaturation}
A very interesting application of deep learning is the desaturation of SDO/AIA data. These
synoptic observations frequently suffer from saturation effects mainly as a consequence
of the occurrence of solar flares. Correcting the saturated regions of the image
is an instance of image inpainting. The aim is to fill the (irregular) holes by leveraging
statistical information from the rest of the image and from the training set. 
To this end, \cite{yu2022} developed a model using a GAN. The 
generator is based on a U-Net that uses partial convolution layers \citep{9903574} instead
of standard convolutional layers. These partial convolution layers are specifically suited
for inpainting tasks. The discriminator is based on a PatchGAN architecture \citep{isola2017image,9318551}.
The results show a promising avenue to provide continuous synoptic observations 
even when energetic events happen in the Sun.

\subsubsection{Farside imaging}
\label{sec:farside_imaging}
Predictions of the active regions currently on the hidden side of the Sun (known 
as farside) are routinely computed using helioseismic 
measurements. They are obtained by solving the inversion problem known as
helioseismic holography \citep{1997ApJ...485..895L}, which uses time series
of waves on the visible surface (nearside) and map them back to the far side. Given the 
dispersive character of the mapping between the nearside and the farside,
the resulting images are quite diffuse. Machine learning has great potential
for the improvement of these inversions. \cite{kim19} trained generative
models to produce farside magnetograms from STEREO extreme ultraviolet (EUV)
images. Since the polarity of the magnetic field is not directly encoded on the
EUV images, it is noteworthy that the correct polarity can be recovered. 
\cite{felipe19} gave the more conservative step
of proposing a CNN (FarNet) that associates the farside maps obtained with helioseismic 
holography with probability maps obtained from magnetograms acquired half a 
rotation later. As a consequence, the aim is to estimate the presence of active
regions, neglecting the polarity, with nearside data.
The neural approach is able to detect much weaker active regions
than those that are detected with the standard technique. Improvements on this approach
will probably require deep architectures directly trained with Doppler maps.
Later, \cite{2021A&A...652A.132B} analyzed the statistical properties of 
FarNet and concluded that for equivalent false positive ratios when compared
with the standard method, it produces $\sim$47\% more true detections. Additionally,
it is able to detect much weaker active regions.
A significant improvement (FarNet-II) was also recently published by 
\cite{2022A&A...667A.132B}, by including attention mechanisms 
and convolutional recurrent layers based on the ConvLSTM approach 
\citep{NIPS2015_07563a3f}. Using temporal information provides a much
improved time consistency of the predicted active regions, also allowing
for a better prediction in the case of weak active regions\footnote{\url{https://github.com/EBroock/FarNet-II}}.

\section{Outlook for the future}
Machine learning has been routinely used in solar physics. However, the recent
deep learning revolution is producing a panoply of new applications that
were never envisioned a few years back, permeating in many subfields of research
inside solar physics. The availability of increasingly larger observational material
is making solar physics transition to the powerful collection of methods that advanced
ML offers to help us understand what we see. We frankly think ML will become an intrinsic
part of our research in the future.

Currently, many applications in solar physics consider the ML model as a very convenient way of 
parameterizing a very flexible mapping that carries out the inverse problem directly. 
This is interesting because we need to accelerate certain complex
operations that cannot be carried out otherwise, especially with the current and
future solar telescopes. However, we are also starting to witness a
huge revolution in solar physics in which deep learning models are informed
with physical models. A good synergy can be obtained if the physical laws
that we currently use to interpret our observations are used together with
neural networks to approximate the most complex parts of the process. We will
surely see new methods for the inversion of the Stokes profiles, new methods
for the extrapolation of magnetic fields, new methods to accelerate MHD simulations,
new methods to understand synoptic observations, and many more. All of them will use deep
learning as a key ingredient.

\begin{appendices}
\section{Glossary}
    \begin{itemize}        
        \item ML: machine learning
        \item GPU: graphical processing unit
        \item TPU: tensor processing unit
        \item RL: reinforcement learning
        \item PCA: principal component analysis
        \item SVD: singular value decomposition
        \item SPoCA: spatial possibilistic clustering algorithm
        \item FCM: fuzzy C-means (FCM)
        \item PCM: possibilistic C-means
        \item DBSCAN: density-based spatial clustering of applications with noise
        \item SVM: support vector machine
        \item RVM: relevance vector machine
        \item CS: compressed sensing
        \item ANN: artificial neural network
        \item CNN: convolutional neural network
        \item RNN: recurrent neural network
        \item LSTM: long short-term memory
        \item FCN: fully connected network
        \item ReLU: rectified linear unit
        \item ELU: exponential linear unit
        \item GD: gradient descent
        \item SGD: stochastic gradient descent
        \item LLE: locally linear embedding
        \item SOM: self-organizing map
        \item t-SNE: Student-t Stochastic Neighbor Embedding
        \item AE: autoencoder
        \item VAE: variational autoencoder
        \item AANN: autoassociative neural network
        \item GAN: generative adversarial network
        \item NF: normalizing flow
        \item DDPM: denoising diffusion probabilistic model
        \item DNN: deep neural network
        \item TL: transfer learning
        \item RBF: radial basis function
        \item MLP: multilayer perceptron
        \item INN: invertible neural network
        \item INR: implicit neural representation
        \item CBR: coordinate-based representation
        \item NeF: neural field
        \item MOMFBD: multi-object multi-frame blind deconvolution
        \item LCT: local correlation tracking        
    \end{itemize}
\end{appendices}

\begin{acknowledgements}
AAR acknowledges financial support from 
the Spanish Ministerio de Ciencia, Innovaci\'on y Universidades through project PGC2018-102108-B-I00 
and FEDER funds. 
 I.C. acknowledges the support of the Coronagraphic German and US Solar Probe Plus Survey (CGAUSS) project for WISPR by the German Aerospace Center (DLR) under grant 50OL1901.
This research has made use of NASA's Astrophysics Data System Bibliographic Services.
R.G. acknowledges to Funda\c{c}\~{a}o para a Ci\^encia e a Tecnologia (FCT) the support through the
research grants UIDB/04434/2020 and UIDP/04434/2020.

\end{acknowledgements}

\phantomsection
\addcontentsline{toc}{section}{References}
\bibliographystyle{spbasic-FS}      

%
%

\end{document}